\documentclass[twocolumn]{aastex61}
\newcommand\uJy{$\mu$Jy}
\newcommand\Fint{F_{\mathrm{int}}}
\newcommand\epse{\epsilon_e}
\newcommand\epsu{\epsilon^u_b}
\newcommand\uG{\mu\mathrm{G}}

\def\ha{{H$\alpha$}}

\def\LUM{\:{\rm ergs\:s^{-1}}}

\def\VEL{\:{\rm km\:s^{-1}}}

\def\sii{[\ion{S}{2}]}

\def\oiii{[\ion{O}{3}]}

\def\hii{\ion{H}{2}}

\newcommand{\MSOL}{\mbox{$\:M_{\sun}$}}

\newcommand{\EXPU}[3]{\mbox{\rm $#1 \times 10^{#2} \rm\:#3$}}  % exponent with units
\newcommand{\POW}[2]{\mbox{$\rm10^{#1}\rm\:#2$}}

\newcommand{\chandra}{{\em Chandra}}
\newcommand{\xmm}{{\em XMM-Newton}}

\usepackage[colorinlistoftodos]{todonotes}
\usepackage{amsmath}
\usepackage{pbox}

\newcommand\pb[2]{\pbox{#1}{\strut #2\strut}}

\def\changed#1{#1}

\defcitealias{gordon99}{G99}
\defcitealias{long10}{L10}
\defcitealias{lee14}{LL14}
\defcitealias{long18}{L18}

\journalinfo{Accepted for publication in the Astrophysical Journal}

\begin{document}


\title{A New, Deep JVLA Radio Survey of M33}

%% Use \author, \affil, plus the \and command to format author and affiliation 
%% information.  If done correctly the peer review system will be able to
%% automatically put the author and affiliation information from the manuscript
%% and save the corresponding author the trouble of entering it by hand.
%%
%% The \affil should be used to document primary affiliations and the
%% \altaffil should be used for secondary affiliations, titles, or email.

%% Authors with the same affiliation can be grouped in a single
\correspondingauthor{Richard L. White}
\email{rlw@stsci.edu}

\author[0000-0002-9194-2807]{Richard L. White}

\affil{Space Telescope Science Institute,
3700 San Martin Drive,
Baltimore MD 21218, USA; rlw@stsci.edu, long@stsci.edu}

\author[0000-0002-4134-864X]{Knox S. Long}
\affil{Space Telescope Science Institute,
3700 San Martin Drive,
Baltimore MD 21218, USA; rlw@stsci.edu, long@stsci.edu}

\affil{Eureka Scientific, Inc.
2452 Delmer Street, Suite 100,
Oakland, CA 94602-3017}

\author{Robert H. Becker}
\affiliation{Physics Department, University of California, Davis, CA 95616; 
bob@physics.ucdavis.edu}

\author[0000-0003-2379-6518]{William P. Blair}
\affiliation{The Henry A. Rowland Department of Physics and Astronomy, 
Johns Hopkins University, 3400 N. Charles Street, Baltimore, MD, 21218; 
wblair@jhu.edu}

\author[0000-0002-6822-4823]{David J. Helfand}
\affiliation{Columbia Astrophysics Laboratory, 550 West 120th Street, New York, NY 10027; djh@astro.columbia.edu}

\author[0000-0001-6311-277X]{P. Frank Winkler}
\affiliation{Department of Physics, Middlebury College, Middlebury, VT, 05753; 
winkler@middlebury.edu}

\begin{abstract}

We have performed new 1.4~GHz and 5~GHz observations of the Local Group galaxy M33 with the
Jansky Very Large Array. Our survey has a limiting sensitivity of 20~\uJy\
($4\sigma$) and a resolution of 5.9\arcsec\ (FWHM),
corresponding to a spatial resolution of 24 pc at 817 kpc. Using a new multi-resolution algorithm, we have created a catalog of 2875 sources, including 675 with well-determined spectral indices.  We detect sources at the position of 319 of the X-ray sources in the T{\"u}llmann et al.\  (2011) \chandra\ survey of M33, the majority of which are likely to be background galaxies.  The radio source coincident with M33 X\nobreakdash-8, the nuclear source, appears to be extended.
Along with numerous \hii\ regions or portions of \hii\ region complexes,
we detect 155 of the 217
optical supernova remnants included in the lists of Long et al.\ (2010) and Lee \& Lee (2014), making this by far the largest \changed{sample} of remnants at known distances with \changed{multiwavelength} coverage. 
The remnants show a large dispersion in the ratio of radio to X-ray luminosity at
a given diameter, a result that challenges
the current generation of models for synchrotron radiation evolution in supernova remnants. 

\end{abstract}




\section{Introduction} \label{sec:intro}

Supernova remnants (SNRs), the long-lasting products of supernovae (SNe), are the working surfaces
at which SNe chemically enrich, and deposit kinetic energy into, the surrounding circumstellar and
interstellar medium.  In extreme cases, the collective action of many SNe can even drive outflows in
starburst galaxies.  SNRs are also almost certainly the dominant source of cosmic rays in a galaxy.
As such, SNRs are central to our understanding of the life cycle of stars and the evolution of galaxies,
as well as for the composition and dynamics of the interstellar and intergalactic
media.

Observations of external galaxies offer the best way to study the class properties of SNRs, since
all of the SNRs are effectively at the same distance and problems associated with dust extinction in
the plane of our Galaxy are minimized.  For example, only about 30\% of the 294 Galactic SNRs in
the catalog by \cite{green14} have been detected at optical wavelengths and just 40\% have been
detected in X-rays, primarily due to limitations imposed by line-of-sight dust and gas extinction.
Also, the distances to individual Galactic SNRs are often highly uncertain. It has thus
been difficult to systematically study unbiased samples of Galactic SNRs at all wavelengths.

Most Galactic SNRs were first discovered as spatially extended, non-thermal radio sources, and the
first studies of the class properties of SNRs were largely based on radio observations alone (see,
e.g., Green 2009 and references therein).  In contrast, most extragalactic SNRs have been identified
from deep optical interference filter imagery, where the ratio of \sii\ $\lambda\lambda$ 6717,6731
to \ha\ provides an effective discriminant between \hii\ regions (\sii:\ha\ $\sim$ 0.1) and SNRs
(\sii:\ha\ $\gtrsim 0.4$), at least for the brighter objects \citep{long18}.  Except in the
Magellanic Clouds, radio and X-ray instrumentation has not hitherto had the appropriate combination of
spatial resolution and sensitivity to detect large numbers of  extragalactic SNRs, let alone
discover new ones.

M33 is \changed{arguably} the best external galaxy in which to study SNRs because it is nearby
\citep[$817\pm58$ kpc, so 1\arcsec\ = 4 pc,][]{freedman01}, relatively face on
\cite[$i=56\pm1\degr$,][]{zaritsky89}, and has relatively low Galactic foreground absorption
\cite[$N_{\mathrm{H}} = \EXPU{6}{20}{cm^{-2}}$,][]{dickey90,stark92}.  
\changed{While the Large and Small Magellanic Clouds are both $\sim 15$ times closer than M33
\citep[50 kpc and 60 kpc, respectively,][]{pietrzynski13, hilditch05} and have similarly low
foreground absorption, the SNR samples in the LMC and SMC are far more limited than that in M33.
Furthermore,} M33 has  been well-studied across the electromagnetic spectrum, providing a wealth of supporting
information.\footnote{M31, which is at a similar distance, is observed at a lower inclination and
lies along a line of sight with more foreground absorption. It is also less well studied, at least
in its entirety, because of its large angular size.}

The first three optical SNR candidates in M33 were identified forty years ago by \cite{dodorico78}, and the numbers
have grown significantly since then, especially after CCDs made interference filter imaging searches
for SNRs more efficient \citep{dodorico80,long90,gordon98,long10}.  The two most recent optical
surveys were carried out using data originally obtained as part of the Local Group Galaxy Survey
\citep[LGGS,][]{massey06,massey07}. Surveys  by \citet[][hereafter L10]{long10} and by
\citet[][hereafter LL14]{lee14} found 137 and 199 optical SNRs or candidates, respectively.  After
overlaps between the two lists are eliminated, there are 217 unique SNR candidates in M33
\citep[][hereafter L18]{long18}.  

The original selection of these nebulae as candidates was made on
the basis of elevated \sii:\ha\ ratios in emission line imagery, but imaging surveys have
limitations due to the potential for contamination of the \ha\ filter image by varying amounts of
[N~II] $\lambda\lambda$6548,6584 emission.  Most of the M33 SNRs have now been shown to have bona
fide high \sii:\ha\ ratios via optical spectroscopy, which resolves the [N~II]--\ha\ complex and
allows a clean \ha\ measurement \citepalias{long18}. Finally, of the 217 SNR candidate nebulae, 112 have been
detected in X-rays, either using \chandra\ \citepalias{long10} and/or {\it XMM-Newton} \citep{garofali17}.

Early radio observations of M33 were limited by the resolution and sensitivity available from
single-dish telescopes \citep{dennison75}, but with the advent of powerful interferometers it became
possible to detect individual objects within the galaxy.  \cite{dodorico82}  were the first to
detect M33 optical SNRs at radio wavelengths; they concluded that the luminosities of SNRs in M33
were similar to those in the Galaxy.  A more detailed study of the radio SNRs in M33 was carried out
by \citet[][hereafter G99]{gordon99} using a combination of the Westerbork Synthesis Radio Telescope
and the (original) Very Large Array at 5~GHz and 1.4~GHz.  They used their maps, which had a
resolution of about 7\arcsec, to construct a catalog of 186 sources in M33, including 53 that they
identified as spatially coincident with one of the 98 optically identified SNRs known at that time
\citep{long90}.  The mean radio spectral index of the radio sources identified as SNRs was $-$0.5,
and the summed radio luminosity of SNRs in M33 comprised 2-3\% of the total synchrotron emission
from the direction of M33. There were a number of other non-thermal sources detected above their
flux density limit of 0.2~mJy along the line-of-sight to M33, but they concluded most of these
were likely background sources.

\begin{figure*}
	\centering
	\includegraphics[height=0.7\textheight]{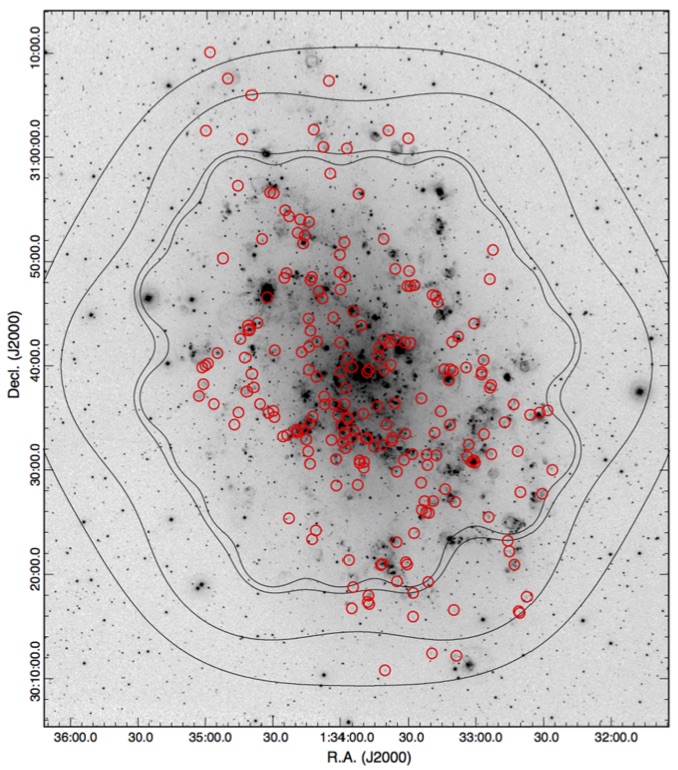}
	\caption{Image of M33 in H$\alpha$, from the 0.6m Burrell Schmidt telescope at Kitt Peak, with the
	locations of 217 SNRs and SNR candidates identified by \cite{long10} and/or \cite{lee14}
	indicated. See \cite{long18} for an optical spectroscopic assessment of
	many of these candidates.   The contours show, from the inside to the outside, the sky
	regions covered by the high- and low-frequency 5~GHz bands and by the high- and low-frequency
	1.4~GHz bands. Most of the SNRs are covered by both frequencies, while only a single SNR
	falls completely outside our radio coverage.
\label{fig:overview}}
\end{figure*}

It has been twenty years since the \citetalias{gordon99} radio survey of M33 was published.  In this
paper, we describe the results of a new, deep, multifrequency radio survey of M33 carried out with
the Jansky Very Large Array (JVLA).  Our primary purpose is to identify radio SNRs in M33 %ideally
in order to better understand the radio properties of SNRs as a class, but of
course many other sources have also been detected, including thermal \hii\ regions, X-ray binaries, background
sources, and even diffuse thermal emission from the ISM of M33.  

We provide an overview of the radio
observations and data processing in the next section \changed{(deferring the full details of the new algorithms
used for source detection and flux density measurements until Appendix~\ref{sec:multires})}. This is followed in
section 3 by an analysis of the detected radio source populations relative to optical and X-ray
source catalogs and the SNR catalogs in particular. Section 4 explores how the class properties
of this unique SNR sample comports with current models for remnant radio emission, while 
Section 5 summarizes our conclusions.

\begin{deluxetable*}{ccccccc}
\tablecolumns{7}
\tablecaption{Grid of M33 JVLA Pointing Centers \label{tab:pointings}}
\tablehead{
	\colhead{Decl. (J2000)} & 
	\multicolumn{6}{c}{R.A. (J2000)}
}
\tablewidth{0pt}\startdata
\cutinhead{1.4 GHz}  
$+$30:24:24 & \nodata  & 01:34:34 & \nodata  & 01:33:11 & \nodata  & \nodata  \\
$+$30:40:00 & 01:35:16 & \nodata  & 01:33:53 & \nodata  & 01:32:30 & \nodata  \\
$+$30:55:36 & \nodata  & 01:34:34 & \nodata  & 01:33:11 & \nodata  & \nodata  \\
\cutinhead{5 GHz}
$+$30:22:00 & \nodata  & 01:34:30 & 01:34:00 & 01:33:30 & \nodata  & \nodata  \\
$+$30:27:00 & \nodata  & 01:34:45 & 01:34:15 & 01:33:45 & 01:33:15 & 01:32:45 \\
$+$30:32:00 & 01:35:00 & 01:34:30 & 01:34:00 & 01:33:30 & 01:33:00 & 01:32:30 \\
$+$30:37:00 & 01:35:15 & 01:34:45 & 01:34:15 & 01:33:45 & 01:33:15 & 01:32:45 \\
$+$30:42:00 & 01:35:00 & 01:34:30 & 01:34:00 & 01:33:30 & 01:33:00 & 01:32:30 \\
$+$30:47:00 & 01:35:15 & 01:34:45 & 01:34:15 & 01:33:45 & 01:33:15 & 01:32:45 \\
$+$30:52:00 & 01:35:00 & 01:34:30 & 01:34:00 & 01:33:30 & 01:33:00 & \nodata  \\
$+$30:57:00 & \nodata  & 01:34:45 & 01:34:15 & 01:33:45 & 01:33:15 & \nodata  \\
\enddata 
\end{deluxetable*}

\section{Observations and Data Processing}

Observations of M33 with the JVLA were obtained at both 1.4~GHz and 5~GHz.  The 1.4~GHz observations
were taken under proposal 12A-403 in A-configuration (October 2012), B-configuration (July--August
2012), and C-configuration (January--April 2012).  A total of 16 hrs, 32 hrs and 16 hrs of
integration time (14.0, 28.2, 12.1 hrs on source) were taken in the A, B and C configurations,
respectively. Seven fields of view were required to cover the central area of M33 at 1.4~GHz. The
field centers of the seven 1.4~GHz pointings are listed in the top portion of
Table~\ref{tab:pointings}.  The field centers are staggered in a hexagonal pattern (as indicated by
the table layout) to produce roughly uniform sensitivity across the galaxy.

The 5~GHz observations were obtained as part of proposal 13A-291 in C-configuration in June and July
of 2013 for a total of 48 hrs integration time (43.2 hrs on target).
\changed{The C-configuration was chosen to give a 5~GHz spatial resolution similar to that of the
1.4~GHz B-configuration, which was utilized for the bulk of the low frequency observations.}
Forty-one pointings were
required at the higher frequency to cover the main body of M33 due to the smaller field of view at 5
GHz.  The 5~GHz field centers are also listed in Table~\ref{tab:pointings} (bottom).

The coverage is indicated on an optical image of the galaxy in Figure~\ref{fig:overview}.  The areal
coverage for the 1.4~GHz observations is significantly larger than for the 5~GHz observations.  The
5~GHz sky area was limited by the allocated observing time, with the observations optimized to cover
the footprint of the \chandra\  M33 survey.  The quality of spectral indices is much better where 5~GHz data are
available, but the entire region (including areas with only low-frequency 1.4~GHz data which thus
lack any spectral index information) has been analyzed for this paper.  The areal coverage is 0.85,
0.64, 0.40, and 0.37 square degrees in the four frequency bands centered at 1.3775, 1.8135, 4.679,
and 5.575~GHz.

\begin{figure*}
    \includegraphics[width=\linewidth]{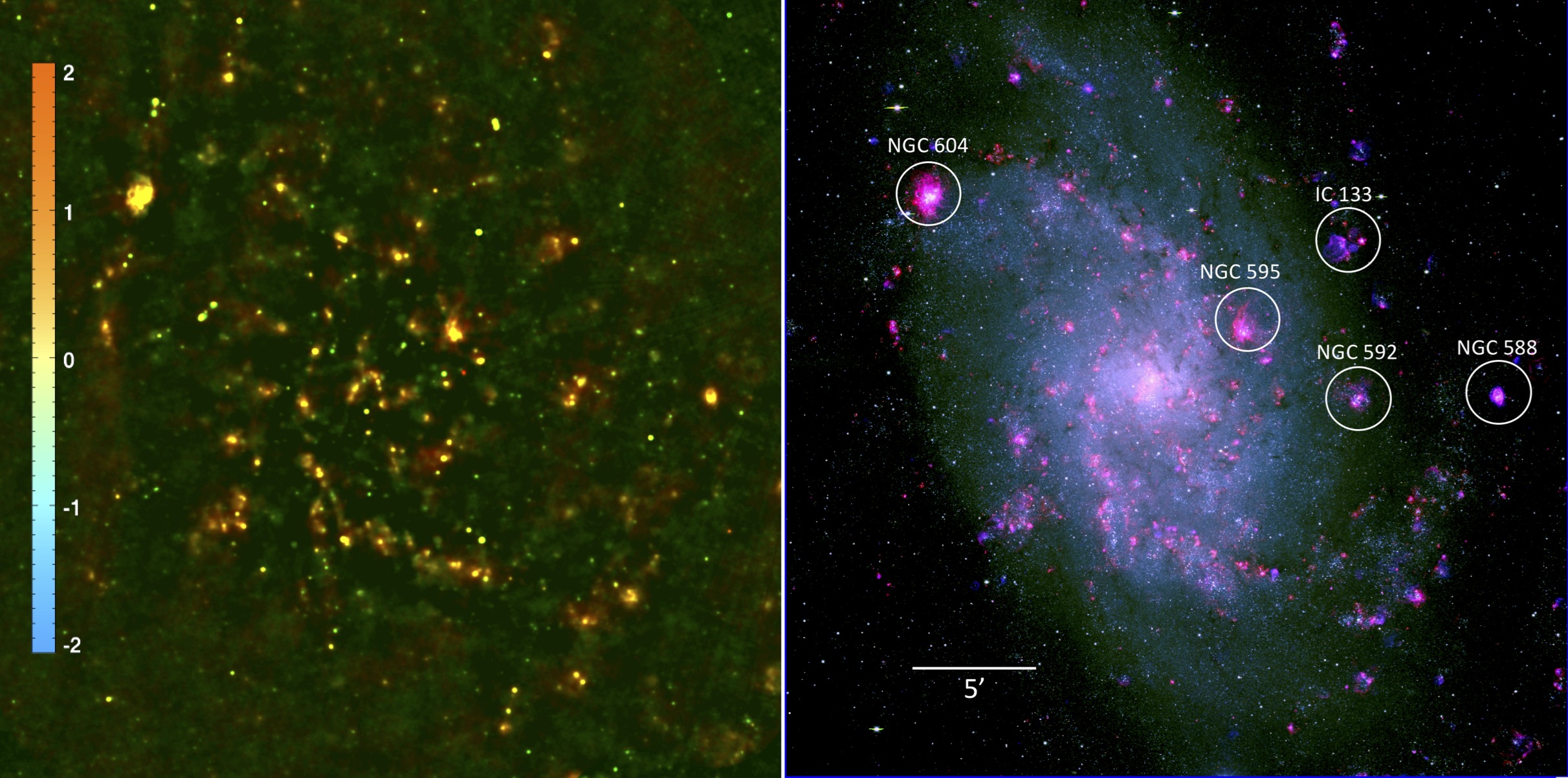}
	\caption{A radio--optical overview of M33.  The left panel shows our radio data over the
	inner galaxy, where the  gradations in color indicate changing spectral indices for the radio
	emission as indicated by the color bar at left. Most \hii\ regions appear yellow, indicating
	a flat spectral index, while many faint green and blue (steep spectrum) sources are either background
	objects or SNRs. The right panel, shown to the same scale, is LGGS optical data for M33,
	including \ha\ in red and V and B continuum images in green and blue to show the galaxy. Some of the
	large, named giant \hii\ regions are labeled for reference. \label{fig:radopt}}
\end{figure*}

The JVLA 1.4~GHz band spans 1 to 2~GHz while the 5~GHz band spans 4 to 6~GHz. These wide bandpasses
present some challenges for imaging, including significant radio-frequency interference (RFI) as
well as large changes to the effective field of view and resolution over the bandpass.  The RFI was
dealt with by severely clipping the affected portions of the uv-data using the AIPS task `CLIP'.
The frequency-dependent primary beam required imaging and self-calibrating at each of 16
intermediate frequencies (IFs) separately. Each IF covered a frequency range of 64~MHz and 128~MHz
at 1.4~GHz and 5~GHz, respectively. Images were cleaned using the AIPS task 'IMAGR'.  At 1.4~GHz,
multiple overlapping fields were simultaneously cleaned to remove distortions created by the
spherical sky, with typically 126 fields used to cover the primary beam area.  At 5~GHz a single
field was sufficient for the cleaning.  The images at different pointings from each IF were coadded
after cleaning using optimal weighting determined by the primary beam to form a single flat image
\citep{becker95}. The images from the individual IFs were then summed in various combinations to
form images with longer integration times.

After processing in AIPS, the images at each frequency were convolved to a fixed round beam with
FWHM 5.9\arcsec. This reduced the resolution of the final images, but using a fixed resolution
resulted in a significantly better catalog than retaining the variable PSF sizes and shapes as a
function of frequency.  The fixed resolution improved both the consistency of the multi-resolution
sky subtraction and filtering step (described below) as well as the accuracy of the flux densities and
spectral indices.

\begin{figure*}
	\centering
	\includegraphics[width=0.60\linewidth]{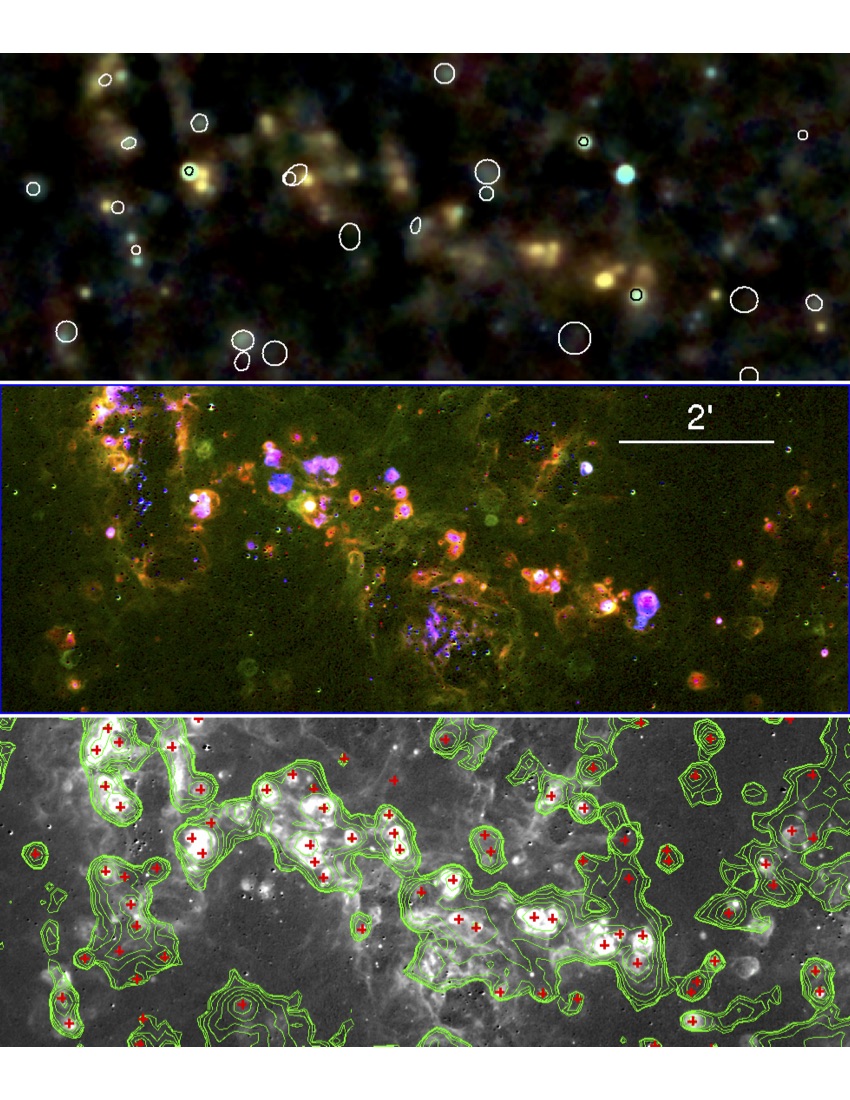}
	\caption{This figure shows a 4.4\arcmin\ by 10.7\arcmin\ region of the southern spiral arm
	in  M33.  The top panel is the color radio spectral index map from Fig.~\ref{fig:radopt}.
	Locations for optically-detected SNRs are shown as white ellipses
	(although the three objects projected against bright radio emission are shown as black); many can be
	identified by cross-referencing to the middle panel that shows a three-color image of LGGS
	continuum-subtracted emission line data, with \ha\ in red, [S~II] in green, and [O~III] in
	blue. SNRs show as greenish areas (strong in [S~II]). (SNRs that are bright in all three
	optical bands will, of course, show as white in the middle panel.) No regions are overlaid on
	this panel so the optical structures can be seen without interference.  The bottom panel is
	a continuum-subtracted \ha\ image from LGGS, shown with log scaling to increase the dynamic
	range. The green overlay shows radio contours from the 1.4~GHz data, with the following
	levels: 1.6, 3.2, 6.4, 12.8, 25.6, 51.2, and 102.8 $\mu$Jy.  Red crosses mark the positions
	of radio sources from our catalog in this region.  Some of the complexity of making
	associations between radio sources and either \hii\ regions or SNRs can be seen here. \label{fig:Sarm}}
\end{figure*}

In Figure~\ref{fig:radopt} (left), we show an overview of the radio data, where the color in the map
represents the spectral index of each source, as shown by the scale bar.  In Figure~\ref{fig:radopt}
(right), we show an optical image of the galaxy at the same spatial scale for comparison. One can
see that many of the brightest radio sources correspond to \hii\ regions in M33, and most of these
have a yellow color in the left panel, indicating fairly flat (thermal) radio spectra.  However, close
inspection shows \ha\ nebulae that align with green radio sources in the left panel, indicative of
steeper non-thermal spectra.  Many of these are optical SNRs and will be discussed below. Many faint
green-to-turquoise (steep) sources do not have obvious optical counterparts and likely represent
background sources.  In some of the larger complexes of \ha\ emission, a more diffuse component of
radio emission is also visible.

These new radio images are deep enough that they are complex and crowded, particularly in the
central regions and southern spiral arm of the galaxy.  To provide more context to this comparison
between the radio maps and the optical data, we show an enlarged section of the complex southern
spiral arm region in Figure~\ref{fig:Sarm}.  The top panel of this figure shows the two-color radio
map of the region as shown in Figure~\ref{fig:radopt} (left), with white ellipses at the positions
of optical SNRs from \citetalias{long18} for reference.  These regions can be referenced to the middle panel of the
figure, where we show a three-color version of the LGGS optical continuum-subtracted emission line
data for the region, with \ha\ shown in red, \sii\ in green, and \oiii\ in blue.  The graytone
background in the bottom panel is a version of the LGGS \ha\ data where we have scaled and
subtracted the continuum and displayed it on a log scale to enhance the dynamic range. Overlaid on
this panel in green are contours from the 1.4~GHz radio map as indicated in the figure caption.  The
red crosses indicate positions of radio peaks in the catalog, as described in the next section.
Correspondences with both SNRs and \hii\ regions are clearly seen, but the difficulty of making
unique associations in some cases is evident. Often the SNRs appear in regions of extended or
adjacent \hii\ emission.

\subsection{Construction of the Radio Source Catalog }

The majority of the emission in our radio maps is in the form of point or discrete sources.\footnote{
	The sum of the 1.4~GHz flux densities of our catalog sources is $\sim1$~Jy, which is similar to
	the sum of the entire JVLA image.  For comparison, \cite{dennison75} measured a flux density
	of $3.3\pm0.5$~Jy for M33 at 1.4~GHz using the original NRAO 91-m telescope.  The JVLA resolves out the
	most highly extended emission.  Note that the current survey is a factor of 100,000 more sensitive
	than the 1975 measurement!
}
Hence, we have constructed a catalog of these sources to permit comparisons between these
new radio data and catalogs at other wavelengths.

The radio source catalog was created by averaging the data in the 1.4~GHz and 5~GHz bandpasses to
produce a single detection image.  For the 1.4~GHz data, the entire bandpass was used, and 14 of the
16 125-MHz channels from the 5~GHz observations were used. (Channels 1 and 2 of the 5~GHz data were
contaminated by RFI.) The 5~GHz and 1.4~GHz bands are equally weighted in the combined image,
resulting in source lists that are not a strong function of the source spectral index. This equal
weighting is not optimal for a typical extragalactic source (which has a power law spectrum
$\nu^\alpha$ with $\alpha \approx -0.7$) but is a reasonably ``spectrum-neutral'' strategy for the
source detection, which is more appropriate for our study.

The complexity of the radio emission causes a significant spatial variation in the background level
with position.  \changed{We developed a multi-resolution (median pyramid) image processing algorithm}, both for use in
source detection and to subtract the background from the images to make the source detection
threshold more uniform.  The multi-resolution median pyramid images were also used for both source
detection and flux density measurements for the sources.  We describe this new and somewhat complex 
algorithm in detail in Appendix~\ref{sec:multires}.

\begin{figure}
    \includegraphics[width=\linewidth]{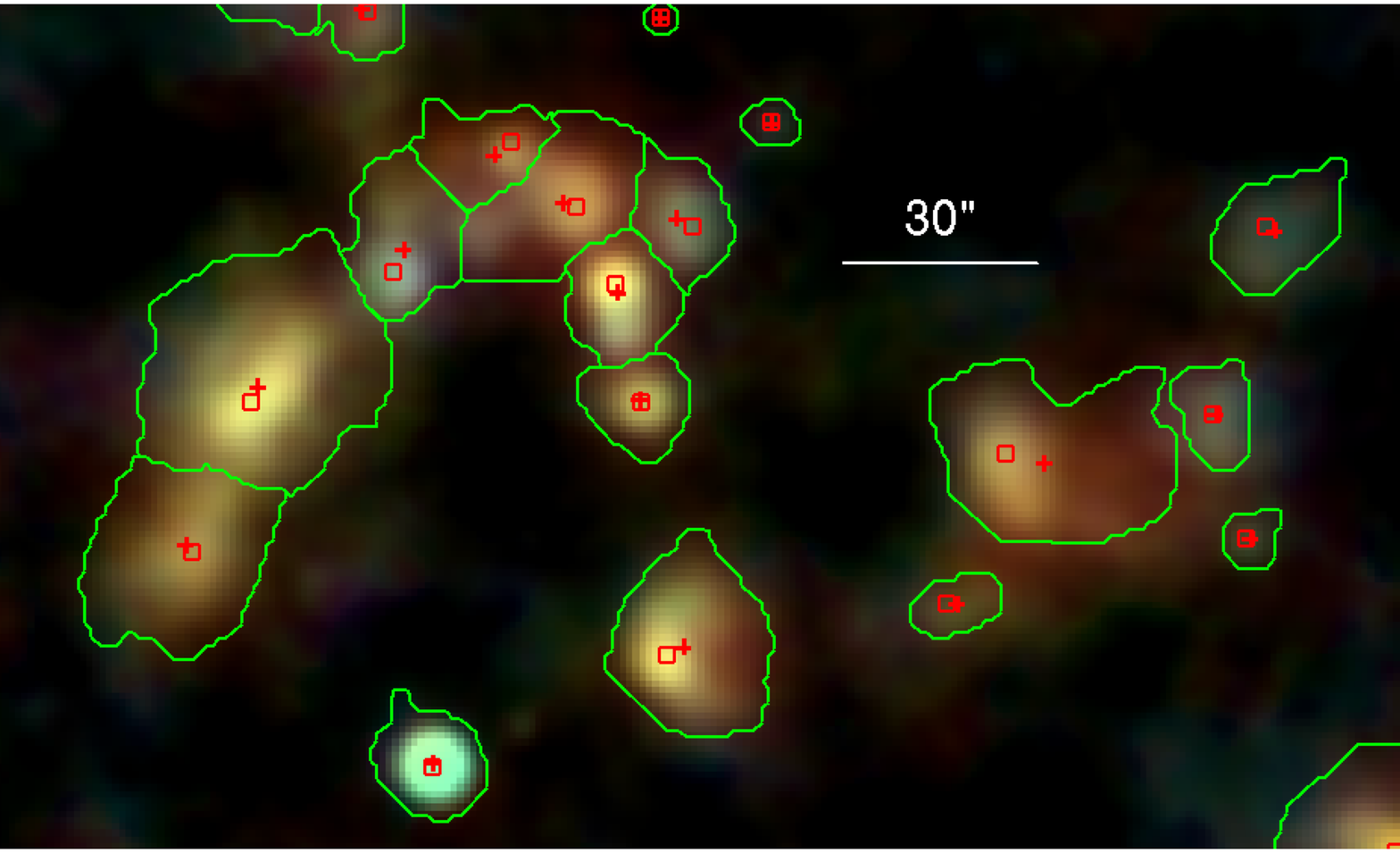}
	\caption{This figure shows a region near the M33 nuclear region (upper left) and just to the
	west, demonstrating the different elements of the radio catalogs. The green outlines
	indicate the radio emission islands described in the text.  The red crosses indicate the
	positions determined for the mean of each island, while the red squares indicate the
	cataloged ``peaks" determined by the algorithm.  These can be slightly offset from each
	other, depending on the complexity of the enclosed emission, although for most isolated
	islands they are nearly identical.
	\label{fig:islands}}
\end{figure}

An island map (segmentation map) technique is used to indicate image regions where sources were detected.  Figure~\ref{fig:islands} shows an enlargement near the complex nuclear region as an example of the results of this process.
%The numbers in the island map correspond to the ``Island'' field in the catalog. 
Here we provide a brief summary of the algorithm used to compute the islands and the positions listed in our catalog:
\begin{itemize}
\item The detection image is separated into a stack of images of equal size that have 
structures on scales ranging from compact to very extended.  The sum of this stack is equal 
to the original image.
\item At each stack level, the local rms noise is estimated, and pixels that exceed that 
noise by a factor of two are included in a segmentation map indicating potential source positions.  
Contiguous pixels are grouped into ``islands'' for different sources.
\item The lowest resolution image from the stack is treated as a variable sky level and is not 
used for source detection or in flux calculations.
\item Overlapping islands at different levels are merged and/or split as necessary to 
group them with islands at other levels.  When this is complete, each pixel of the image 
is assigned to a single island in one or more levels of the stack or is not assigned to 
any source.
\end{itemize}

As seen in Fig.~\ref{fig:islands}, the resulting islands can be irregular in shape and sometimes
include what to one's eye might appear as two or more sources (or a point source with adjacent more
diffuse emission). However, in complex regions the islands usually fit together snugly and thus
capture all of the radio flux from the region.

Each island corresponds to a single source in the catalog.  The fluxes for sources are calculated by
computing similar multi-resolution image stacks for each frequency band image.  Pixels that are
assigned to an island at a given resolution level are included in the summed flux for the source.
Note that the island size may change at each resolution level, which effectively leads to an
adaptive weighting of the image pixels that depends on the source profile.  There is a PSF-dependent
correction factor (close to unity) for flux that spills outside the island.  The island shapes are
identical for all the different frequency bands (as are the PSFs after the beam sizes have been
matched), which leads to accurate and reliable spectral indices.  See Appendix~\ref{sec:multires} for further
details.

The moments of the mean detection-image flux densities in each island are used to determine the
centroid position and a representative elliptical morphology (FWHM major and minor axis and position
angle) for each source, as listed in the catalog.  The size quoted in the catalog has not been
corrected for the round 5.9\arcsec\ FWHM beam size.  For an unresolved source the major and minor
axes will be approximately equal to 5.9\arcsec.  Note that the size is not constrained to be larger
than this value; sources with sizes smaller than the beam width (due to noise fluctuations) can have
integrated flux densities smaller than their peak flux densities.  In such cases the peak flux
density may be a more reliable estimate of the source brightness.

The integrated flux densities for each of four bands (two from 1--2~GHz and two from 4--6~GHz) are
fitted with a power law to determine the spectral index $\alpha$ for each source, $F_\nu =
\Fint(\nu/\nu_p)^\alpha$. The noise in the band fluxes is used to determine the errors in $\Fint$
and $\alpha$.  The pivot frequency $\nu_p$ is chosen as the frequency where the covariance between
$\Fint$ and $\alpha$ is zero, which also is the frequency of the optimal signal-to-noise $F_\nu /
\sigma(F_\nu)$.  The spectral index fit is constrained to $-3 \le \alpha \le 3$; sources with
spectral indices at these limits have $\sigma(\alpha) = 0$.

\begin{figure}
    \includegraphics[width=\linewidth]{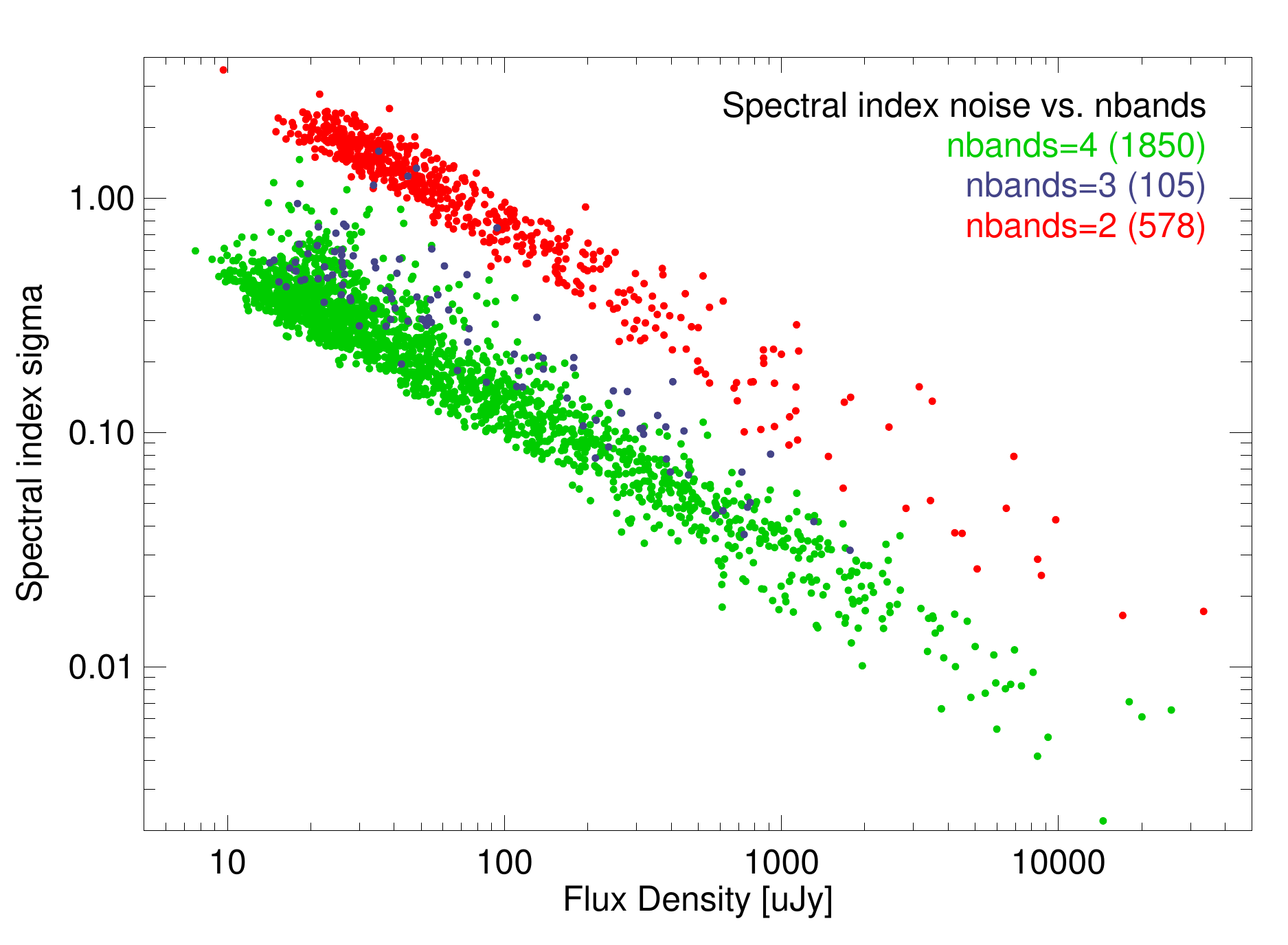}
	\caption{A comparison of the noise in the computed spectral index as a function of the source
		brightness and the number of frequency bands with data for a given source.  In the outer parts of
		the galaxy, where only two frequency bands are available, the spectral index uncertainties are six times
		larger than in the central regions, where four frequency bands are measured.
		Only the brightest sources with
		two-band detections have accurately measured spectral indices.
	\label{fig:spnoise}}
\end{figure}

The accuracy of the spectral indices degrades in the outer region where fewer frequency bands are
available (Fig.~\ref{fig:spnoise}).  \changed{In the center of the galaxy, sources are measured
across the full 1--6~GHz bandpass.} The
frequency-dependent field of view reduces the sky area covered at higher frequencies, so outer
regions of the galaxy are covered by fewer frequency bands (Fig.~\ref{fig:overview}).  The
uncertainty in the spectral index is determined by both the signal-to-noise of the flux density measurements
and by the frequency lever-arm of the data.  The small region with three bands has only slightly
worse noise than the four-band data, but the two-band data (with measurements only in the 1--2~GHz
bandpass) has spectral index errors that are six times larger.  As a result, only the brightest
sources with two-band detections have accurately measured spectral indices. (Obviously the outermost
region with only a single band detected has no spectral index information at all.) Fortunately, the
majority of the sources are found in the inner region of the galaxy where there is full frequency
coverage.

In addition to the island centroids, the catalog also includes the peak flux density and the
position of the peak for each island.  These values are computed from the detection image, just as
for the flux-weighted positions. The peak is simply the brightest pixel in the island after
background subtraction.  The position of the peak (which may differ from the flux-weighted mean in
asymmetrical sources) is determined in the sharpest channel of the multi-resolution image stack
where the source is detected. This means it represents the position of the peak for the most compact
source component.  We found that this choice gives more accurate peak positions in crowded regions.

The results are summarized in Table \ref{tab:radio_cat}\footnote{A one page sample of the Table is
shown here, where we have selected the first twenty lines of the table that have a coincidence with
a SNR, an X-ray source, or an \hii\ region. \changed{The sample also includes one of the (rare) subthreshold sources
as an example.} The full table is available electronically.}, which includes all sources detected 
at a signal-to-noise $\Fint/\sigma(\Fint) \ge
4$.  It also includes seven sources that fall below that detection threshold but that have
associations in external X-ray or \hii\ region catalogs; those sources are flagged as sub-threshold
objects.  A full description of the columns in the table is found in Table~\ref{tab:radio_cat_columns}.

\begin{figure}
    \includegraphics[width=\linewidth]{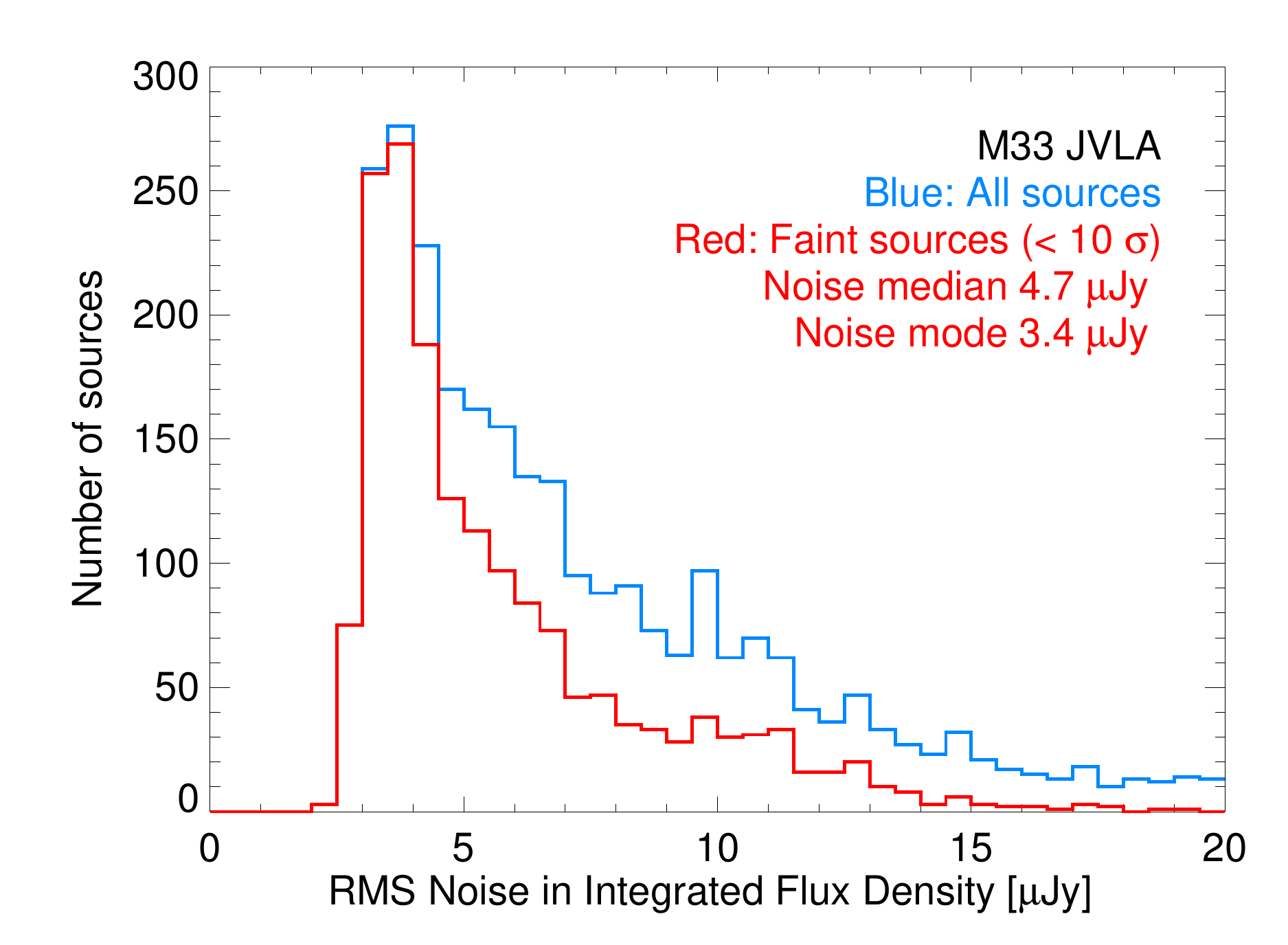}
	\caption{The distribution of noise values in the integrated flux density in the JVLA M33 catalog.
		The distribution has a peak near 3--4~\uJy\ with an extended tail of higher noise values.
		However, the tail is contributed mainly by bright sources (that are well above the detection
		threshold).  For sources detected at $< 10\sigma$, the noise mode is 3.4~\uJy\ and the median is 4.7~\uJy.
		Over most of the survey area our catalog is complete to $\sim20$~\uJy ($4\sigma$).
	\label{fig:rms}}
\end{figure}

Our final catalog contains 2875 sources with 1.4~GHz flux densities ranging from about 2.2~\uJy\ to
0.17~Jy.  (The faintest 1.4~GHz fluxes are for objects with inverted spectra that are brighter at
5~GHz.) Of these, 670 have spectral indices determined with an accuracy $\sigma(\alpha) \le$ 0.15.
Essentially all sources with 5~GHz data (2/3 of the catalog) and fluxes in excess of 150~\uJy\ have
well-measured spectral indices.  The typical noise in the integrated flux density is $<5$~\uJy,
although there are a few crowded regions where the noise is significantly higher (Fig.~\ref{fig:rms}).

\begin{figure*}
	\centering
    \includegraphics[width=\linewidth]{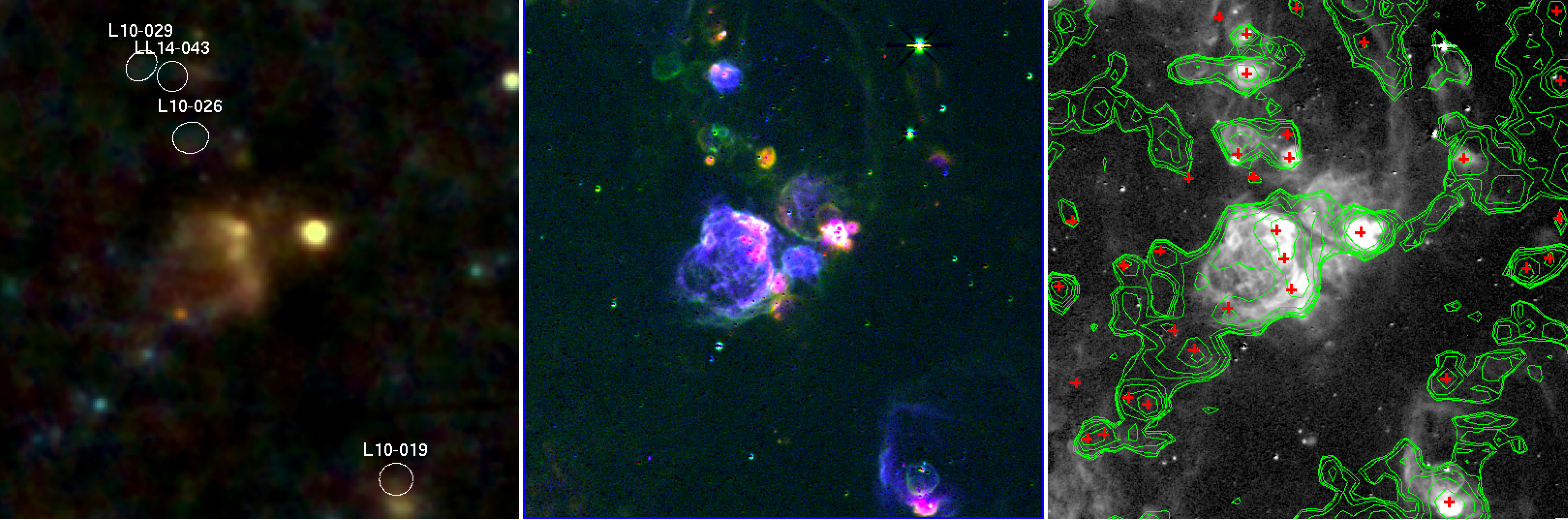}
    \vspace{1mm}
    \includegraphics[width=\linewidth]{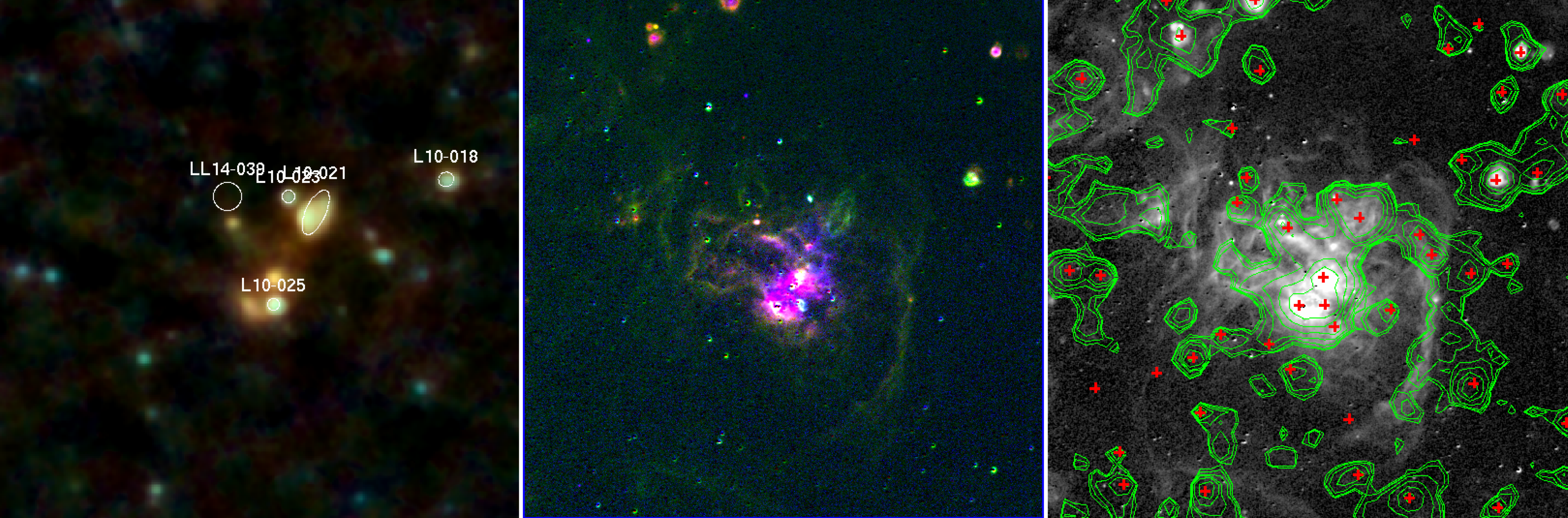}
	\caption{Two 5\arcmin\ regions centered on giant H~II regions in M33. (top) IC~133, and
	(bottom) NGC~592. The panels and labeling are similar to Fig.~\ref{fig:Sarm}. Note that both regions include multiple radio peaks (red +'s in the right panel), and
	the radio contours show connections from the main optical H~II region to more extended
	structure with additional radio peaks.  \label{fig:giant}}
\end{figure*}

With the sensitivity and resolution of these new data, some of the larger optical emission
structures like giant \hii\ regions or \hii\ emission complexes incorporate numerous individual
radio emission peaks. Figure~\ref{fig:giant} shows two examples, IC~133 and NGC~592, where the color
coding and labeling are similar to those in Figure~\ref{fig:Sarm}.  IC~133 shows a bright,
relatively high-excitation (strong \oiii-emitting) bi-lobed structure in the optical with five radio
peaks on portions of the main nebula.  The radio contours extend to the lower left, incorporating
another eight radio peaks that are listed in the catalog.  Several optical SNRs are indicated nearby
but are not directly related to IC~133.  In contrast, NGC~592 has a very bright core of
optical emission with a number of fainter loops of surrounding emission that are mapped reasonably
well by the radio contours, including a faint arm of emission to the west. Several SNRs are
associated with this region as well. Several radio peaks correspond with the bright core region,
while the extended radio contours incorporate a dozen or more additional radio peaks.

\begin{figure}
    \centering
    \includegraphics[width=\linewidth]{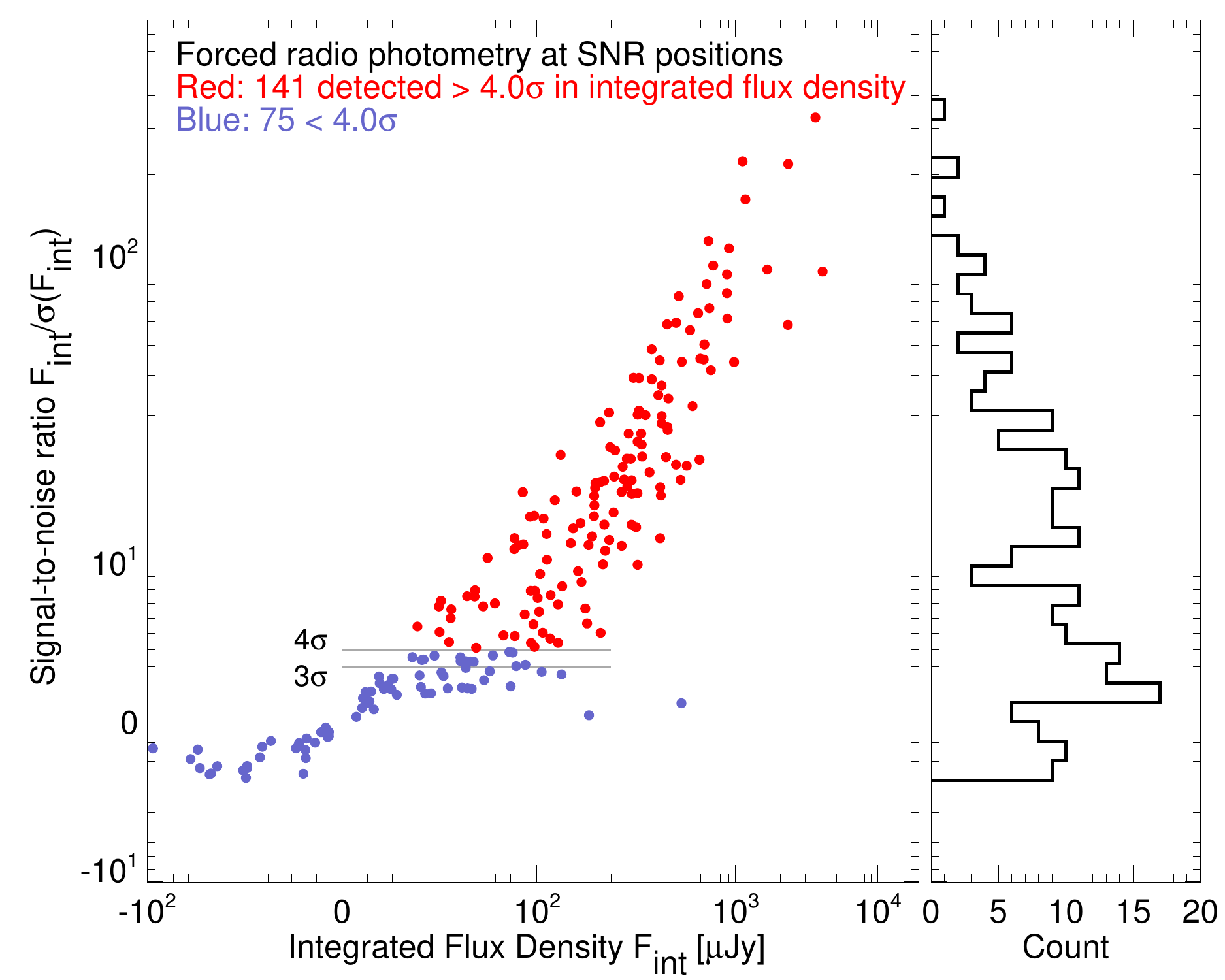}
    \includegraphics[width=\linewidth]{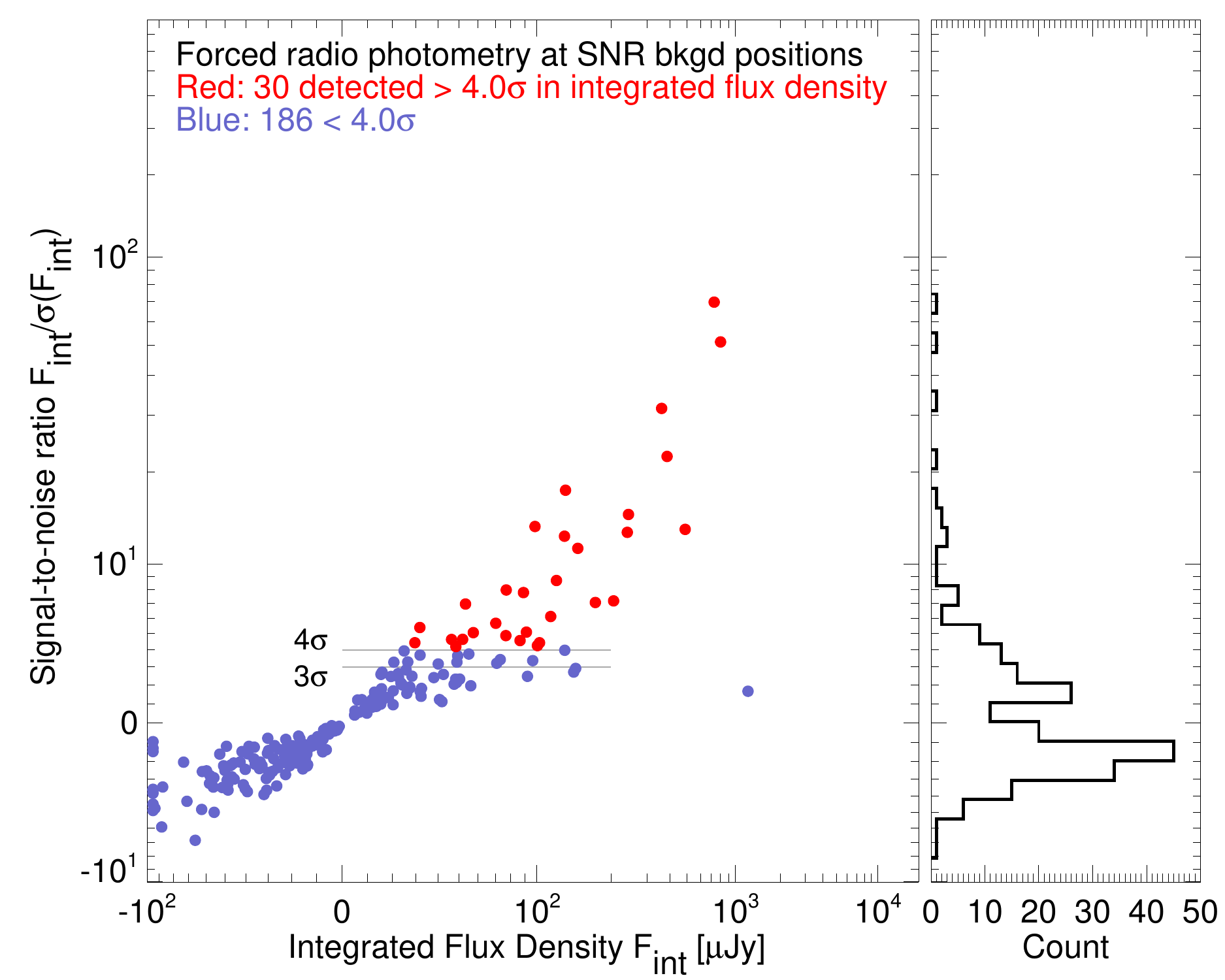}
	\caption{The signal-to-noise as a function of the integrated flux density $\Fint$ for the
		SNR forced photometry (top) and background photometry using SNR regions shifted by
		1~arcmin (bottom). Note that the measured flux densities can be negative for non-detections.
		The histogram on the right shows the distribution of signal-to-noise values.  The
		false positions are obviously much fainter, but sometimes sources are ``detected''
		due to random coincidences with radio sources in the crowded field.  Eight
		simulations \changed{like the one at the bottom} were performed to determine the
		signal-to-noise distribution of chance matches.
	\label{fig:falsematch}}
\end{figure}

% In the catalog, we {\bf [have attempted to/ will attempt to]} indicate many of the individual sources that represent portions of larger structures by {\bf [doing what?]}. However, this is difficult to do systematically and is only intended as a guide to those interested in particular objects or regions.

\subsection{Forced Photometry Catalog at Optical SNR Positions}

The radio catalog described above was constructed without any reference to the positions of known
optical SNRs in M33 (although many of the remnants were detected).  However, we have also
constructed a separate catalog of radio properties at the positions of known optical SNRs by
integrating the radio images over the elliptical region determined for each optical SNR, as
measured previously by either \citetalias{long10} or \citetalias{lee14}.  (When a SNR appeared in both lists, we somewhat
parochially adopted the regions of \citetalias{long10}.) This method allows us to establish either measured flux densities
or appropriate upper limits for the radio emission from all SNR candidates, whether or not a
radio source was independently detected at the SNR position.  We will refer to this version of the SNR radio
data as the ``forced photometry'' SNR extraction catalog.

The calculations of flux densities, spectral indices, and flux-weighted positions proceeds much as
described above for the radio catalog.  The island map for the forced photometry catalog is
determined from the elliptical regions in the optical SNR catalog rather than from the radio map itself.
Since the island map makes no reference to the radio morphology, we do not have multi-resolution
islands but instead sum the radio flux density over the entire region using the single background-subtracted
image in each radio band.  The catalog does not include the radio peak information (which was not
found to be useful) but does include the input positions for the regions along with the
radio-flux-weighted centroid positions.

Associations between SNRs and objects in our master radio catalog are determined by the overlap between
the radio island map and the SNR island map.  Many SNRs have unambiguous matches with cataloged
radio sources, but there are also ambiguous cases where a single SNR island overlaps several radio
islands and vice versa.  The information on the overlapping sources and a flag that captures information
on the ambiguity of the association are also included in the table.

Of the 217 SNRs and SNR candidates in the merged list of \citetalias{long18}, 216 are in the region covered by our
radio images (see Fig.~\ref{fig:overview}).  The only object falling completely outside the radio
region is LL14-195. There are 188 SNRs in the central region covered by both the 1.4~GHz and 5~GHz
frequency bands plus two more that have data from 1.4~GHz and the low-frequency 5~GHz bandpass only.  
Of the 26 sources that have only 1.4~GHz data, only one (L10-003) has a reasonably accurate spectral index; 
we have little spectral index information in the outer region of the galaxy.

We find that 155 of the 216 SNRs and candidates are detected above 3$\sigma$ at radio wavelengths, a
detection rate of 72\%.  That detection rate rises to 76\% (145 of 190) in the central region where
we have full frequency coverage.  The remainder have radio upper limits from the forced photometry
exercise.  (Some of the sources below the $3\sigma$ threshold are also likely to be detected: there
are, for example, 12 $2\sigma$ detections discussed further below.) Of the 155 $3\sigma$
radio-detected SNRs, 126 came from the list of 137 candidates in \citetalias{long10}, while 29 came from the 79 SNRs
that were first suggested as candidates by \citetalias{lee14}. The fact that a much lower percentage (37\% versus
92\%) of the extended list of SNR candidates identified by \citetalias{lee14} were detected in the radio is
undoubtedly related to the fact that many of the \citetalias{lee14} candidates are optically fainter than those
in the \citetalias{long10} list.

\begin{figure}
    \includegraphics[width=\linewidth]{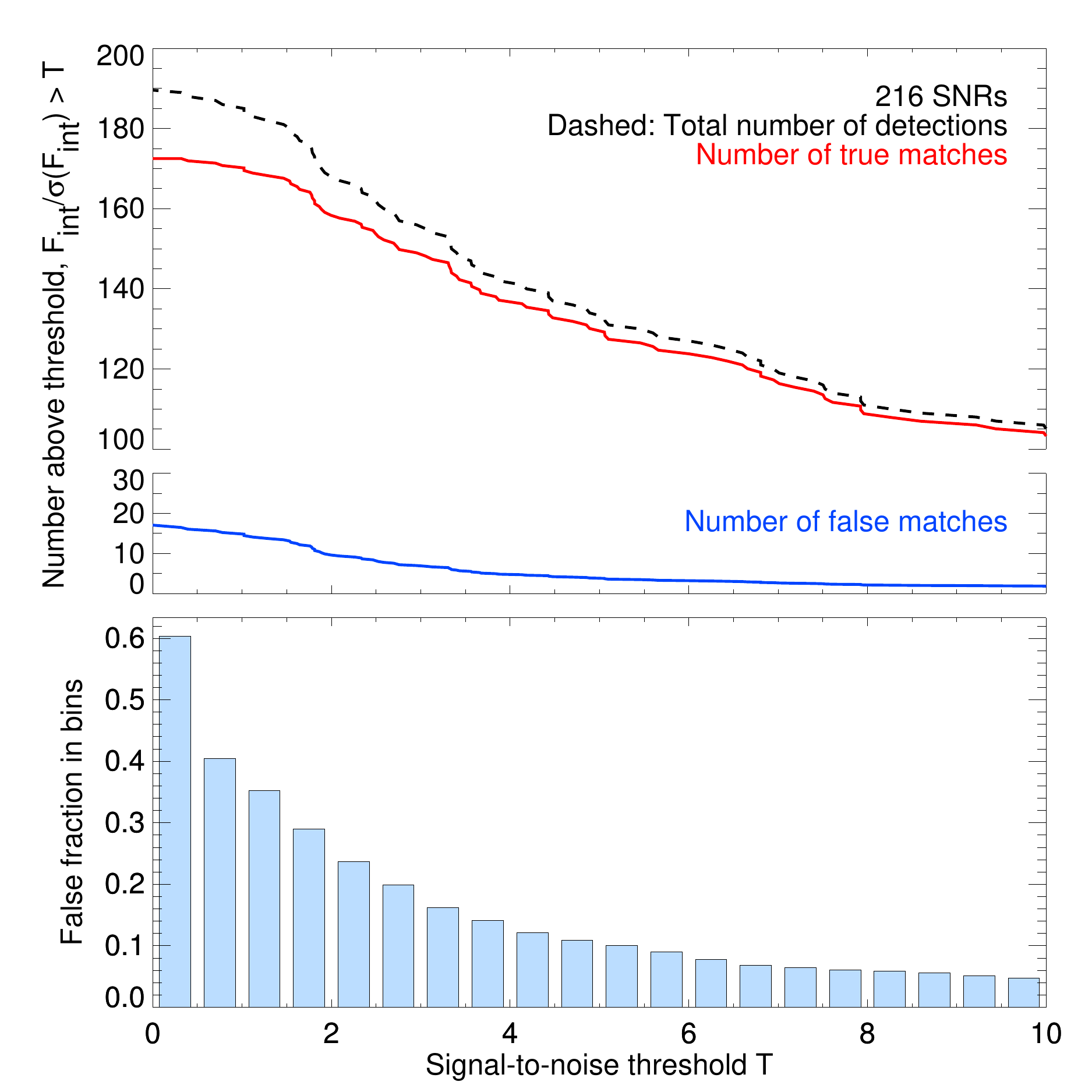}
	\caption{The number of false detections in the SNR forced-photometry table
		as a function of signal-to-noise.  The top panel shows the cumulative number of
		detections (dashed line), the estimated number of false detections (blue line),
		and the number of remaining true matches (red line) as a function of the
		signal-to-noise.  The bottom panel shows the fraction of false detections
		in signal-to-noise bins.  Using a $3\sigma$ detection threshold, the contamination
		rate is low: fewer than 15\% of the SNRs with
		$3\sigma < \Fint/\sigma(\Fint) < 3.5\sigma$ are expected to be the result of chance.
	\label{fig:snrfalse}}
\end{figure}

An important consideration in setting the detection threshold at $3\sigma$ is the expected rate of
chance coincidences between the SNR islands and the radio map.  The radio images are sufficiently
crowded that randomly placed ellipses will sometimes fall on radio sources.  This is particularly
true in the center of the galaxy and in the spiral arms.  We estimated the rate of chance radio
detections by shifting the SNR islands by $\pm1$~arcmin in RA and Declination.  The size of the
shift was chosen to be larger than the largest island's major axis.  The shift is large enough to
move off the local object while still being small enough to be sampling the same environment (spiral
arm, galaxy center) as the true SNR position.  The shifted calculations also use the same
distribution of region sizes as the real catalog, which has a significant influence on the
background rate.  Eight shifts were done, including shifts only in Declination, only in RA, and in
both RA and Dec.  For every shift, forced photometry was performed on all the regions using exactly
the same approach as for the real positions.

A comparison between the measured flux densities for the real SNR regions and a typical example of
the shifted SNR region fluxes is shown in Fig.~\ref{fig:falsematch}.  While positive detections are
much rarer at the shifted positions, they are common enough that they cannot be ignored.

All eight shifted catalogs for each object class were combined to compute the distribution of the
signal-to-noise, $\Fint/\sigma(\Fint)$, for random positions on the sky.  With eight simulations of
the random detections, the simulated noise distribution is reasonably well determined.  Given the
cumulative distribution of the number of sources exceeding a given signal-to-noise ratio, we then
compute for every source in the real catalog the probability that a random source of equal or
greater flux would be detected.

Finally we compare the cumulative distribution of the actual number of sources detected as a
function of the detection threshold with the cumulative distribution of the random detection
probability above that threshold (Fig.~\ref{fig:snrfalse}).  That gives an estimate of the number of
false detections that are included as a function of the detection threshold.

We conclude that a detection threshold of $3\sigma$ leads to acceptable contamination by false
detections.  Of the 141 SNRs above $4\sigma$, $\sim5$ are likely to be due to false matches.  Adding
the SNRs between $3\sigma$ and $4\sigma$ increases the number detected by 14 while adding only 2
additional false matches.  Even detections between $2\sigma$ and $3\sigma$ are likely to be fairly
reliable, with $\sim75$\% of the 12 detections in that range expected to be true detections.  (These
coincidence rates are lower than intuition might suggest because the high true detection rate for
SNRs means that there have been relatively few ``rolls of the dice'' at empty sky positions.)

\begin{figure*}
    \includegraphics[width=\linewidth]{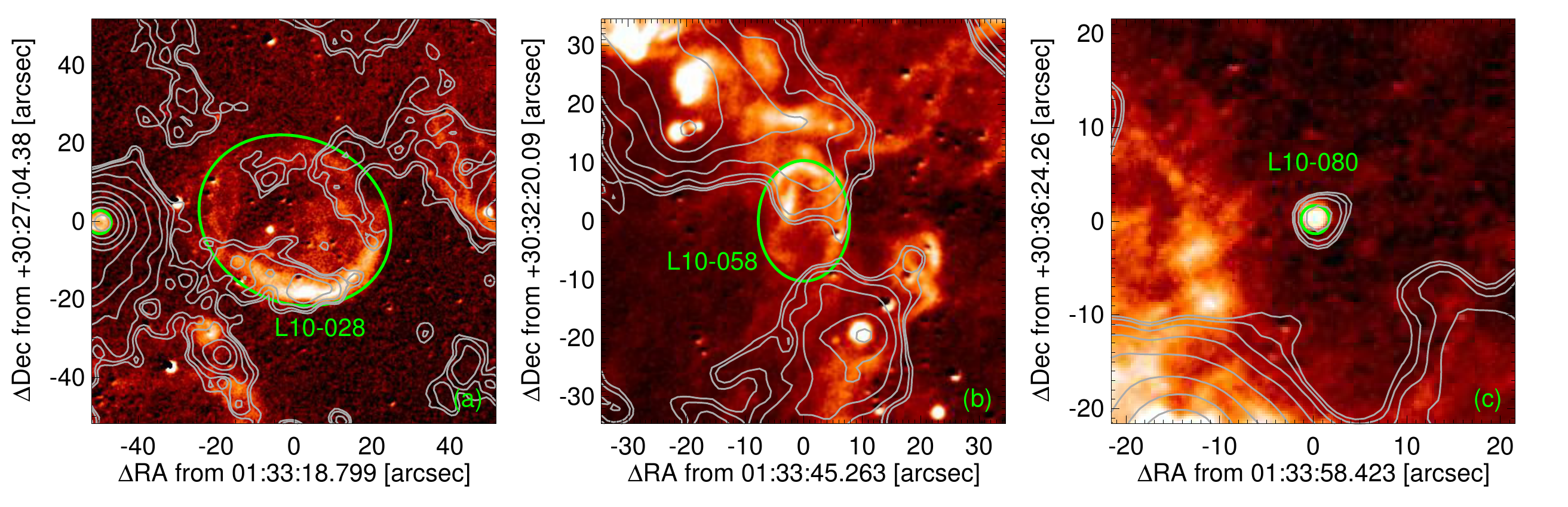}
	\caption{
		The H$\alpha$ and combined radio images for several
		SNRs for which coincident radio emission is
		present but fails to 
		rise above the forced photometry detection threshold.
		The image is the LGGS H$\alpha$ data, the green ellipses show
		SNRs, and the contours are for the combined 1.4 plus 5~GHz image
		at levels 1.6, 3.2, 6.4, ..., 819.2 \uJy.
		(a) In L10-028, only the brightest part of the
		optical shell has a radio counterpart.
		(b) L10-058 shows a close morphological match of
		the optical and radio images but the higher background
		induced by the nearby bright source
		hampers detection.
		(c) Source L10-080, the smallest diameter remnant
		in the catalog, has a clear sub-threshold counterpart that 
		falls just below the $3\sigma$ detection threshold (at $2.7\sigma$).
\label{fig:snrfaint}}
\end{figure*}

The basic characteristics of the SNRs as well as results of our forced photometry extraction are
presented in Table \ref{tab:forced}\footnote{A one page sample of the table is shown here; the full
table is available electronically.}, where the non-radio data are from \citetalias{long18}. The \sii:\ha\ column in
this Table indicates whether optical spectra exist and whether these showed a \sii:\ha\ ratio of 0.4
or greater.  The X-ray column indicates whether an object was detected at 3$\sigma$ or greater with
\chandra\  by \citetalias{long10} or with \xmm\ by \cite{garofali17}.
A full description of the columns in the table is found in Table~\ref{tab:forced_columns}.
Our forced photometry radio catalog of SNRs
contains 178 SNRs that have been optically confirmed by the \sii:\ha\ criterion; there are another
27 objects that have spectra but show lower \sii:\ha\ ratios, and 12 for which there are not yet any
optical spectra.  A total of 112 of the objects are also detected in X-rays.

The SNR forced-photometry table also identifies radio sources from the master catalog
(Table~\ref{tab:radio_cat}) that have detection islands that overlap the SNR regions.  In cases
where more than one radio source overlaps the SNR, all the matches are listed.  Multiple matches are
sorted by the number of overlapping pixels, so the radio source sharing the largest number of pixels with
the SNR region is listed first.  Similarly information is included in the radio catalog for cases
when a radio source overlaps one or more SNRs, and the same information is available for radio matches
to the forced photometry X-ray catalog (described below).  The Sflag (Table~\ref{tab:radio_cat}) and Rflag
(Tables~\ref{tab:forced} and \ref{tab:forced_xray}) columns are bit flags with values that help determine whether a match is
reliable.  For the Sflag column, for example, the bit values $b$ indicate that
\begin{itemize}
	\item $b=1$: the SNR has a radio match,
	\item $b=2$: the match is unambiguous, so only one radio source matches the SNR, and 
	\item $b=8$: the match is mutually good, meaning that the best radio source for this SNR
	also has this SNR as its own best match, where ``best'' is determined by the number of
	overlapping pixels between the islands.
\end{itemize}
The most reliable matches will have flag bit 8 set.  All of those will have bit 1 set as well, and
most of them will also have bit 2 set.  Flag values of 9 and above indicate good matches, with the
best having flag values of 11.  The distributions of the values of Sflag (in the radio catalog) and
Rflag (in the SNR and X-ray catalogs) are shown in
Table~\ref{tab:flags}.

\begin{deluxetable}{crrrl}[b!]
\tablecolumns{4}
\tablecaption{Flag Value Counts in the Radio and SNR Catalogs \label{tab:flags}}
\tablehead{
	\colhead{Value} &
	\colhead{\pb{6ex}{Sflag\\Radio\tablenotemark{a}}} &
	\colhead{\pb{6ex}{Rflag\\SNR\tablenotemark{b}}} &
	\colhead{\pb{7ex}{Rflag\\X-ray\tablenotemark{c}}} &
	\colhead{Meaning}
}
\tablewidth{0pt}\startdata
0  & 2728 &  75 & 342 & \pb{10em}{No match} \\
1  &    0 &   1 &   1 & \pb{10em}{Ambiguous matches} \\
3  &   18 &  11 &  20 & \pb{10em}{Single match but not mutually good} \\
9  &   17 &  19 &  22 & \pb{10em}{Mutually good match} \\
11 &  112 & 110 & 277 & \pb{10em}{Single, mutually good match} \\
\enddata 
\tablenotetext{a}{The Sflag column in the radio catalog
indicates the quality of SNR associations.}
\tablenotetext{b}{The Rflag column in the SNR forced photometry catalog
indicates the quality of radio source associations.}
\tablenotetext{c}{The Rflag column in the X-ray forced photometry catalog
indicates the quality of radio source associations.}
\end{deluxetable}

Of the 155 SNRs detected at 3$\sigma$ in the forced photometry catalog, 134 are associated with a
source in the master radio catalog.  Of these, there are 19 sources in the SNR catalog associated
with two (or three) radio sources.  There are 21 SNRs detected as faint sources in the forced
photometry catalog that are not found in the master catalog.  Since the master radio catalog has a
$4\sigma$ detection limit (compared with $3\sigma$ for the forced photometry), it is not surprising
that there are sources below the radio catalog limit.

In some cases where the forced photometry does not yield a $3\sigma$ detection, radio emission is
still apparent at the optical remnant location. These cases arise from the incommensurability of the
flux density distribution in the optical and radio bands, the presence of a nearby bright radio
source, or radio emission just below the detection threshold. Examples of these cases are shown in
Figure~\ref{fig:snrfaint}.  In some cases (L10-028, L10-030, LL14-038, and L10-131), only the
brightest part of the optical shell has coincident radio emission, and the overall source falls
below threshold (Fig.~\ref{fig:snrfaint}a).  In other cases (L10-058, LL14-096, L10-132, LL14-142) a
bright, nearby (and apparently unrelated) source raises the local background sufficiently that the
radio counterpart falls below threshold (Fig.~\ref{fig:snrfaint}b).  In yet other cases (LL14-125,
LL14-128, L10-048, L10-080, L10-092, LL14-058), a clear radio counterpart is apparent but falls just
below the threshold (Fig.~\ref{fig:snrfaint}c); note that the division between the latter two
categories is somewhat arbitrary. In summary, over one-quarter of the remnants not listed here as
having radio emission do have modest levels of radio flux associated with them that a deeper and/or
higher resolution survey would definitely have detected.

\subsection{Forced Photometry Catalog at Chandra X-ray Positions}

The list of X-ray source positions from the ChASeM33 catalog of \cite{tuellmann11} was also used to
construct a forced-photometry catalog with radio flux densities.  The procedure followed was very similar to
the SNR forced-photometry catalog described above.  The resulting table (shown as a sample in
Table~\ref{tab:forced_xray}) has a $3\sigma$ threshold for radio detections (but includes the radio
measurement regardless of the signal-to-noise).  It includes information from the X-ray catalog
including the diameter and count rates in various energy bands.  The Rflag column is defined exactly
as for the SNR forced photometry table, with values of 8 or greater indicating confident (or at
least unconfused) associations.  The format is similar to Table~\ref{tab:forced}; a
full description of the columns is found in Table~\ref{tab:forced_columns}.

Of the 662 X-ray sources, 319 are detected in our radio survey.  All of these X-ray sources fall in
the region that has both 1.4~GHz and 5~GHz radio data, and 102 of the detections have radio spectral
indices determined to an accuracy $\sigma(\alpha) < 0.15$.

\subsection{Identification of H~II regions}

Many of the radio sources in the catalog are associated with star-forming \hii\ regions, with
optically thin thermal bremsstrahlung radiation having a spectral index $\alpha \sim 0$.  In order
to determine which sources in our radio catalog are associated with \hii\ regions, we have
computed the \ha\ surface brightness over the radio island regions for each source in the catalog using the \ha\
image obtained by one of us (PFW) using the Burrell Schmidt camera at Kitt Peak in Arizona (cf.
\citetalias{long10}).  The tables for both the radio catalog and the forced photometry catalogs include the total
\ha\ flux, $F(\mathrm{H}\alpha)$, and the average flux surface brightness (per square arcsec), $\Sigma(\mathrm{H}\alpha)$.

\begin{figure}
    \includegraphics[width=\linewidth]{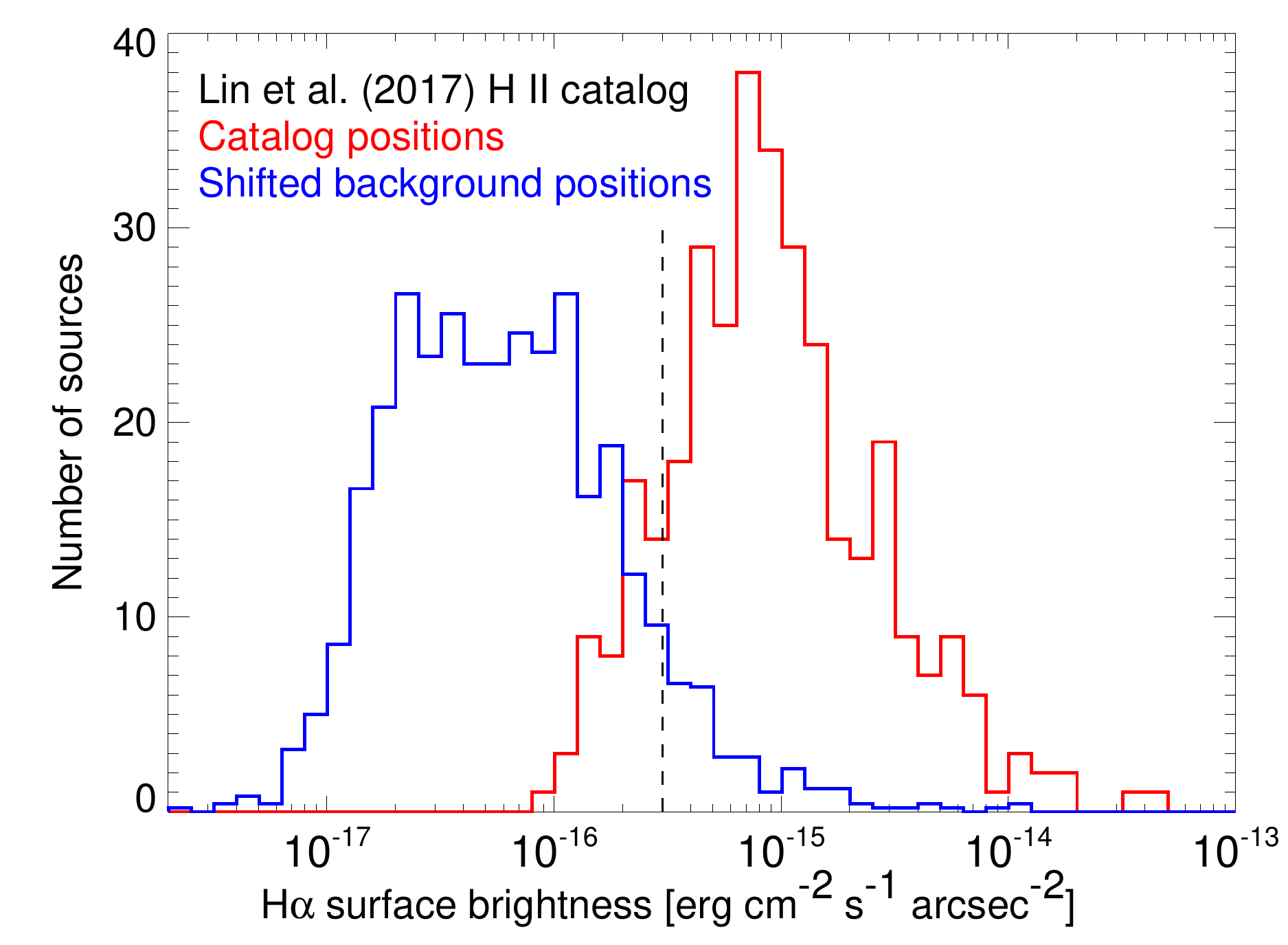}
    \caption{A comparison of the \ha\ surface brightnesses for the \cite{lin17} \hii\ region sample (red) with the background 
        \ha\ surface brightness distribution (blue).  The surface brightnesses are averaged from our \ha\ image over the region in the catalog.
        The background distribution was determined by shifting the true catalog positions by 1~arcmin to five different
        nearby positions.  We conclude that a surface brightness threshold
		$\Sigma(\mathrm{H}\alpha) > 3\times10^{-16}\,\mathrm{erg\,cm^{-2}s^{-1}arcsec^{-2}}$
		(dashed line)
        separates \hii\ regions from background sources with a contamination rate of 10\% by background sources.
    \label{fig:halpha_threshold}}
\end{figure}

To determine an appropriate $\Sigma(\mathrm{H}\alpha)$ threshold to identify \hii\ regions, we have used the
\cite{lin17} catalog with spectroscopic observations of 413 star-forming regions in M33.  This catalog is not a complete compilation of \hii\ regions in M33 because its sample was shaped
by the constraints of the fiber-fed spectrograph used for these observations.  Despite that limitation,
it represents a good subset of M33 ionized regions for our analysis.

Figure~\ref{fig:halpha_threshold} compares the \ha\ surface brightnesses for the \cite{lin17} fiber positions with
the background distribution computed from a set of shifted regions.  Five different shifts of
1~arcmin were used to determine the background distribution. We conclude that a
surface brightness threshold
$\Sigma(\mathrm{H}\alpha) > 3\times10^{-16}\,\mathrm{erg\,cm^{-2}s^{-1}arcsec^{-2}}$
separates \hii\ regions from
background sources, generating a reasonably reliable and complete sample.
The background rate is not negligible (about 10\% of the background positions
exceed this threshold), but some contamination is unavoidable in a galaxy this crowded with
star formation.

\section{Analysis}

\begin{figure}
    \includegraphics[width=\linewidth]{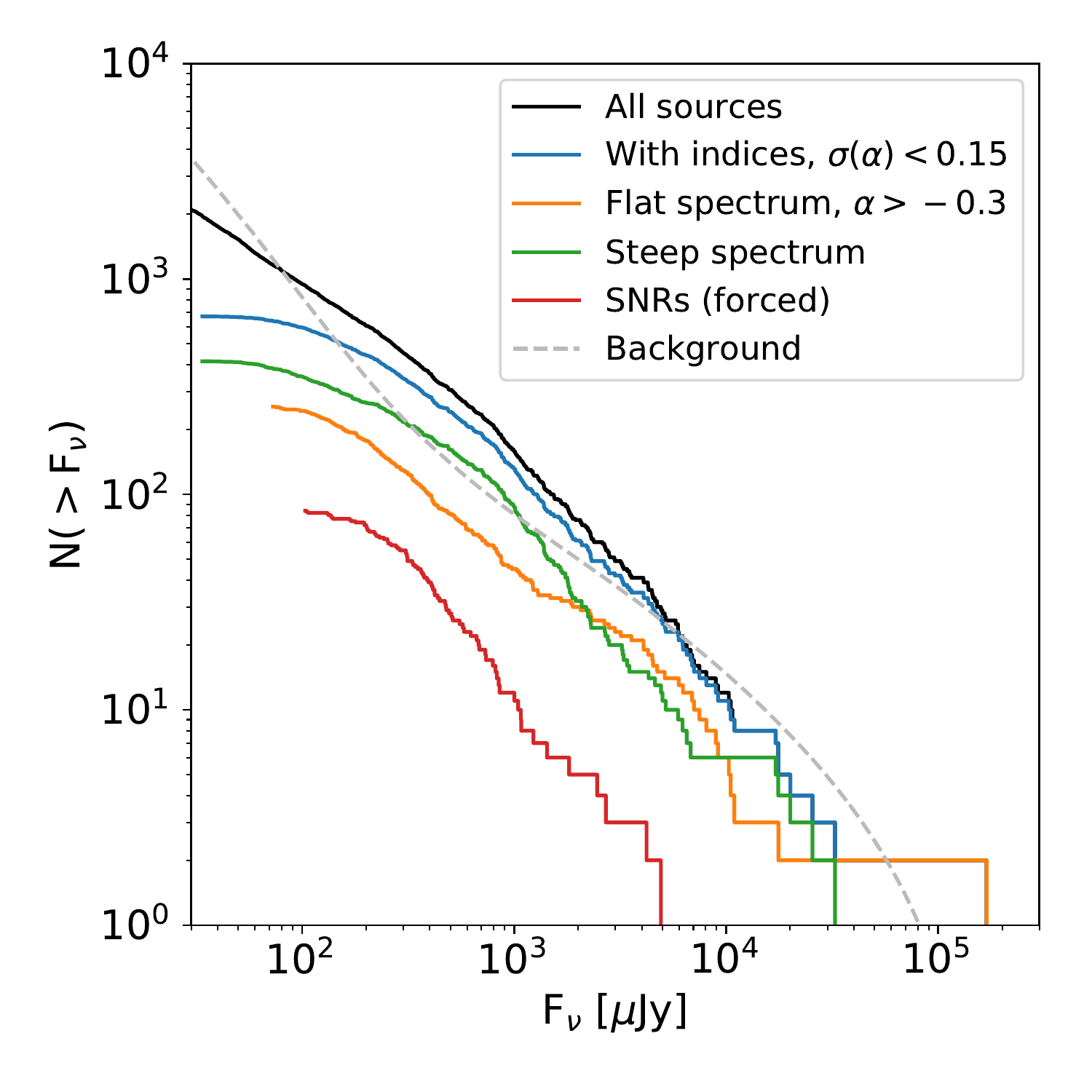}
	\caption{The number of sources with flux densities greater than F$_\nu$ for the entire catalog (in
	black) and various subsamples in other colors. Here flat-spectrum sources are defined as
	sources with spectral index $\alpha>-0.3$ and steep-spectrum as $\alpha \leq -0.3$.  Only
	sources with accurately measured spectral indices, $\sigma(\alpha) < 0.15$, are included
	in the flat/steep samples.  The flux distribution for the SNR forced photometry is also shown.
	The dashed gray line shows the expected background counts \citep{smolcic17}.
	\label{fig:lum_func}}
\end{figure}

In Figure~\ref{fig:lum_func}, we show cumulative luminosity functions for the catalog and for
various sub-sets thereof.  The higher luminosity portions of each curve can be fitted with a
straight line, but all curves show apparent ``breaks'' indicating limits to the completeness of the
sample being plotted.  It is difficult to set a single limiting flux for the survey because the
background levels change with position and with the complexity of emission surrounding individual
sources.

In this and other plots of the paper, we select samples of sources with accurately
determined spectral indices using $\sigma(\alpha) < 0.15$.  One third of the radio sources
fall in the outer parts of the survey where only 1.4~GHz data is available (Fig.~\ref{fig:overview}).
Excluding those sources, which do not have accurate indices, 90\% of objects with $\mathrm{F(1.4\,GHz)} > 100$~\uJy\ have
accurate spectral indices (see Fig.~\ref{fig:spnoise} for details).

Background radio galaxies and AGN contribute significantly to the source counts.  Figure~\ref{fig:lum_func} also shows
the background counts expected based on \cite{smolcic17}.  Many of the brightest sources ($> 10\,$mJy) are likely to be background
objects, but the bulk of fainter objects belong to M33.

\begin{figure*}
	\centering
	\includegraphics[width=0.9\linewidth]{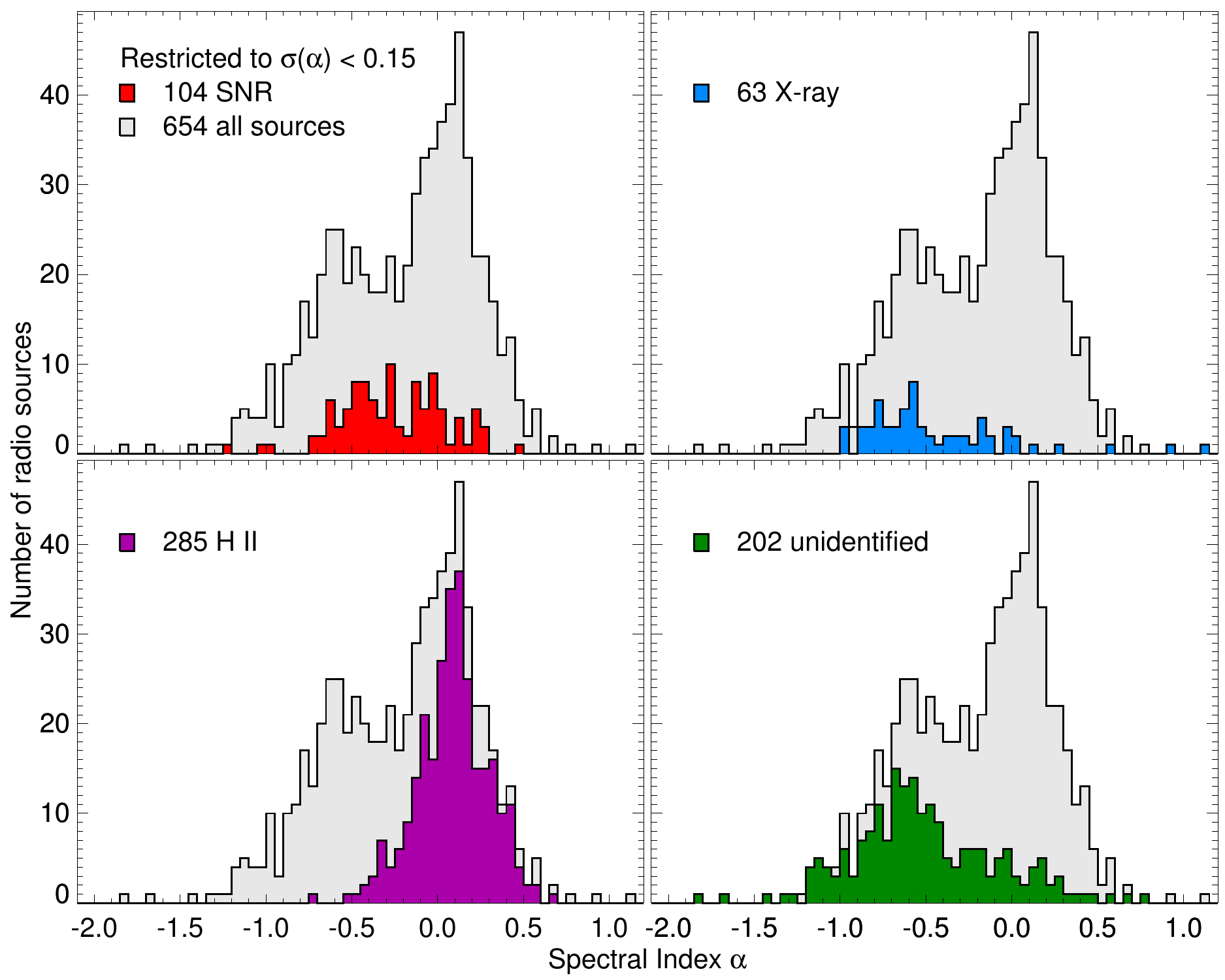}
	\caption{The distribution of spectral indices for sources
	whose spectral index was measured  to an accuracy of 0.15
	or better.  The gray shaded histogram shows the entire source
	distribution and is the same in each panel.  The four panels
	highlight sources associated with SNRs, \hii\ regions, X-ray sources
	(excluding SNRs and \hii\ regions), and unidentified sources not
	associated with known objects.  Potentially confused objects with
	flags less than 9 (Table~\ref{tab:flags}) are excluded.
    More than
		40\% of radio sources with $-0.5 < \alpha < -0.2$ are associated with
		SNRs.
	\label{fig:spectral_index}}
\end{figure*}

The distribution of measured spectral indices for the sources in the catalog is shown in
Figure~\ref{fig:spectral_index}.  It is double peaked, with one peak near spectral index 0 and one
near $-$0.6, as one would expect if there were two populations of sources, one dominated by \hii\
regions and one dominated by extragalactic sources and SNRs.  The four panels of the plot show the
overall distribution in gray (identical in each panel) along with the distributions for objects
identified as SNRs, \hii\ regions, X-ray sources (not including SNRs or \hii\  regions), and unidentified
sources.  There are clear differences among the different populations.  Essentially all of these
sources associated with \hii\ regions have spectral indices $\alpha \sim 0$ that one would associate
with optically thin, thermal free-free emission.  By contrast, the sources associated with SNRs and
with X-ray sources have spectral indices which are mostly negative and therefore associated with
non-thermal emission.  Most objects that are not SNRs or \hii\ regions are background radio galaxies
and AGN, although there are also a few radio sources associated with X-ray binaries in M33.

\begin{figure*}
    \centering \leavevmode
    \includegraphics[width=0.35\linewidth]{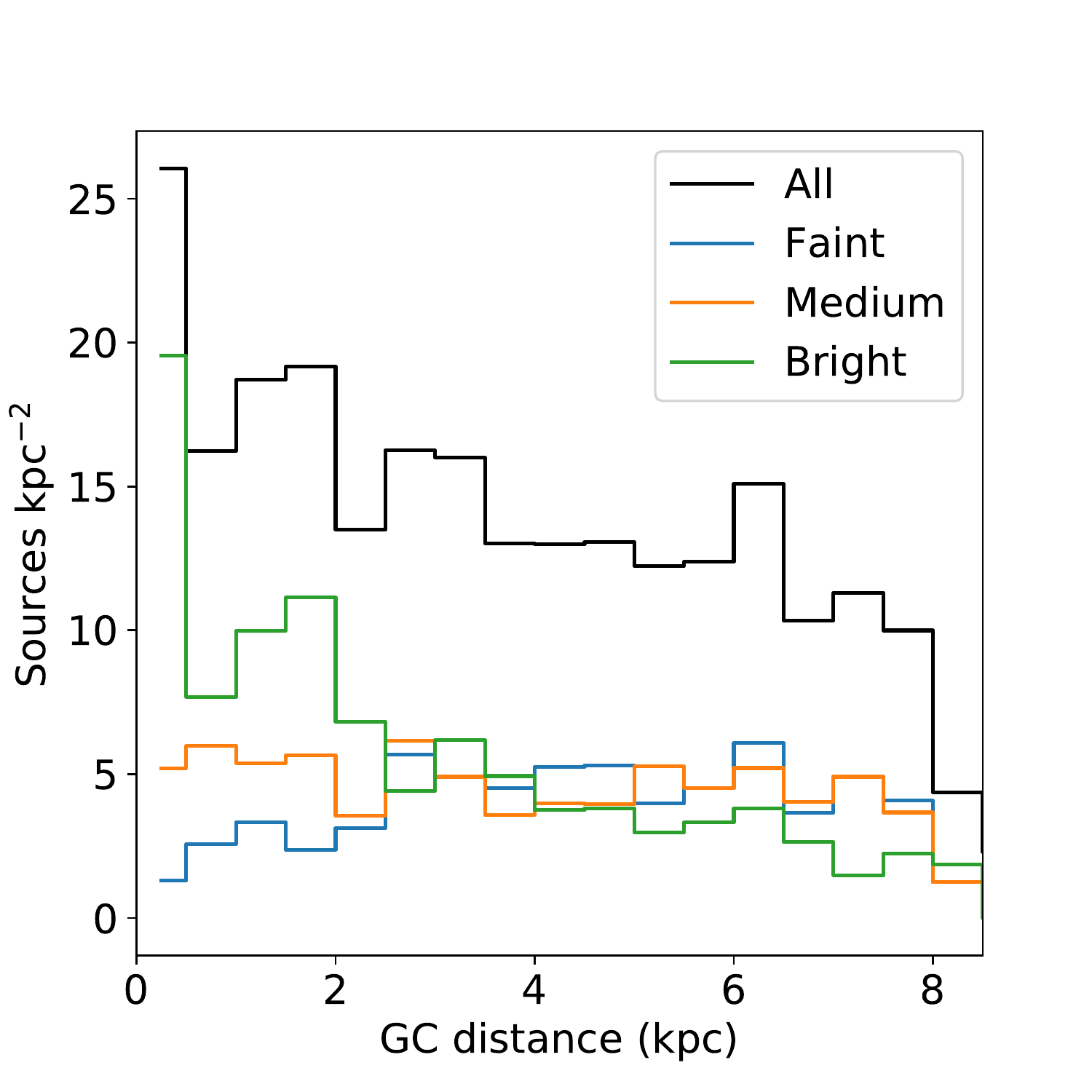}\hfil
    \includegraphics[width=0.35\linewidth]{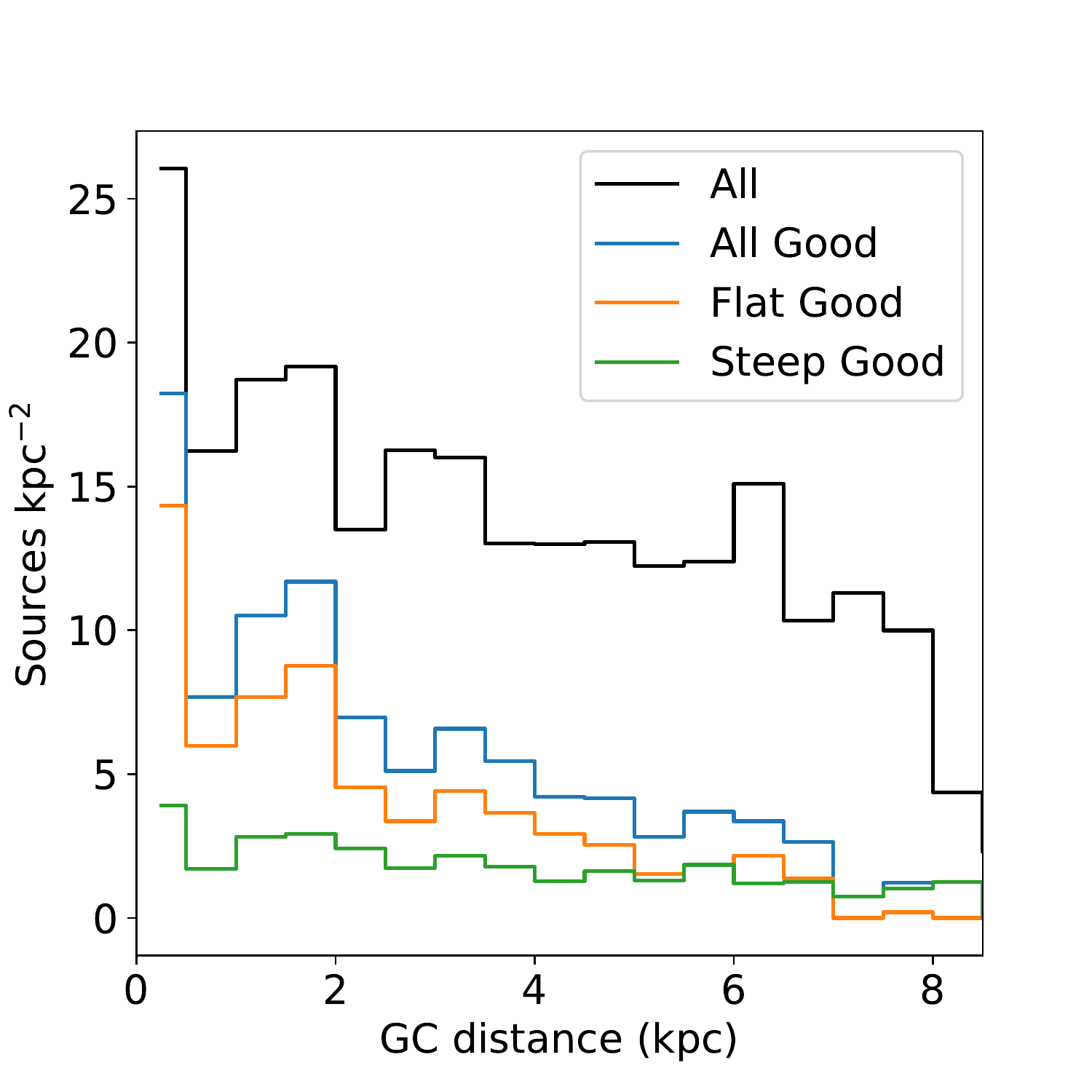}
    % \plottwo{f14a.pdf}{f14b.pdf}
	\caption{The surface density  of sources as a function of deprojected galactocentric
	distance (GCD).  Only sources with 5~GHz data are included.
	Beyond about 5 kpc the densities have been corrected for the limited portions of M33 that were observed.
	% Beyond about 5 kpc only portions of M33 were observed and so the densities
	% have been corrected for this.
	Both panels show the source densities for all of the sources
	in the catalog in black.  The left panel shows the distribution of the brightest third of
	the sources (in green), the middle third (in orange) and the faintest third (in blue).  Bright
	sources tend to appear closer to the center, as expected.  The right panel shows the surface
	density for all sources with well measured spectral indices (in blue), 
	flat-spectral indices (in orange) and steep indices (in green).  The
	steep-spectrum source distribution is nearly flat, consistent with the majority of these sources 
    being background AGN. \label{fig:gc_distance}}
\end{figure*}

The distribution of sources as a function of deprojected galactocentric distance (GCD) is shown in
Figure~\ref{fig:gc_distance}.
GCDs were calculated assuming standard M33 values for the major axis
position angle (23$^\circ$) and inclination (56$^\circ$) \cite[see][]{zaritsky89}.  Beyond about 5 kpc, our radio map is not
complete, so all of the distributions were corrected for the fraction of the galaxy observed at each
GCD.  As shown in the figure, the bright sources are more concentrated (at least in terms of
density) at smaller GCDs than the faint sources.  This is what one would expect if a larger fraction
of the faint sources were extragalactic. Indeed, there appears to be a deficit of faint sources near
the center, most likely due to source crowding there.  When the sources are split according to
spectral index (-0.3), those with flat spectra are more concentrated toward the center, consistent with the
hypothesis that most of the flat spectrum sources are due to \hii\ emission in M33.  By contrast the
non-thermal sources have a flat distribution, indicating that the majority of these are background
sources.

\subsection{Supernova Remnants Detected in Our Radio Survey}

Of the list of 217 SNRs and SNR candidates compiled by \citetalias{long18}, 216 are in the region covered
by our radio survey and 155 are detected above 3$\sigma$ via forced photometry, including 84 that have radio spectral indices with an
uncertainty of less than 0.15.  There are 122 of the detected
objects where spectroscopy has confirmed \sii:\ha\ flux ratios exceeding 0.4, the
usual optical criterion that an emission nebula is a SNR, and of these 92 have X-ray detections.

\begin{figure*}
	\plottwo{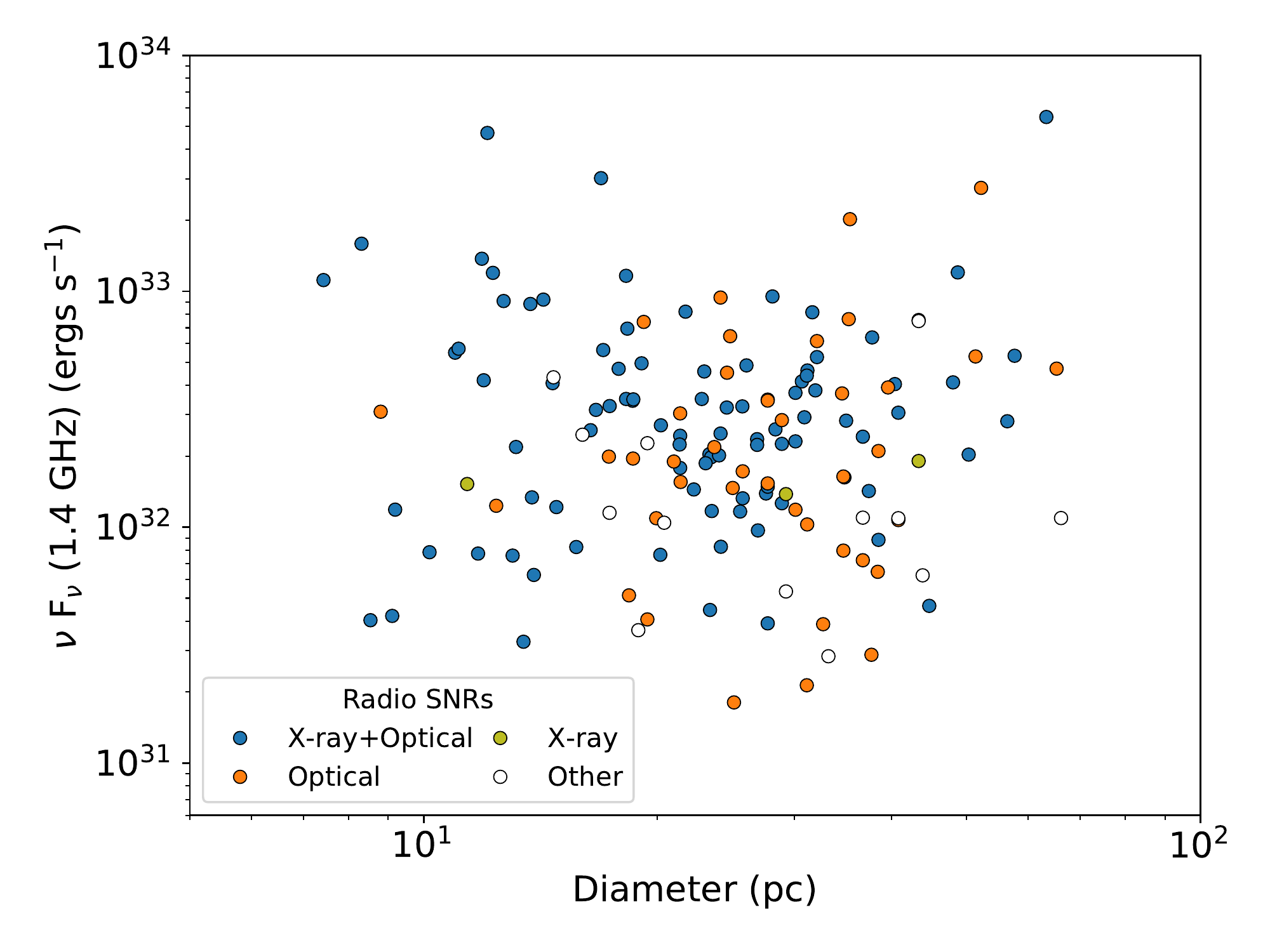}{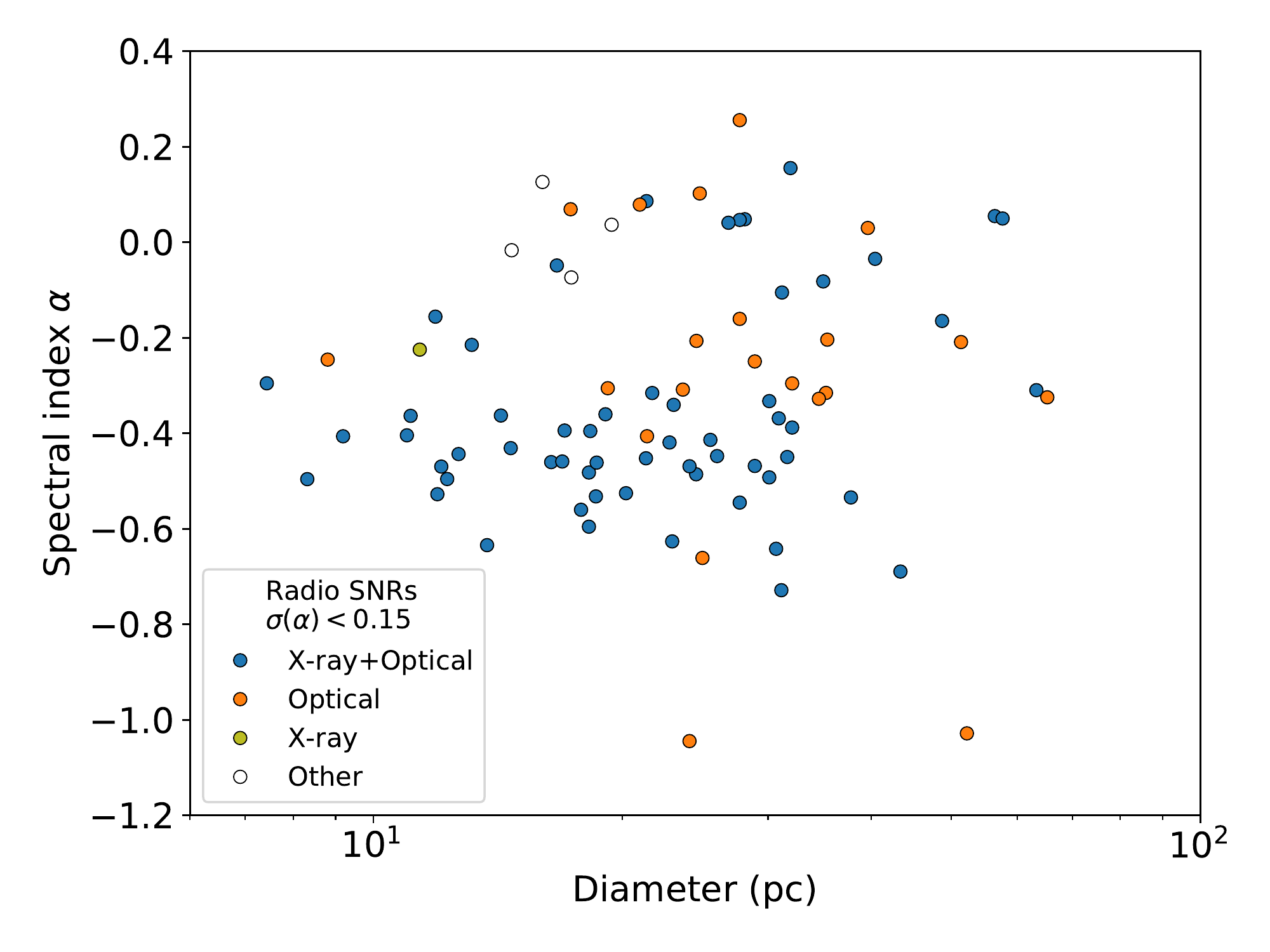}
    % \centering \leavevmode
    % \includegraphics[width=0.35\linewidth]{f15a.pdf}\hfil
    % \includegraphics[width=0.35\linewidth]{f15b.pdf}
	\caption{Left: The 1.4~GHz radio luminosity of SNRs detected through forced photometry as a
	function of SNR diameter.  Right: Accurately determined radio spectral indices for the SNRs.  The
	sample has been divided into objects with an X-ray detection plus optical confirmation of high
	[S~II]:\ha\ ratio (blue), objects with only optical spectroscopic confirmation (orange), objects with
	only an X-ray detection (green), and
	objects identified as possible SNRs through optical imaging, but without confirmation through spectroscopy or
	X-ray emission (open circles).  \label{fig:snr_dia}}
	% objects without confirmation in the X-ray or the optical (open circles).  \label{fig:snr_dia}}
\end{figure*}

The derived radio luminosities at 1.4~GHz and spectral indices of these objects are shown as a
function of SNR diameter in Figure~\ref{fig:snr_dia}.  Neither the luminosity nor the spectral
indices show a significant correlation with diameter.  Radio luminosities for the entire sample and
the various subsamples show very large dispersions at all diameters.  Smaller diameter SNRs are not
significantly brighter or fainter than those at large diameter.  Similarly, radio spectral index
does not appear to evolve with diameter.  \changed{In contrast, a correlation between spectral index and diameter has been
reported for the LMC \citep{bozzetto17}; we discuss this difference further in section~\ref{sec:othergals}.}

\begin{figure*}
	\plottwo{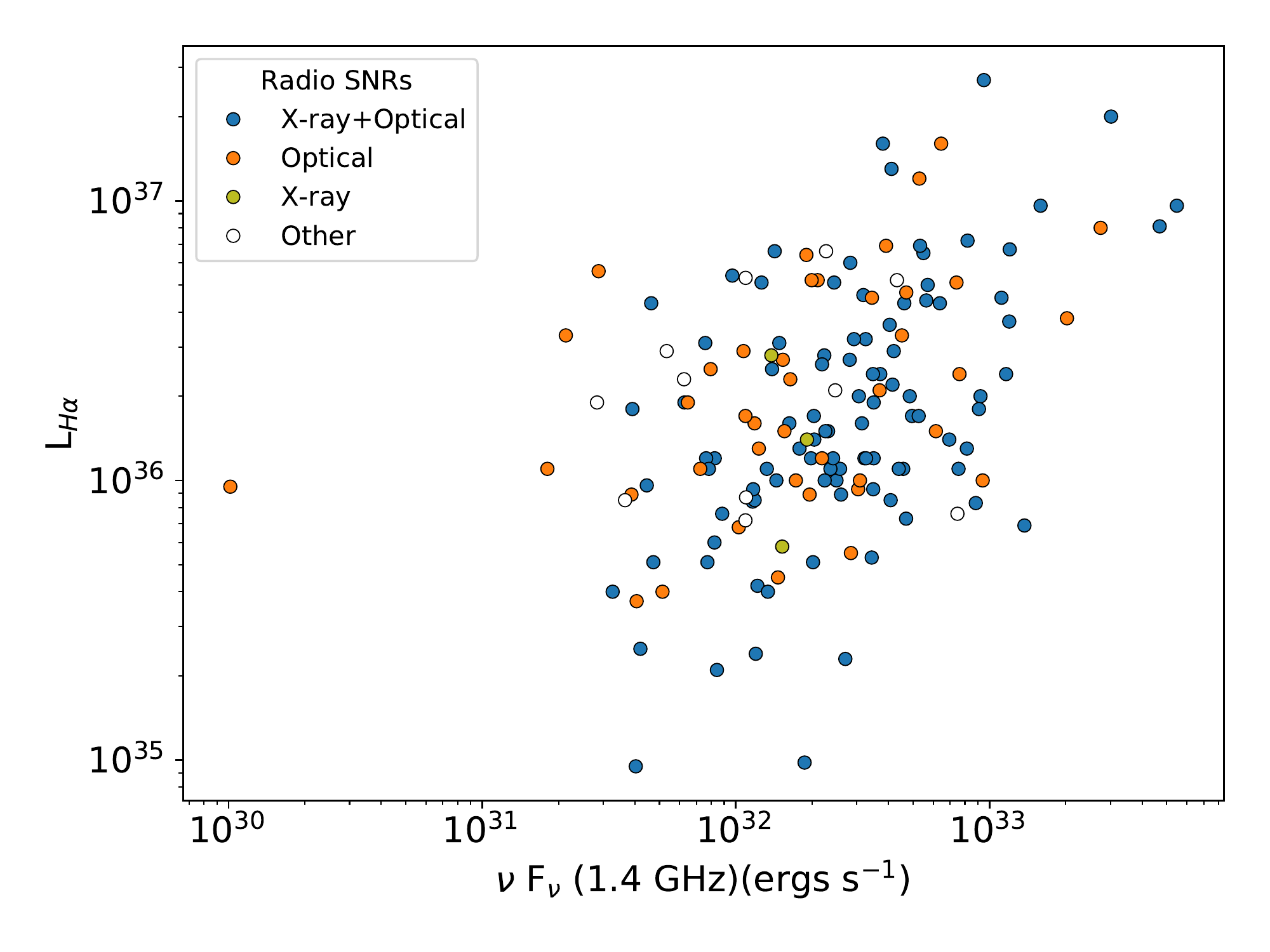}{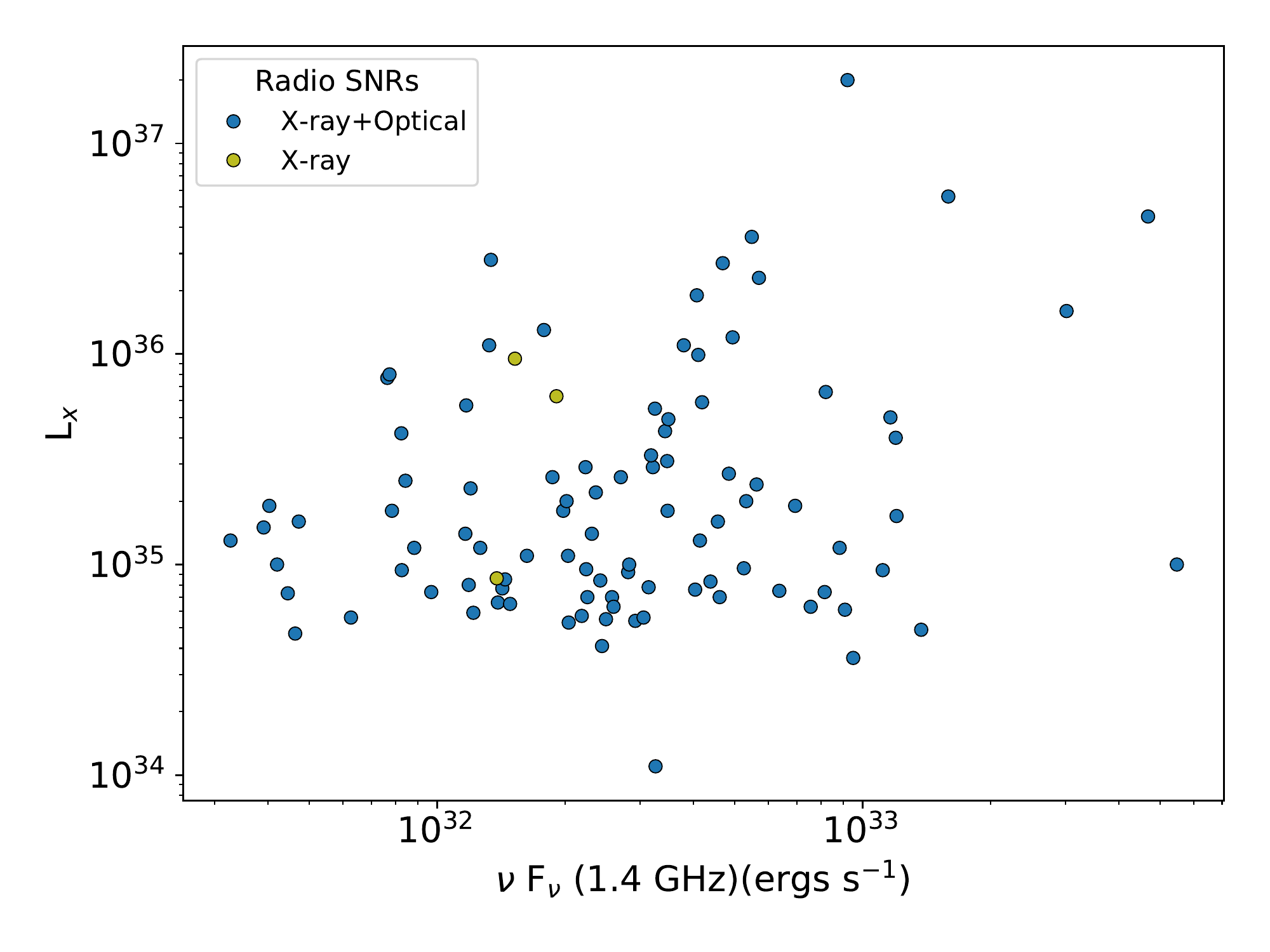}
	\caption{Left: \ha\ luminosity as a function of 1.4~GHz luminosity for SNRs detected through
	forced photometry.  Right: X-ray luminosity as a function of the 1.4~GHz luminosity.  The
	points are color coded as in Fig.~\ref{fig:snr_dia}.   \label{fig:xray_ha_corr}}
\end{figure*}

Comparisons between the 1.4~GHz luminosity and the \ha\ and X-ray luminosities are shown in
Figure~\ref{fig:xray_ha_corr}.  Although there is considerable scatter, there appears to be a trend
between \ha\ and radio luminosity.  More luminous SNRs in \ha\ appear to be more luminous at radio
wavelengths.  Whether this trend is physical is hard to determine because the scatter is large.  The
SNRs in M33, like those in essentially all other galaxies, were initially identified through optical
interference filter imaging, and are thus surface brightness limited.  Larger SNRs tend to have higher
\ha\ luminosities as a result.  By contrast, the X-ray sample is luminosity limited.   This accounts
for the fact that there are almost no SNRs with L$_x$ less than \EXPU{4}{35}{\LUM}, the approximate 
limit of the \chandra\ survey.  That said, it
is interesting that the objects with highest X-ray luminosities also tend to have the highest radio luminosities.

\begin{figure}
    \includegraphics[width=\linewidth]{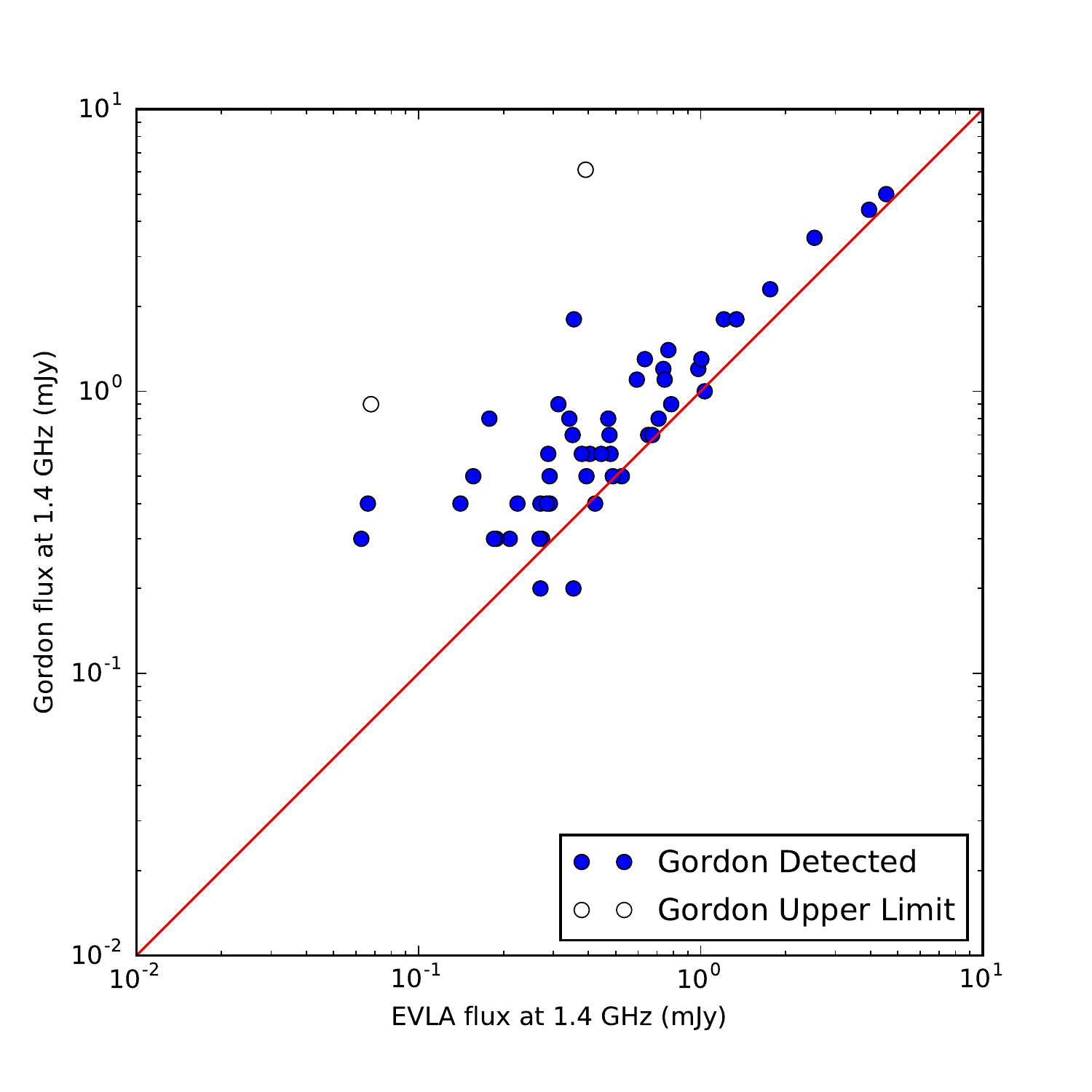}
	\caption{A comparison between our measured 1.4~GHz flux densities in the forced photometry catalog
	and those reported by \citetalias{gordon99}. The open circles show sources that \citetalias{gordon99}
	reported as detections at 5~GHz, but only upper limits at 1.4~GHz. The points are plotted at
	the upper limit values. \label{fig:gordon}}
\end{figure}

\subsubsection{Comparison to Previous M33 Radio Detections of SNRs}

As noted in the Introduction, the last detailed radio study of SNRs in M33 was carried out by \citetalias{gordon99},
who reported the radio detection of 53 out of 98 optically-identified SNR candidates that were known
at the time.  Of these 53 objects, 52 are in the region covered by our radio
survey.\footnote{G98-03=L10-003 was not in the region we surveyed.  Note that \citetalias{gordon99} adopted
identifications from the optical SNR catalog in G98.} All of these are detected at 3$\sigma$ or
greater in our forced photometry catalog.  As shown in Figure~\ref{fig:gordon}, there is a fairly
good correlation between the flux densities we measure at 1.4~GHz (by constructing a weighted average of our
two 1.4~GHz bands) and those measured by \citetalias{gordon99}.   The \citetalias{gordon99} flux densities tend to be higher, and in a few cases
considerably higher. Many of the objects that are higher were identified by \citetalias{long10} as likely to be
associated \hii\ emission rather than emission from the SNR (or the SNR only).  But the median of
the ratio of the flux densities they measured is only 50\% higher than those we measure suggesting that we and
they are identifying the same radio sources as SNRs, but with differing amounts of
contamination by thermal emission.

\begin{figure*}
	\plottwo{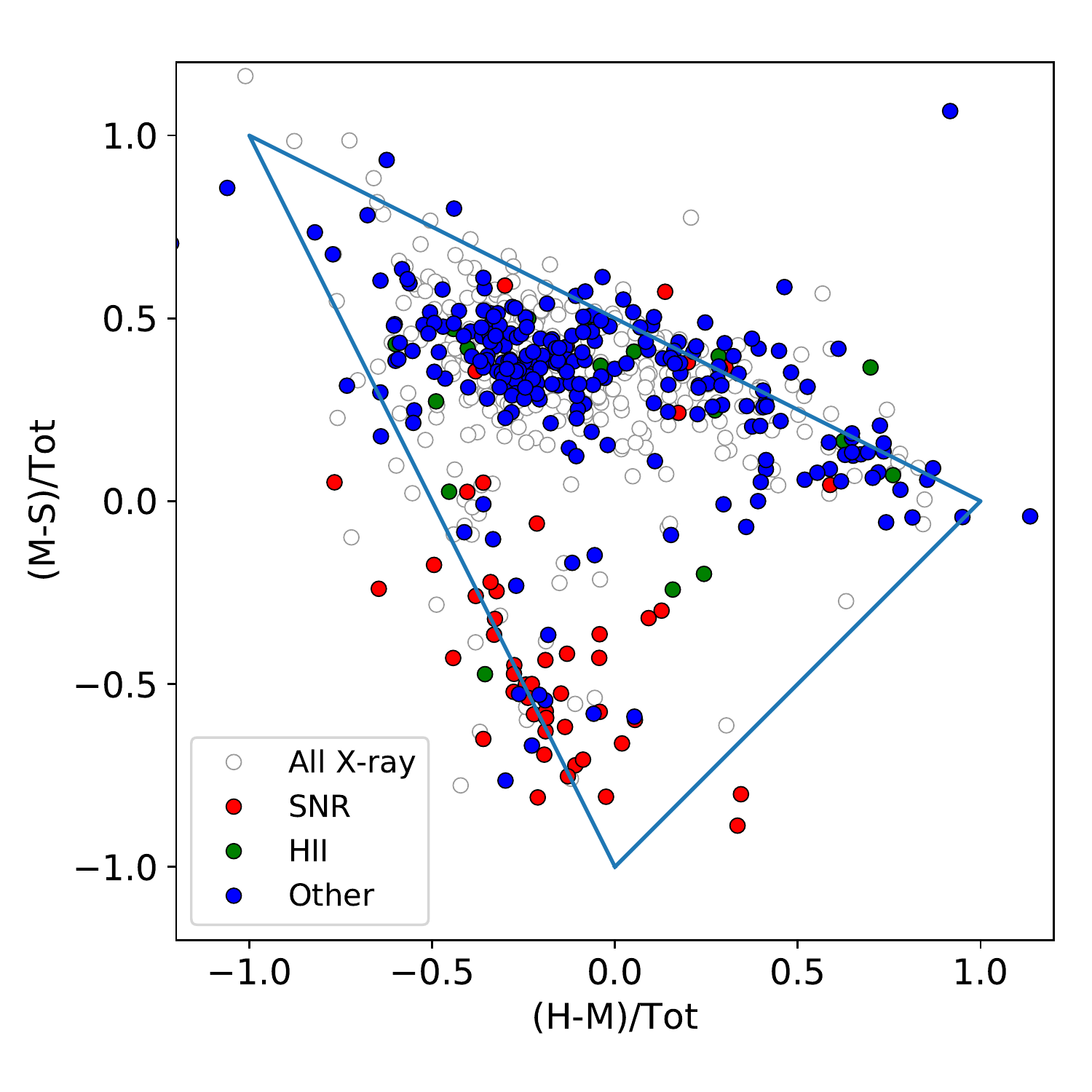}{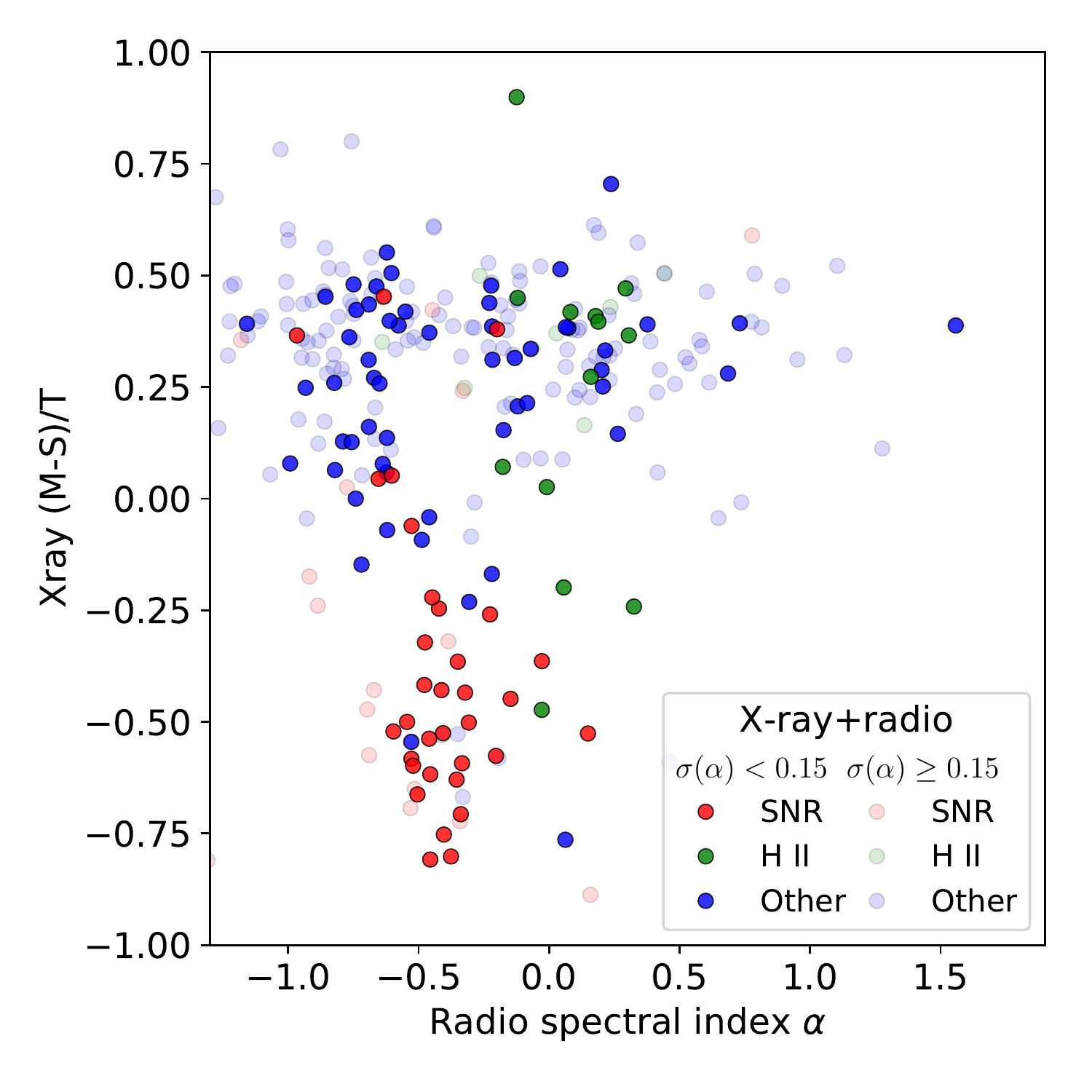}
	\caption{Left:  Hardness ratio diagram of X-ray sources from \cite{tuellmann11} that are
	detected via forced photometry in our radio map.  Sources identified with SNRs are
	shown in red, and \hii\ sources in green.  Open circles are X-ray sources that fall below
	the $3\sigma$ radio detection threshold.  Right:  X-ray hardness for the sources as
	reflected by the $(M-S)/T$ ratio as a function radio spectral index. Sources with
	noisy spectral indices are shown as pale symbols.
\label{fig:xray}}
\end{figure*}

\subsection{X-ray Sources Detected in Our Radio Survey}

The \cite{tuellmann11} catalog of X-ray  sources in M33 contains 662 X-ray sources, the vast
majority of which are due to background sources (galaxies and AGN).  We detect radio emission at the
position of 319 of these sources at 3$\sigma$ or greater via forced photometry.  Of these, 55 are
positionally coincident with SNRs and 25 with \hii\ regions observed by \cite{lin17}.  As shown in
the X-ray hardness ratio diagram in the left panel of Figure~\ref{fig:xray}, most of the X-ray
sources we detect lie in the region of the diagram where background sources and X-ray binaries are
expected to fall.  As expected, the sources identified with SNRs mostly have soft X-ray
spectra.  The X-ray sources associated with \hii\ regions are not concentrated in a particular region of the
diagram, suggesting a mix of source types (X-ray binaries in M33, unrecognized SNRs, and background objects).  The right hand panel of the figure shows X-ray hardness ratios as function of the radio
spectral index.  The SNRs form a well defined region in the diagram; this suggests that one might be
able to identify SNRs directly from such a diagram with sufficiently sensitive X-ray and radio
observations, even without a prior optical identification.

%\todo[inline]{Need to check that sources not already identified as SNRs in the lower right part of  Figure~\ref{fig:xray} are not SNRs.  Maybe we can find a new one.  Results above need to be updated for July 20th catalog, and possibly for the fact that we may ultimately define H~II regions by surface brightness.}

\begin{figure}
    \includegraphics[width=\linewidth]{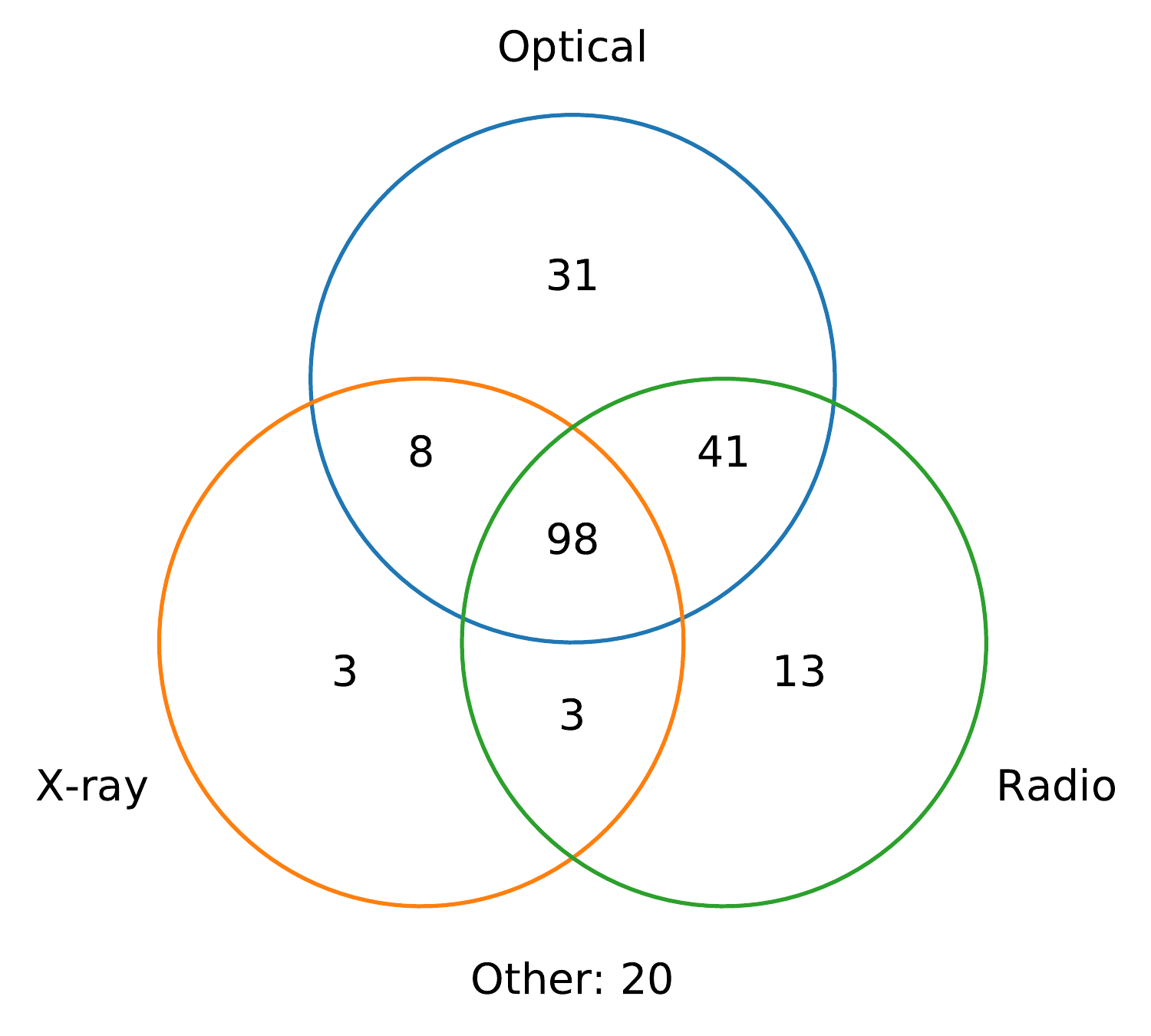}
	\caption{A Venn diagram showing the number of SNRs that have been detected in various
	wavelength bands and combinations thereof.  For this comparison, the optical SNRs must have a
	spectrum showing that the ratio \sii:\ha\ $\ge$ 0.4.  Optical SNR candidates for which no spectra 
    have been obtained or for which spectra failed to confirm a high ratio could still align with a radio
	or X-ray source.  These account for the sources shown as X-ray, radio, or radio--X-ray only.
    There are an additional 20 optical SNR candidates from imaging surveys that have no confirming optical 
    spectra and no radio or X-ray emission.  See text section 3.3 for full details. \label{fig:venn}}
\end{figure}

\subsection{Multi-wavelength Comparisons}

A Venn diagram showing the number of objects detected in the various wavelength bands is shown in
Figure~\ref{fig:venn}.  To be counted as a confirmed optical SNR, we require not only that a nebula
was suggested either by \citetalias{long10} or \citetalias{lee14} as a candidate based on interference filter imagery, but that a
spectrum exists that confirms the \sii:\ha\ ratio is at least 0.4, the conventional criterion for
declaring that an emission nebula is a valid optical SNR candidate.  There are 98 emission nebulae
satisfying this criterion, that also show up as positive detections in our radio forced-photometry catalog, and that have
been detected at 3$\sigma$ or greater with either \chandra\ or {\it XMM-Newton}. Additionally, there are 31
optical SNRs with spectral confirmations but without detected X-ray or radio counterparts, 41 that
have a radio counterpart but no X-ray, and 8 objects with optical--X-ray detections but no radio
counterpart.

The small numbers in the non-optical parts of the diagram deserve specific discussion.  The location
of an individual object in the Venn diagram depends on many factors, not the least of which is our
application of the optical spectral confirmation criterion above and beyond the imaging candidate
status.   There are optical SNR candidates that were identified from imaging but have no optical
spectra, and there are optical candidates for which the spectra obtained showed \sii:\ha\ ratios
somewhat below 0.4. The objects that lie in the X-ray (3), Radio (13), and radio--X-ray (3) portions of the
Venn diagram align with some of these objects. 

For the 13 ``radio-only'' objects, five do not have spectra, two were too contaminated for accurate
spectral measurements, and six have spectra with derived \sii:\ha\ ratios below 0.4. Three of these
objects had \sii:\ha\ ratios between 0.36 and 0.39, and the other three were near 0.25. However, 
for fainter objects or objects with nearby \hii\
contamination, even the spectrally-derived ratios can have significant errors bars.  If the background
\ha\ is even slightly under-subtracted in the spectra, the derived \sii:\ha\ ratio can be
artificially lowered.  The presence of associated radio emission for these eight objects
significantly enhances the probability that these nebulae are SNRs.

Of the three `X-ray-only' objects, one has no optical spectrum, one is too contaminated to make a good
measurement, and one has a spectrum with a \sii:\ha\ ratio of 0.27, significantly below the threshold 
(LL14-008).  The presence of an X-ray source at the optical position strengthens these as possible
SNRs as well.

The three objects shown as having both radio and X-ray emission would already seem to be strong
candidates.  All three have optical spectra with derived \sii:\ha\ ratios of 0.37 (L10-035), 0.33
(L10-040), and 0.37 (LL14-005).  Again, slight \ha\ background subtraction problems could have
caused these to fall below our normal threshold ratio, so these are quite likely SNRs.

There are 20 emission nebulae that, other than their original selection as nebulae with enhanced
\sii:\ha\ ratios from imaging, have no additional supporting information to suggest that they are
SNRs, and thus do not appear in this diagram.  This does not necessarily mean that they are not 
SNRs since some objects have not been
observed spectroscopically, and \changed{one other was} simply outside of the region surveyed with the JVLA.
Fourteen of the 20 have spectra that show ratios below the normal threshold, so without further
corroboration the majority of these are probably not SNRs.

\begin{figure*}
    \includegraphics[width=\linewidth]{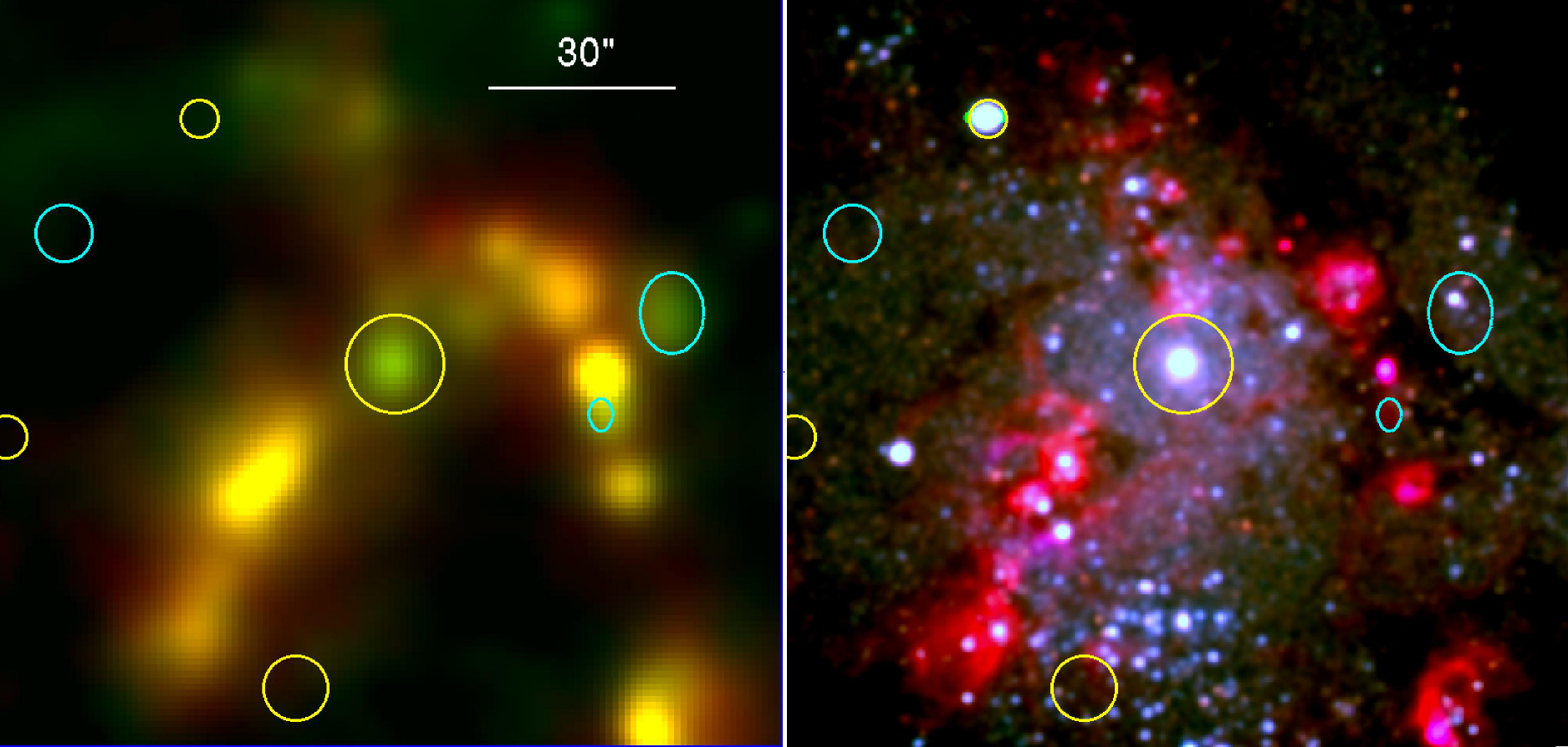}
	\caption{A view of the 2\arcmin\ region near the center of M33.  The same data shown in
	Fig.~\ref{fig:radopt} have been rescaled to just highlight the nuclear region. A steep-spectrum
	(green) source in the left panel corresponds to the bright optical source in the nucleus;
    numerous nearby \hii\ regions are bright radio sources as well. Yellow region overlays
	indicate X-ray source positions from Long et al. (2010) and the cyan regions indicate
	optical SNR detections nearby. \label{fig:nuclear}}
\end{figure*}

\section{Discussion}

\subsection{The Nuclear Source and other X-ray binaries}

With an X-ray luminosity of $\sim$\POW{39}{\LUM},  M33 X\nobreakdash-8, the source coincident with the the
nucleus of M33, is one of the brightest persistent X-ray sources in the Local Group
\citep{long81}.  The nuclear star cluster is compact but otherwise unprepossessing optically; its spectrum can be modeled
in terms of two bursts of star formation, one with an age of 40 Myr involving stars with a total mass of
9000\MSOL, and another of age 1 Gyr with a mass of 76,000 \MSOL \citep{long02}. If a black hole
exists at the center of the cluster it must have a mass $<$1500\MSOL \citep{gebhardt01}.  It has
been suggested (although not confirmed) that the nuclear source has a period of 109 days \citep{dubus97}. Taken together,
these observations imply the source is not an AGN but is instead an $\sim$10\MSOL\ X-ray binary radiating
near the Eddington limit \cite[see, e.g.][]{foschini04,middleton11}.   

As shown in Figure~\ref{fig:nuclear}, a non-thermal radio source, W19-0683, is coincident with the nucleus. 
The spectral index is $-0.39\pm0.11$  and the flux at 1.4~GHz is $0.61\pm0.04$~mJy.
The radio source is extended, with a ratio of peak to integrated flux density of 0.35 (compared with
unity for a point source), and a nominal major axis of 18\arcsec, or 70 pc.
In the forced photometry catalog, where source T11-318 corresponds to the nuclear source,
the 1.4~GHz flux is similar, $0.60\pm0.04$~mJy, with a slightly steeper spectral index of $-0.68\pm0.17$.
(The spectral indices in the main catalog and the X-ray catalog are consistent, differing by $1.4\sigma$.)
The nuclear source was also observed by \citetalias{gordon99} who
reported a consistent flux density ($0.6\pm0.2$~mJy at 1.4~GHz) and spectral index  ($-0.8\pm0.3$).
For comparison, SgrA* in our Galaxy, which arises from the accretion disk of a $4 \times 10^{6}$\MSOL\
black hole (BH), is a 0.7~Jy source at 1.4~GHz, corresponding to $\sim 0.07$~mJy at the distance of M33, and has a
spectral index of $-0.3$ \citep{duschl94}.

Whether M33 X\nobreakdash-8  is very low mass BH at the nucleus of M33 or simply, as is generally thought, a BH
binary, the existence of radio emission is not surprising, given that BH systems on all mass scales
produce radio emission \cite[see, e.g.,][]{plotkin12}.   The 1.4~GHz radio luminosity of M33 X\nobreakdash-8 is
about \EXPU{6.7}{32}{\LUM}, less than \POW{-6} the X-ray luminosity.   Exactly how to fit M33 X\nobreakdash-8
into the zoo of BH binaries remains unclear. It has properties that resemble a microquasar trapped
in the so-called high soft state, where disk accretion is near the Eddington rate
\citep{long02,foschini04}.  Most microquasars, however,  exhibit major outburst events, and M33 X\nobreakdash-8
has not been seen to vary appreciably.  During the rise to X-ray maximum, microquasars show active
radio jet emission (with flat spectral indices) and hard X-ray spectra.  At outburst maximum, however, the
X-ray spectra transition to something similar to that of M33 X\nobreakdash-8 and the radio spectral indices
steepen.  The idea is that active jet generation has ceased (or ebbed at least), and the residual
emission arises from the interaction with the surrounding medium \citep{fender04}.

Because of its X-ray luminosity and the fact that the luminosity is persistent, M33 X\nobreakdash-8 has also
been characterized as an ultraluminous X-ray source (ULX). It is at the low luminosity end of
sources classified as ULXs that also have X-ray spectra fit in terms of steady state accretion onto
a massive BH.  Other ULXs in this state observed at radio wavelengths have spectral indices
characteristic of optically thin synchrotron emission. The radio emission for a number of the
nearest ULXs (and microquasars) has been resolved, and appears to arise from a jet interaction with
the surrounding ISM \citep{pakull03}. The typical sizes of the resolved bubbles around these sources
are 100-300 pc.   At the resolution of our observations (about 20 pc), the bulk of the radio
emission in M33 X\nobreakdash-8 appears point-like, although, as mentioned above, the source is extended at the $\sim$70 pc level.  Thus, source extent does not rule out the jet-interaction scenario
in M33 X\nobreakdash-8. The prototypical high-mass X-ray binary Cygnus X-1 is surrounded by a 5 pc diameter radio
ring that appears to have been inflated by a jet from the central source \citep{gallo05}.  Clearly,
sensitive, higher resolution radio observations are needed to clarify the situation for M33 X\nobreakdash-8.

%\todo[inline]{Note that the nominal major axis for this source, 18~arcsec, translates to 70~pc.  That
%does seem reasonably consistent with the ULX interpretation. rlw}

% \todo[inline]{There have been suggestions in the past that this is a microquasar and this seems to justify that.  Note that the source was also detected by Gordon, although they did not remark onthis much. But microquasars are usually variable, and this source does not appear to be.  microquasars show flat spectra in the hard low state, and then they tend to have steeper spectra later.  This is usuall interpreted as the ject becoming optically thin.  For ULX's many show radio emission, which fairly standard non-thermal spectral indices.  Often the emission is extended though and we do not see that here.   KSL.  Note the fluxes and spectra index given here are from the full radio catalog, which is not what I think we want.  From the x-ray forced photometry catalog, we have     T11-318     603.8037--53.252003  and a spectral index of -0.6809709+-0.1745642.   There is a fundamental relationship between X-ray luminosity and radio luminosity for BH that is said to work at all mass ranges, see Plotkin et al 2012.  We need to see where }

In addition to detecting a nuclear radio source, radio sources were also detected
at the positions of several other known X-ray binaries in M33. Among these is a source with a
1.4~GHz flux density (in the forced photometry catalog) of $0.63\pm0.02$~mJy at the position of M33 X\nobreakdash-7,
the well-studied eclipsing BH binary, comprised of a 16\MSOL\ BH orbiting a 70\MSOL\ O-star
every 3.45 days \citep{orosz07}.  However, the spectral index of this source is $0.08\pm0.02$ and
the \ha\ surface brightness at this position is high, about \EXPU{3.4}{-15}{ergs~cm^{-2}s^{-1} arcsec^{-2}},
suggesting that most if not all of the radio emission arises from \hii\ regions along the line of
sight.   A similar situation obtains for M33 X\nobreakdash-4, another bright X-ray source thought to be an X-ray
binary \citep{grimm07}, where the forced photometry 1.4~GHz flux density is $0.17\pm0.01$~mJy, but
the apparent spectral index is $+0.3\pm0.1$. This source is also located along a line of sight
with considerable \ha\ flux, \EXPU{1.1}{-15}{ergs~cm^{-2}~s^{-1} ~arcsec^{-2}}.  Higher resolution
observations are required to determine whether either of these sources has any intrinsic radio
emission.

\subsection{The M33 SNR Sample: Models vs.\ Data}

Our large sample of SNRs with both radio and X-ray luminosities enables some useful comparisons to models for
SNR radio emission.  \cite{sarbadhicary17} describes a relatively simple analytical model for the radio
luminosities of SNRs as a function of their radius, evolutionary phase, ISM density,
explosion energy, etc.
According to equation~(A12) in \cite{sarbadhicary17}, the radio luminosity density $L_{\nu}$ at 1.4~GHz is given by:
\begin{equation}\label{eqn:sar}
\begin{aligned}
	L_{1.4} = {}& \EXPU{2.2}{24}{ergs~s^{-1}Hz^{-1}}
\left( \frac{R}{10\,\mathrm{pc}} \right)^3 \\
& \left(\frac{\epse}{10^{-2}} \right)
\left(\frac{\epsu}{10^{-2}} \right)^{0.8}
\left( \frac{v_s}{500~\mathrm{km~s}^{-1}} \right)^{3.6}
\end{aligned}
\end{equation}
where $R$ is the shock radius, $v_s$ is the shock velocity and $\epse$  and $\epsu$
are the fractions of the post shock energy density $\rho v_s^2$ converted into relativistic electrons
and the amplified upstream magnetic field in a diffuse acceleration model, respectively.

If this model were correct, and if all other factors except the shock velocity were the same,
$L_{1.4}$ would be a strong function of $v_s$.  Indeed, a factor of two difference in shock velocity
would correspond to a factor of 12 difference in $L_{1.4}$.  The radio luminosity also increases in
proportion to the SNR volume, $L_{1.4} \sim R^3$. As a result, during the SNR free expansion phase,
when the shock velocity is constant, the radio luminosity increases rapidly as $R^3 \sim t^3$.  The
radio luminosity peaks at the beginning of the Sedov phase, declines through the adiabatic expansion
phase as the shock slows, and finally decreases rapidly in the radiative phase as $v_s$ drops off
and the SNR expansion slows.

\cite{sarbadhicary17} assume that  $\epse$  is constant with time, citing \cite{soderberg05} and
\cite{chevalier06}.
However, $\epsu$ is not a constant
in this equation but depends on the magnetic field, shock velocity, and other parameters.
Following the discussion in their appendix, the definition of $\epsu$ is:
\begin{equation}
\epsu=\frac{B^2/8\pi}{\rho v^2_s}
\end{equation}
Citing \cite{bell04}, \cite{sarbadhicary17} argue that at high Alfv\'en Mach number ($M_A$), magnetic
amplification saturates at a value given by:
\begin{equation}
\frac{B^2}{8\pi} \sim \frac{1}{2} \frac{v_s}{c} \xi_{cr} \rho v^2_s
\end{equation}
where $\xi_{cr}$ is the fraction of shock energy that goes into relativistic particle acceleration (including electrons and ions).
They then assume $\xi_{cr}$ is a constant value of 0.1 for high Alfv\'en Mach numbers but declines for smaller $M_A$:
\begin{equation}
	\xi_{cr} =
	\begin{cases}
		0.1 & \text{when } M_A > 30 \quad , \\
		10^{-2} \left( 0.15\,M_A + 6 \right) & \text{when } M_A < 30 \quad .
	\end{cases}
\end{equation}
Here $M_A<30$ is an approximate limit -- $\xi_{cr}$ increases in proportion to $M_A$ until it
saturates at a value $\xi_{cr}=0.1$.

The final equation that \cite{sarbadhicary17} use for $\epsu$ is
\begin{equation}\label{eqn:epsu}
\epsu=\frac{\xi_{cr}}{2} \left(\frac{v_s}{c}+\frac{1}{M_A}\right)
\end{equation}
That makes a transition from values that are proportional to the shock velocity $v_s$ (at high $M_A$
values) to values that are proportional to $1/M_A$, meaning $\epsu$ scales as $1/v_s$, at lower
shock velocities.  Coincidentally, Figure~A1 in \cite{sarbadhicary17} shows that the transition
between these scalings occurs close to the beginning of the Sedov phase of evolution.  In the
example they show, the value of $\epsu$ varies by only a factor of three from the initial SN
explosion to an age of $\EXPU{2}{4}{yr}$; given the modest dependence of $L_{1.4}$ on $\epsu$
this implies that $\epsu$ contributes only a factor of $\sim$2 to the variation in $L_{1.4}$ over a
SNR lifetime.  That is why the treatment of $\epsu$ as a constant in equation~(\ref{eqn:sar}) is
approximately correct.

The $\epsu$ term also introduces dependencies on the density $n_H$ and magnetic field $B_0$ in the
ISM. And the magnetic field itself depends on the ISM density since it is coupled to the ISM state.
Equation~(A4) of \cite{sarbadhicary17} can be rewritten to give the ISM magnetic field as
\begin{equation}\label{eqn:bscaling}
	B_0 = 9\,\uG\, n_H^{0.47} \quad ,
\end{equation}
where $n_H$ is the ISM hydrogen number density.\footnote{This is a corrected version of the equation in the paper,
which has a normalization error (Sarbadhicary, private communication).}

To see what this model for the radio emission means, it is useful to recall the Sedov equations:
\begin{equation}
R = 5.2\,\mathrm{pc} \left(\frac{E_{51}}{n_H}\right)^{1/5} \left(\frac{t}{1000\,\mathrm{yrs}}\right)^{2/5}
\end{equation}
\begin{subequations}
\begin{align}
	v_s &= 2000 \VEL  \left(\frac{E_{51}}{n_H}\right)^{1/5} \left(\frac{t}{1000\,\mathrm{yrs}}\right)^{-3/5} \\
		   &= \frac{2}{5} \frac{R}{t} \\
		   &= 3900 \VEL \frac{R/10\,\mathrm{pc}}{t/1000\,\mathrm{yr}}
\end{align}
\end{subequations}

In the Sedov phase, the radio luminosity density $L_{1.4}$ from equation~(\ref{eqn:sar}) can be expressed as a function of $t$:
\begin{equation}
	\begin{aligned}
		L_{1.4} = {}& \EXPU{4.8}{25}{ergs~s^{-1}Hz^{-1}}
\left(\frac{t}{1000\,\mathrm{yr}}\right)^{-0.96} \\
& \left(\frac{\epse}{10^{-2}} \right)
\left(\frac{\epsu}{10^{-2}} \right)^{0.8}
\left( \frac{E_{51}}{n_H} \right)^{1.32} \quad .
	\end{aligned}
\end{equation}

For this relationship and assuming $\epsu$ is approximately constant, then according to
\cite{sarbadhicary17}, we should expect $L_{1.4} \propto t^{-0.96}$. A SNR in the Sedov phase should
be brightest when it enters the Sedov phase and should decline linearly with time (ignoring the 
variation in $\epsu$).

The radio luminosity can also be expressed as a function of $R$:
\begin{equation} \label{eqn:LR}
	\begin{aligned}
		L_{1.4} = {}& \EXPU{1.0}{25}{ergs~s^{-1}Hz^{-1}}
\left(\frac{R}{10\,\mathrm{pc}}\right)^{-2.4} \\
& \left(\frac{\epse}{10^{-2}} \right)
\left(\frac{\epsu}{10^{-2}} \right)^{0.8}
\left( \frac{E_{51}}{n_H} \right)^{-9/5} \quad .
	\end{aligned}
\end{equation}
Thus we expect small diameter SNRs to be significantly brighter, all other things being the same
($L_{1.4} \propto R^{-2.4}$).   Clearly we do not see our observations reflecting this dependence (cf. Fig.\ \ref{fig:snr_dia}).  

On the other hand, this 
relationship does allow for significant variations in the radio luminosity at a specific diameter if the ratio $E_{51}/n_H$ varies significantly.  A SNR with an
explosion energy of $10^{51}\,\mathrm{erg}$ expanding into an ISM with density 1~cm$^{-3}$ would be 63 (4000) times brighter than one expanding into an 
ISM with density 0.1 (0.001) cm$^{-3}$, assuming both are in the Sedov phase.  \cite[See also][who reached a similar conclusion about radio luminosity variations associated with variations in $E_{51}/n_H$ from models made with slightly different assumptions.]{asvarov06}

If the variations in radio luminosity are primarily due to variations
in $n_H$, then we would expect that to have consequences at X-ray
wavelengths, where emission arises from the same shock that powers
the radio emission.  Simply put, SNRs expanding into a dense ISM
are brighter at the same diameter in soft X-rays than those expanding
into a tenuous one, so we would expect radio-bright SNRs to also
be X-ray bright and vice versa.  \cite{berkhuijsen96} argued that
a correlation of this type does exist using a collection of Galactic
and extragalactic SNRs; here we discuss the specific case of M33.

The X-ray luminosity in the Sedov phase has a complex dependence on the physics involved (e.g.,
non-equilibrium ionization, electron conduction, etc.).  However, \cite{white91}
give a simple approximation based on the observation that the X-ray
emissivity function $\Lambda_x(T)$ is constant to within 50\% over a wide range of temperature
\citep{hamilton83}.  From equation~(21) in \cite{white91}:
\begin{equation} \label{eqn:Lx}
	\begin{aligned}
		L_x = {}& \EXPU{3.0}{37}{ergs~s^{-1}} \left(\frac{R}{10\,\mathrm{pc}}\right)^3 n_H^2 \\
		 & \left( \frac{\Lambda_x}{\EXPU{1}{-22}{erg~cm^3 s^{-1}}} \right) \quad ,
	\end{aligned}
\end{equation}
where we set the contribution of evaporating clouds to zero and are using an ISM density
$n = \rho/m_H = 0.75\,n_H$.  Combining equations~(\ref{eqn:LR}) and~(\ref{eqn:Lx}),
we can calculate the ratio of the radio luminosity $L_R = \nu L_\nu$
to the X-ray luminosity:
\begin{equation}\label{eqn:rxratio}
	\begin{aligned}
		\frac{L_R}{L_x} = {}& \EXPU{4.6}{-4}{}
\left(\frac{R}{10\,\mathrm{pc}}\right)^{-5.4}
n_H^{-1/5}
E_{51}^{-9/5} \\
& \left( \frac{\Lambda_x}{10^{-22}} \right)^{-1}
\left(\frac{\epse}{10^{-2}} \right)
\left(\frac{\epsu}{10^{-2}} \right)^{0.8}
\quad .
	\end{aligned}
\end{equation}
Note that this decreases rapidly with radius but depends very weakly on the ISM density.  It
predicts that large-diameter SNRs should have much lower $L_R/L_x$ ratios, regardless of the local
ISM density (which may vary widely among remnants).

\begin{figure*}
	\centering
	\includegraphics[width=0.6\linewidth]{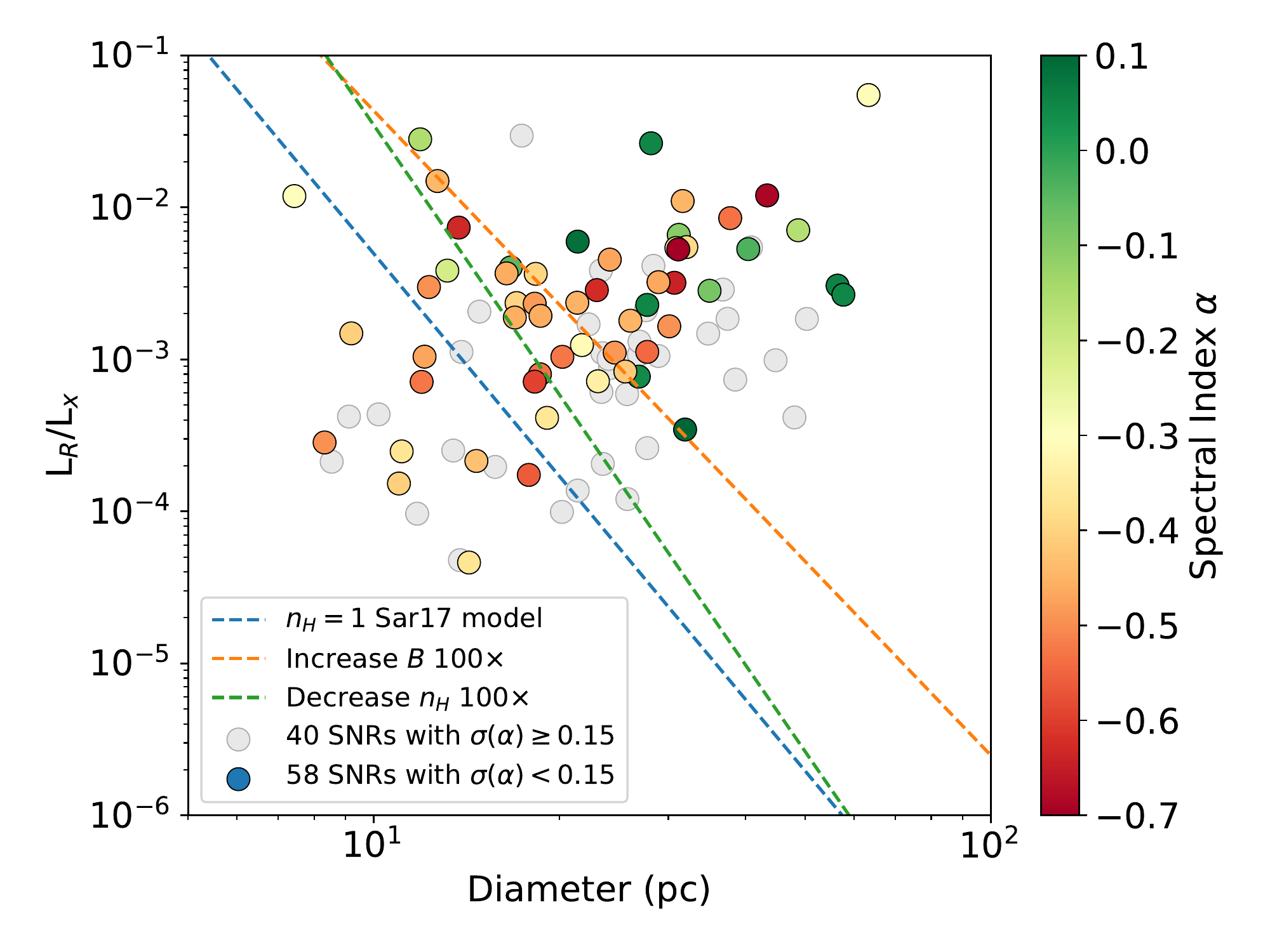}
	\caption{The ratio of the radio luminosity $\nu L_\nu$ at 1.4~GHz to the X-ray luminosity as a function of
		diameter $D$ for SNRs in M33.
		The \cite{sarbadhicary17} radio luminosity model predicts that in the adiabatic expansion (Sedov) phase,
		this ratio should decline rapidly with diameter.  The blue dashed line is for nominal model parameters with
		$n_{H} = 1\,\mathrm{cm}^{-3}$. While the model predictions depend on the ISM density and magnetic field,
		even changes by a factor of 100 (red and green lines) do not affect the conclusion that the observed variation of $L_R/L_x$
		with diameter is much flatter than the predictions.
		The SNRs that have well-determined spectral indices have symbols color-coded by $\alpha$; objects that
		are too faint to have an accurate $\alpha$ (including those in regions that lack 5~GHz data) are
		shown in gray.
	\label{fig:lx_lr_diam}}
\end{figure*}

This appears to be in sharp conflict with the observations (Fig.~\ref{fig:lx_lr_diam}).  There is a
wide range of radio-to-X-ray ratios at all SNR diameters, and there is no indication of a marked
decline for larger SNRs.  The dashed lines show the model prediction from
equation~(\ref{eqn:rxratio}).  We have included in the model calculation the additional dependence
on $B$ and $n_H$ that comes through the $\epsu$ factor by using equation~(\ref{eqn:epsu}) rather
than assuming $\epsu$ is constant.  But even large changes in the ISM density and the $B$ field
scaling in equation~(\ref{eqn:bscaling}) have little effect on the general trend.  Moreover, 
\cite{elwood19} found that the scatter in $n_H$ for SNRs in M31 and M33 is relatively small
($\log n_H \sim -1.35 \pm 0.7$), which makes explanations that rely on a wide variation in the ISM
density problematic.  The predicted
$L_R/L_x$ ratios from the \cite{sarbadhicary17} model are too low at $D\sim30\,\mathrm{pc}$ by one
to two orders of magnitude compared with the observations, even using extreme parameter values.

We have considered several possible explanations for the discrepancy in the $L_R/L_x$ ratio predictions compared 
with the observations.  Our calculation assumes that the SNR is in the Sedov phase.  That is likely to be a good 
assumption since the radio and X-ray luminosities both decline dramatically at the onset of the late radiative 
phase \citep{asvarov06}, while the early free-expansion phase is short ($\sim 10^2$~yr) and can only be relevant 
for small-diameter remnants \citep{elwood19}.

Another possible explanation is that the forced photometry SNR radio fluxes are contaminated by emission from 
nearby \hii\ regions, leading to elevated values of $L_R/L_x$.  To test this hypothesis, 
Figure~\ref{fig:lx_lr_diam} shows (via the symbol colors) the spectral indices of the SNRs.
If \hii\ contamination were important, all the large diameter SNRs would have the thermal spectral indices 
typical of \hii\
regions, $\alpha\sim0$, rather than the non-thermal $\alpha\sim-0.5$ indices observed for most SNRs.
While there is an indication that some of the largest diameter SNRs have thermal contamination, there are
also large diameter SNRs with non-thermal spectra. 
Care is required to account for the overestimated radio luminosities for SNRs embedded in \hii\
regions, but that effect cannot explain the large $L_R/L_x$ ratios in M33.

As shown In Figure~\ref{fig:lum_func}, the luminosity function for the SNRs we detect can be represented as a power law, as was pointed out by \cite{chomiuk09} for M33 (and a number of other galaxies) based on earlier data.  \cite{chomiuk09} sought to interpret this fact in terms of a model that was similar to that of \cite{sarbadhicary17}.  While it may well be true that the radio luminosity function of SNRs can be represented as a power law, our analysis makes it clear that a physical connection between the observations and the theory is lacking.

Although we have chosen to focus on the specific model for radio emission in SNRs described by \cite{sarbadhicary17}, this model embodies the same elements as most other models of radio emission in SNRs.  We conclude that the $L_R/L_x$ ratio is a strong diagnostic for testing models of radio emission from SNRs in the presence of ISM density variations.  Apparently current models still have significant shortcomings when confronted by the radio and X-ray observations.

\subsection{Comparison to Radio Studies of Other Galaxies} \label{sec:othergals}

\subsubsection{Large Magellanic Cloud}

As a result primarily of their proximity (and low Galactic extinction), more is known about the SNRs
in the Magellanic Clouds than in any other galaxies.\footnote{The
Small Magellanic Cloud (SMC) is at a similar distance to the the LMC, but because of its very low
mass, has fewer SNRs.  The existing radio data were summarized by \cite{filipovic05} and used by
\cite{bozzetto17} in their analysis of radio trends.}
As recently summarized by \cite{bozzetto17},
there are now 59 confirmed  SNRs in the Large Magellanic Cloud (LMC), of which 42 have radio fluxes
at 1~GHz and 51 have X-ray spectra \citep{maggi16}.   The existence of high quality X-ray spectra
(not possible to obtain for most of the SNRs in M33) has allowed most of the SNRs to be typed as the
products of core-collapse or Type Ia explosions.  There are also another 15 candidates, most of
which are larger objects with typical diameters of 100 pc or more.  The median diameter for the SNRs
in the LMC is about 40 pc, very similar to that in M33, but  unlike M33, the LMC does have several
SNRs known to have ages less than a few thousand years.  As noted by \cite{long10} and more recently
by \cite{maggi16}, the LMC has more very luminous X-ray SNRs than M33.
There
is a rough correlation between radio and X-ray brightness that is particular evident in the
subsample of core-collapse remnants, where ejecta interactions are still important \citep{bozzetto17}.

% \changed{
% In the LMC, the radio data suggest that the radio spectral index of SNRs flattens with time and diameter, especially among the SNRs associated with core-collapse SNe, with larger and older objects having spectral indices around −0.5 \citep{bozzetto17}. That is in contrast with the lack of such a correlation in our M33 data (Fig.~\ref{fig:snr_dia}b). We do not have an explanation for the difference, although it would clearly be of interest to replicate the effect seen for the LMC in other datasets. Certainly the astrophysical properties of the galaxies, including elemental abundances, ISM characteristics, and star formation histories, are very different between the LMC and M33. There are also significant technical challenges in measuring accurate spectral indices for the larger LMC SNRs from interferometric radio data, which could bias spectral indices as a function of size. The more severe crowding in M33 leads to greater contamination by H II region emission, which could also be affecting the spectral indices (although that could also have an effect in the LMC, particularly for the core collapse SNe that show the strongest correlation). We simply do not know whether the difference in the $\alpha$-D correlation is telling us about the physics of SNRs or about the difficulties of radio data analysis.}

\changed{
In the LMC, the radio data suggest that the radio spectral index of SNRs flattens with time and
diameter, especially among the SNRs associated with core-collapse SNe, with younger 10 pc diameter
SNR having a spectral index of about $-0.6$~pc and older 40~pc  objects having spectral indices around
$-0.5$ \citep{bozzetto17}.  The correlation is considerably less apparent in the Galactic sample, but as
\cite{bozzetto17} point out, one would expect a sample from an external galaxy to be cleaner as a
result of the fact that all of the SNRs are at the same distance and, as a result of this, have
similar apparent sizes.}

\changed{Firmly establishing a correlation between spectral slope and other properties of a SNR
sample is clearly important, since it suggests that the structure of the shock front and/or the
nature of the physical processes that generate radio emission are changing with time or with shock
velocity. Indeed, various suggestions for what might be changing have been proposed, based either on
earlier hints at spectral slope changes, or by noting that specific types of SNRs (such as those
interacting with molecular clouds) had flatter spectral indices than the general populations of SNRs
\cite[see, e.g.][]{bell11,urovsevic14}.}

\changed{ We do not see a similar correlation in M33 (Fig.~\ref{fig:snr_dia}b). We do not have an
explanation for the difference.  Although the properties of the LMC and M33 certainly differ in
terms of star formation history and ISM properties (including magnetic field strengths and cosmic
ray densities), it is not clear why any of these properties would lead to a different trend.  From
an observational perspective, if an object has a flux of 1~mJy at 1.4~GHz it will have a flux of
0.53 or 0.47~mJy at 5~GHz, depending on whether the spectral index is $-0.5$ or $-0.6$, respectively, so
obtaining accurate spectral slopes requires accurate flux measurements from interferometric
observations.  In the case of our JVLA observations, we have taken a lot of care to match beam sizes
and to extract accurate fluxes, but even so these are technically challenging observations.  The
radio spectral indices compiled by \cite{bozzetto17} included measurements made by various
investigators with different instrumental setups.  As result, we believe it is very important that
more attempts be made to establish whether the correlation seen by \cite{bozzetto17} can be
replicated with either newer observations of the LMC with a single observational setup or in
other galaxies.  }

\subsubsection{M31}

With a limiting sensitivity of 20~\uJy\ ($4\sigma$)
or $\nu L_{\nu}$ = \EXPU{2.2}{31}{\LUM}, and a resolution of 5.9\arcsec\ or 23.6 pc at a distance of 817 kpc, this JVLA survey of M33 is the deepest radio survey of any extragalactic spiral galaxy.  In contrast, its neighbor M31 has not been surveyed to anything approaching this depth, in part, of course, due to its large angular size.   \cite{galvin14} used archival data from the VLA to construct a catalog of radio sources in M31, but their 1.4~GHz catalog has a limiting sensitivity of only $\sim$2~mJy and contains some 916 sources. Of these, just 98 have spectral indices, most of which are steep and characteristic of either background AGN or SNRs.  Of the 156 optical nebulae identified as SNR candidates in M31 by \cite{lee14b}, only 13 lie within 5\arcsec\ of sources in the radio catalog, a much lower fraction than is observed in M33.  Presumably this is a result of the much lower sensitivity of the M31 radio catalog.  

\subsubsection{M83}

\cite{long14} carried out a deep \chandra\ observation of M83 supplemented by radio observations with ATCA at 5 and 9~GHz.  The radio catalog consisted of 102 sources, the faintest of which is $\sim$50~\uJy\  at 5~GHz, corresponding to  $\nu L_{\nu}$ of \EXPU{6.3}{33}{\LUM} at the assumed 4.6 Mpc distance of M83.  Many of the radio sources were in the spiral arms of M83.   Of the 102 sources, 21 were identified with one of the 225 optical SNR candidates known in M83 at the time \citep{blair12}.  Another 15 were identified with other \chandra\ X-ray sources; some of these may be SNRs but most are likely to be background sources in the field or \hii\ regions in which an X-ray binary is embedded.

\subsubsection{NGC 7793}

\cite{galvin14_ngc7793} used archival ATCA data on the southern Sculptor group galaxy NGC~7793 obtained at 1.4, 5 and 9~GHz to create a catalog of 76 sources brighter than  11 \uJy\ per beam (3~$\sigma$) or $\nu L_{\nu}$ = \EXPU{1.3}{33}{\LUM} at 5~GHz, using an assumed distance of 3.9 Mpc.
They classified 57 sources; 37 are most likely \hii\ regions. The microquasar known as NGC~7793--S26 was also detected.  They claimed 14 SNR identifications, but they applied an unusual definition of a SNR candidate as a radio source with a steep radio spectral index associated {\it with an X-ray source}.  Many such sources in the M33 catalog are clearly background AGN, and so this way of identifying SNR candidates is fraught with concern. 
An earlier paper by \cite{pannuti02} listed five radio SNR candidates in NGC~7793 identified using the  criterion of a nonthermal radio source aligned with optical \ha\ emission, as used by \cite{lacey01}, but again, this is suspect without the use of the \sii:\ha\ criterion.  Of the 28 optical SNR candidates in the survey by \cite{blair97}, only two objects had radio counterparts in \cite{pannuti02}, one of which was the aforementioned NGC~7793--S26.  \cite{galvin14_ngc7793} found only five of the ATCA sources that aligned with any SNR candidate from the earlier lists.  Clearly very few radio SNRs have been detected in NGC~7793.  
%{\bf Bill - you may want to work on this a bit.  As you remember the Pannuti paper had an interaction with our old NGC7793 paper; most of the SNRs in his catalog had been identified by us first}

% \todo[inline]{At some point this discussion has to move to the question of how SNRs are identified from the radio.  The usual criterion is a non-thermal spectral index plus Ha emission (without explaining what constitutes Ha emission.  This leads into Chomiuk's radio SNR studies and universal luminosity function.  The galaxies we may want to consider in this context include NGC7793, NGC300, M101 perhaps, and NGC6946.  Also we probably have to think about what to say about the LMC and SMC where most of the SNRs are said to be detected in the radio.  Also, are we missing something by not making a clearer comparison with Galactic SNRs.  }

% \todo[inline]{There is an argument that can be developed here.  Lacey and other argue radio SNRs are non-thermal sources with \ha\ emission.  They do not define the surface brightness for \ha\ to use, but we can take our value and see how many non-thermal sources that are not SNRs we see associated with \ha.  This is a first order estimate of contamination since we do not believe the non-thermal sources are really SNRs.  But we can improve this estimate somewhat since in NGC6946 etc the surveys are not as deep as indicated above.  We can then restrict the question of contamination to objects that are brighter than a certain flux.  It's possible that we will agree that the radio objects in other galaxies really are in other galaxies.}

\subsubsection{Efficacy of SNR identification using radio emission}

Most extragalactic SNRs and SNR candidates have been identified optically based on \sii:\ha\ emission line ratios.  However, some attempts have been made to identify SNRs in other galaxies on the basis of their radio properties, usually coupled with more limited optical data.  For example, \cite{lacey01} identified 35 radio-selected SNRs/SNR candidates in NGC6946 on the basis of their non-thermal radio spectral indices and associated \ha\ emission.  These SNRs were concentrated in the spiral arms of NGC~6946.  They compared this sample to one obtained by \cite{matonick97} optically and found very little overlap between the two samples.  The \cite{matonick97} sample was distributed in both arm and interarm regions of the galaxy and so  \cite{lacey01} argued that the fact that the samples were largely disjoint was a selection effect.  Optical SNRs are simply difficult to pick out against the light of an \hii\ region.  Most SNe occur in star forming regions of a galaxy and hence most SNRs should exist in the spiral arms.   Star forming regions are also the regions of a galaxy with the most \ha\ emission and therefore the fact that optical SNRs are not concentrated there is prima facie evidence that they are being missed in optical surveys.   

The general argument made by \cite{lacey01} that there are portions of a galaxy where optical SNR detection is difficult has to be correct at some level, but it raises a number of questions, especially as many of the spectral index measurements in radio maps are not precisely determined and that, with current instrumentation, it has been difficult to obtain any confirming evidence that a radio SNR candidate in an external galaxy based on radio criteria alone is indeed a SNR.  On the theoretical side, these questions include how important it is that after the first SN explodes in a star cluster, subsequent SNe will explode into the region evacuated by the first SN.  On the observational side, one of the questions is at what \ha\ surface brightness will what fraction of SNRs actually be missed, and how many background interlopers will be counted as SNRs.  Unfortunately, there has very little systematic work to investigate any of these problems.   

\begin{figure}
	\centering
	\includegraphics[width=\linewidth]{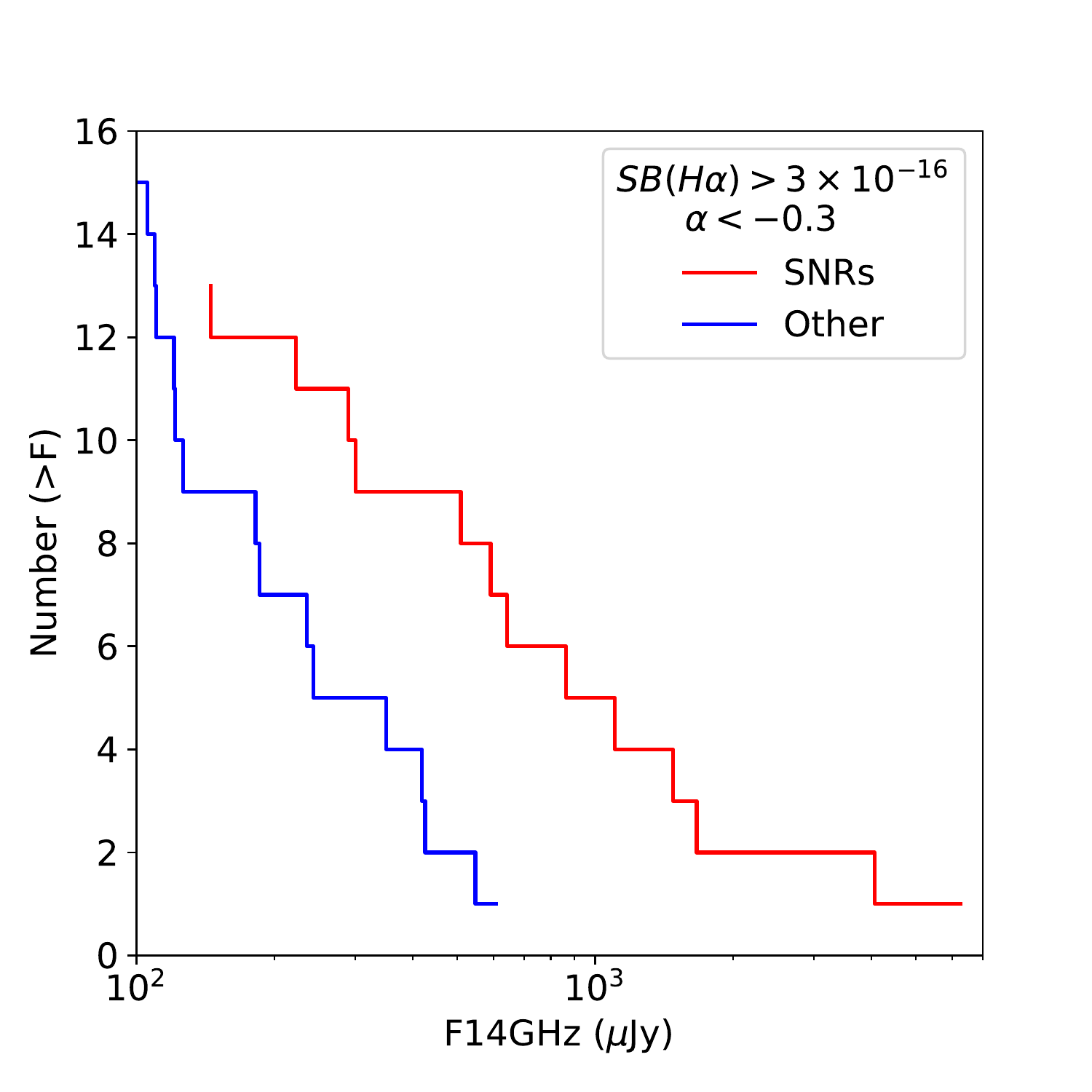}
	\caption{The number of sources with non-thermal spectral indices that lie on regions of high
	\ha\ surface brightness.  Radio sources identified with known SNRs and those that are not
	are plotted separately.  The sample is restricted to objects with 5~GHz fluxes, so
	sources with $F(1.4\,\mathrm{GHz}) > 100\,\mu\mathrm{Jy}$ have well-determined spectral
	indices.  The sources not known to be SNRs are almost certainly background objects.
	\changed{There are very few sources that are not SNRs above 0.6~mJy. That fact suggests that most radio
	sources identified in nearby galaxies as SNRs on the basis of a non-thermal radio spectral index
	in regions with \ha\ emission are indeed good candidates, since most such surveys are not
	sensitive below 1~mJy.}
	\label{fig:snr_interlopers}}
\end{figure}

We can use the observations of M33 to gain some insight into this. There are 284 non-thermal sources
(with a spectral index of less than $-0.3$) in our survey that are brighter than 0.1~mJy and are
in the region covered by our 5~GHz data.  Essentially
all of those sources have well-determined spectral indices.  Of these, 28 lie at positions where the \ha\ surface brightness exceeds
\EXPU{3}{-16}{\LUM ~ arcsec^{-2}}.  Of these sources, 14 are known SNRs.  However, most of the
sources that are not known to be SNRs are quite faint.  As shown in Figure
\ref{fig:snr_interlopers}, the number of interlopers drops quite rapidly as a function of the radio
source brightness.  At 200~\uJy, the number of interlopers and known SNRs is about the same, and at
600~\uJy\ essentially all of the interlopers have disappeared.  At 200~\uJy\, the typical spectral
index error is 0.05 (1$\sigma$) in our survey, so confusion with \hii\ regions is not a problem;
nearly all of the non-SNRs are background sources.  As our results are not very sensitive to the specific \ha\
surface brightness limit chosen, and since the apparent distribution of fluxes of background sources
does not depend on the distance to the foreground galaxy, this indicates that as long as a radio SNR
candidate is relatively bright ($\gtrsim$ 0.5~mJy), radio SNR identifications should be fairly
reliable, as \cite{lacey01} suggested on the basis of their location.

\subsection{Prospects for Future Radio Observations}

The JVLA's combination of high spatial resolution and wide frequency coverage provides an
excellent tool for surveys of M33 and other nearby northern galaxies.  Future observations that are
higher resolution might enable algorithms that separate SNR emission from \hii\ region emission using
the morphology and spectral index of the radio flux.  Higher angular resolution would also allow one to separate background objects from SNRs. 

Improved algorithms for processing the very wide-band 
JVLA data could also enable more information to be extracted from our current data.  Although we have put considerable effort in to development of new methods for analyzing multi-resolution radio images, we have deliberately
degraded the resolution to generate images that are matched across the full frequency bandpass.  One can imagine methods that might avoid this step.

Elsewhere, the LOFAR radio telescope is now being utilized for low frequency (0.15~GHz) surveys of the northern
sky with a resolution of 6\arcsec\
\citep{shimwell19}.  While this data alone will not produce accurate spectral indices from its 0.12--0.17~GHz bandpass,
it should be readily combined with our similar-resolution, higher frequency JVLA observations.  There are plans to do
LOFAR observations of nearby galaxies including M33 to an rms sensitivity of 15~\uJy\ at 0.15~GHz \citep{vanhaarlem13}; for a SNR with a spectral index 
$-0.5$, that corresponds to an equivalent rms $\sim5$~\uJy\ at 1.4~GHz, which is also well-matched to our survey.

The SKA pathfinder telescopes (ASKAP, MeerKAT) are located in the southern hemisphere and so are not directly relevant to
studies of M33, but they should contribute interesting observations of other nearby galaxies.  According to their specifications
they will generate very deep images with noise comparable to or better than our JVLA data \citep{johnston08,norris11,norris13}.
Their lower resolution of 10--15\arcsec\
will make studies in the vicinity of \hii\ regions more challenging, and their initial configurations have a more limited
frequency range than the JVLA.  Nonetheless, with a flood of high quality radio data on the horizon, we can expect dramatic
advances in our understand of radio emission from SNRs over the next decade.

\section{Summary and Conclusions}

We have carried out a radio survey of M33 using the JVLA at 1.4~GHz and 5~GHz.  After adjusting all
of the various radio bands to a common angular resolution of 5.9\arcsec, our observations have a
limiting sensitivity of 20~\uJy\ (4~$\sigma$).  Using a new multi-resolution algorithm, we detect 2875
sources in the radio images.   Additionally, we have extracted radio fluxes at the positions of
several types of objects, notably SNRs and discrete X-ray sources in the \chandra\ X-ray survey of M33,
in order to explore the radio properties of these objects.   Our principal results are as follows:

\begin{itemize}
\item
Most of the sources in the catalog can be categorized as (portions of) \hii\ regions,  SNRs, or
background AGN.  A few are aligned with X-ray binaries in M33 although some are buried in \hii\ emission, making an assessment of intrinsic radio emission difficult.  Catalog sources associated
with \hii\ regions have flat (thermal) spectral indices, whereas those associated with background galaxies and
SNRs tend to have steep (non-thermal) spectral indices.  Excluding the SNRs, the number and shape of the luminosity function of the
steep spectrum sources is roughly consistent with that expected from radio surveys 
of high latitude fields.

\item
Of the 217 emission nebulae suggested to be SNRs in M33 as a result of optical interference imagery,
we detect 141 in the catalog and 155 via forced photometry.  As there is only one SNR outside the field of view of
the radio maps we used for this analysis, we have detected (in forced photometry)  72\% of
the SNRs that we could have detected.  Neither the radio luminosity nor the spectral index of SNRs
in M33 is correlated with SNR diameter.  There does appear to be a weak correlation between radio
luminosity and \ha\  luminosity, but the scatter is quite large. As discussed in Section 4.2, the
radio luminosities of SNRs are not easily interpretable in terms of the current generation of models
of radio synchrotron emission in SNRs.

\item
Of the 662 X-ray sources reported by \cite{tuellmann11}, 320 are positionally coincident with sources
in our radio catalog, and 319 are detected via forced photometry.  Most of these have steep radio spectra and hard X-ray
spectra.  Most of these radio sources that are not SNRs are likely to be
background AGN, seen in projection.  One of the brightest radio sources in the survey is
associated with the nuclear X-ray source X\nobreakdash-8 in M33; the fact that there is radio emission is consistent
with its identification as a microquasar, although the conclusion is not definitive.

\end{itemize}

\acknowledgments

This project would never have begun without the support of NRAO Director Fred Lo, who decided that the project, though challenging, was important to pursue, despite an initial negative technical review.
The authors thank Sumit Sarbadhicary for useful information about the radio models in \cite{sarbadhicary17}.
PFW acknowledges support from the NSF through grant AST-1714281. 
WPB thanks the JHU Center for Astrophysical Sciences for partial support during this work.
The National Radio Astronomy Observatory is a facility of the National Science Foundation operated under cooperative agreement by Associated Universities, Inc.

% \todo[inline]{Add additional acknowledgments.}

\appendix

% \todo[inline]{If this gets to be too long, I guess it could be split into a separate paper.  [RW]}

\section{Multi-resolution source detection algorithm} \label{sec:multires}

This appendix describes a new multi-resolution algorithm that has been used to subtract the
background from the radio images, to detect sources and identify regions with significant emission,
and to measure the flux densities of those sources.

The traditional approach to measuring fluxes and source sizes from radio images is to fit elliptical
Gaussians.  That is the approach that was used for the FIRST survey \citep{becker95,white97} and for
many other radio catalogs.  However, Gaussian fits simply do not work well in fields as complex as
our M33 JVLA images.  The biggest problem is that in these deep, crowded images there may be many
sources close together that must be simultaneously fitted.  Also, many of the sources are extended
with broad halos or tails around compact cores.  Such objects are not well modeled by Gaussians.
Even for sources that are roughly Gaussian, in crowded regions the Gaussian fits sometimes fail
catastrophically, with a typical failure mode leading to source center positions that migrates to the
edge of a shared boundary between adjacent islands.  When that happens, the Gaussians are no longer
distinct, and the fitted flux densities are poorly constrained.  In principle, it might be possible to
improve the Gaussian fitting procedures to handle these issues better, but it will be very
difficult to produce good results for the full range of morphological variations in the radio sources.

Another approach is to integrate the map flux densities over the regions identified for each object.  This
has the major advantage that it does not depend on any assumptions about the source morphology.  The
disadvantage is that the uncertainty in the derived integrated fluxes is considerably greater than
the uncertainties for Gaussian fits.   This failing can be traced to the ability of the Gaussian
fit to weight down pixels in island regions that are essentially empty, with pure noise, while giving
higher weight to pixels where the object is detected.  The unweighted sum of the island pixels weights each
pixel equally and adds extra noise from the faint or empty regions in the island that may
contribute little flux but still dominate the total noise.

We have developed a novel approach to this problem.  We decompose the image into a stack of components
of varying resolution using a median pyramid transform.  The resulting stack has four levels (in this
case), with the highest resolution (sharpest) level containing objects that are about the size of
the PSF, the second level having objects that are twice that big, the 3rd level having objects that
are $2^2 = 4$ times as big, etc.  If the 4 levels are summed, the original image (including all
sources and noise) is recovered exactly.  But in any single level the noise is lower because higher
frequency variations have been removed.  The spatially varying background is also subtracted, making
it easier to do photometry.
The median pyramid is used to detect sources on a range of scales (which improves the detection of low surface
brightness objects).  Finally, the multi-resolution source detection maps are integrated over regions that
adapt to include just the necessary pixels, reducing the noise while preserving the flux.

These steps are described in more detail below.

\subsection{Median Pyramid}

The median pyramid calculation begins with a median filter over a circular region of radius $R=8$
pixels (8\arcsec\ for the full-resolution image). The median filtered image is the ``smooth''
channel, and the difference between the original image and the median-filtered image is the
``sharp'' channel.  We also estimate the local rms noise as a function of position in the sharp
channel by computing a second median filter, operating on the absolute value of the sharp channel
and using an annular region with an inner radius of $R$ pixels and an outer radius of $2R$ pixels.
The median absolute value is multiplied by 1.4826, a factor that converts the value to an equivalent
Gaussian sigma when the noise is purely Gaussian.  (This robust noise estimation approach is
insensitive to outliers in the image.)

To create additional levels in the pyramid, the process is then repeated using the smooth image,
with the radius of the filtering region expanded by a factor of two.  This is iterated as many times
as desired, with the filtering radius doubling at each new level.  The choice of the number of
levels determines the largest scale structure that remains in the image. For M33 we compute $L=4$
levels of the pyramid.

The median-filtered smooth channel has only faint residuals remaining at the positions of sources
that are small compared with the filter area.  We further reduce the contamination by sources in the
smooth channel by doing a second filtering pass after removing pixels that are found to be larger
than twice the local rms estimate.  Those pixels are replaced by interpolating values from nearby
pixels (using yet another median filter).  Then the smooth channel is computed again.  This single
iteration of source rejection leads to very good removal of sources from the smooth channel.

We also improve the algorithm's performance, which would be slow for higher levels of the pyramid when the
filtering radius becomes large, by reducing the size of the smoothed image at each level by a factor
of two in each dimension.  We simply bin the smoothed image by averaging blocks of $2\times2$
pixels.  To compute the sharp image, we then re-expand the smoothed image back to the original size
using bilinear interpolation.  The result is a stack of smoothed images of decreasing size.  With
this modification, the iterated smoothing uses a median filter region that is the same size (in
pixels) as used for the initial iteration.  The effective radius is increased by a factor of two
because the smoothed image was shrunk by that factor.  The result is that the algorithm is actually
faster for the higher levels of the pyramid.

When the median pyramid is complete, we are left with a stack of $L=4$ images ($L-1$ sharp images
and the remaining smooth background image) along with $L-1$ local rms estimates for each sharp
image.  There are several adjustable parameters in the algorithm. The filter radius $R$ should be
slightly larger than the PSF FWHM so that the first sharp images includes the point sources.   The
number of levels $L$, combined with $R$, determines the scale of the largest sources in the lowest
resolution sharp image.  Larger structures are in the final smooth image and are treated as
background.

\subsection{Source Detection}

A single background-subtracted image can be constructed by summing the $L-1$ sharp channels.
However, it is more effective to do source detection directly on the pyramid levels.  At each level
we use the FellWalker clump-finding algorithm \citep{berry15} to identify islands of source
emission.  FellWalker identifies distinct peaks in the image and assigns nearby pixels to the
nearest peak that is reachable by going uphill.  It has parameters that prevent noise bumps from
being included, that determine when two neighboring peaks have a significant valley between them,
and that merge islands that are consistent with being noisy features of the same peak.  The output
of FellWalker is an island (or segmentation) map that divides the image into empty regions and
contiguous regions at the locations of detected sources.  We morphologically filter the FellWalker
maps to eliminate islands that are smaller than expected given the resolution at that pyramid level.

There is one FellWalker island map for each level of the pyramid.  We combine the independent island
maps by merging islands that are detected in more than one level (which is a frequent occurrence).
When necessary, islands that have multiple high-resolution peaks with overlapping low-resolution
emission are split among the peaks using rules that attempt to maintain the morphological
characteristics of the existing islands.

The merging process starts by creating a merged island map $M$ from the highest resolution image.
The island numbers at each higher level are then updated using these rules:
\begin{enumerate}
	\item{Islands that do not overlap an existing island in $M$ get a new number larger than any existing island number.}
	\item{Islands that overlap a single existing island in $M$ are assigned the same number as that island.  This is the most
		common case (e.g., when a core-halo source is detected at several levels of the pyramid.)}
	\item{Islands that overlap more than one existing island in $M$ are split and are assigned
	numbers chosen from among those islands:}
	\begin{enumerate}
		\item{Pixels already assigned in $M$ to an island number retain their values.}
		\item{Unassigned pixels are assigned to the closest island using a distance metric that weights the
			Euclidean distance to the island center by the elliptical moments of the island shape in $M$.}
		\item{The extended islands are checked to ensure that they are contiguous.  If non-contiguous regions
			are found, those pixels are reassigned to the second closest island.}
	\end{enumerate}
	\item{After renumbering is complete, $M$ is updated by filling blank pixels with the corresponding values
		in the new level's island map.}
\end{enumerate}

There are other algorithms that could be used for island splitting (case 3), but our approach is
simple and has been found to produce reasonable results.  The effect of our choice is to assign a
pixel that is near two different islands to the island that reaches most closely toward that pixel,
either because it is large or because it is elliptical and is extended in the direction of the
pixel.

There are several advantages to this source detection approach.  The detection of sources at each
level is simplified because more extended background structures have already been subtracted.
Sources that are extended with low surface brightnesses are much more easily recognized in the lower
resolution levels of the pyramid.  The island merging process provides a clear and simple approach
for combining the source lists at different resolution levels.  And finally, having a
multi-resolution set of detection maps enables the computation of higher signal-to-noise flux density
estimates for sources with complex morphologies that include sharp components embedded in broad
halos.  That algorithm is described below.

\subsection{Extracting Fluxes}

It is possible to combine the multi-resolution detection maps into a single map that assigns every
pixel of the image to a particular source (or to the empty background).  We use that map,for example, to
display the locations of detected sources.  In some applications it might be sufficient to sum the
fluxes within islands in order to determine source brightnesses. In our images, however, that leads to
relatively high noise for core-halo sources which have a sharp component surrounded by a lower
surface brightness halo that can spread across many pixels.

To reduce the noise, we instead use the segmentation maps at each level of the image stack along with
the image from that level of the stack.  For each level we sum only the pixels that are detected in
the island at that level.  For a sharp core embedded in a broad halo, the sharp channel island
(``level~0'') only includes the few pixels needed to cover the PSF.  But the broad channel island
(``level~1'') with the halo includes many more pixels.  The flux for the object is the sum of the
few pixels from level~0 plus the many pixels from level~1.  It does \textit{not} include all the
level~0 pixels that are covered by the level~1 island.  That decreases the noise in the sum by not
adding in a bunch of empty pixels that contribute only noise.  Moreover, the level~1 image is
significantly smoothed, reducing the rms noise in that level.  As a result, the pixels that
\textit{are} included over a large region for the smooth halo contribute significantly less noise to
the sum.

Note also that sources that have only a sharp component (e.g., point sources) typically have islands
that have no components at any of the lower resolutions.  Omitting those pyramid levels from the sum
also reduces the noise for compact sources.

The end result is that this approach effectively creates an adaptive set of weights for the image
pixels that adjusts itself depending on the actual distribution of flux in the object.

\subsection{Performance of the New Algorithm}

\begin{figure}
	\includegraphics[width=\linewidth]{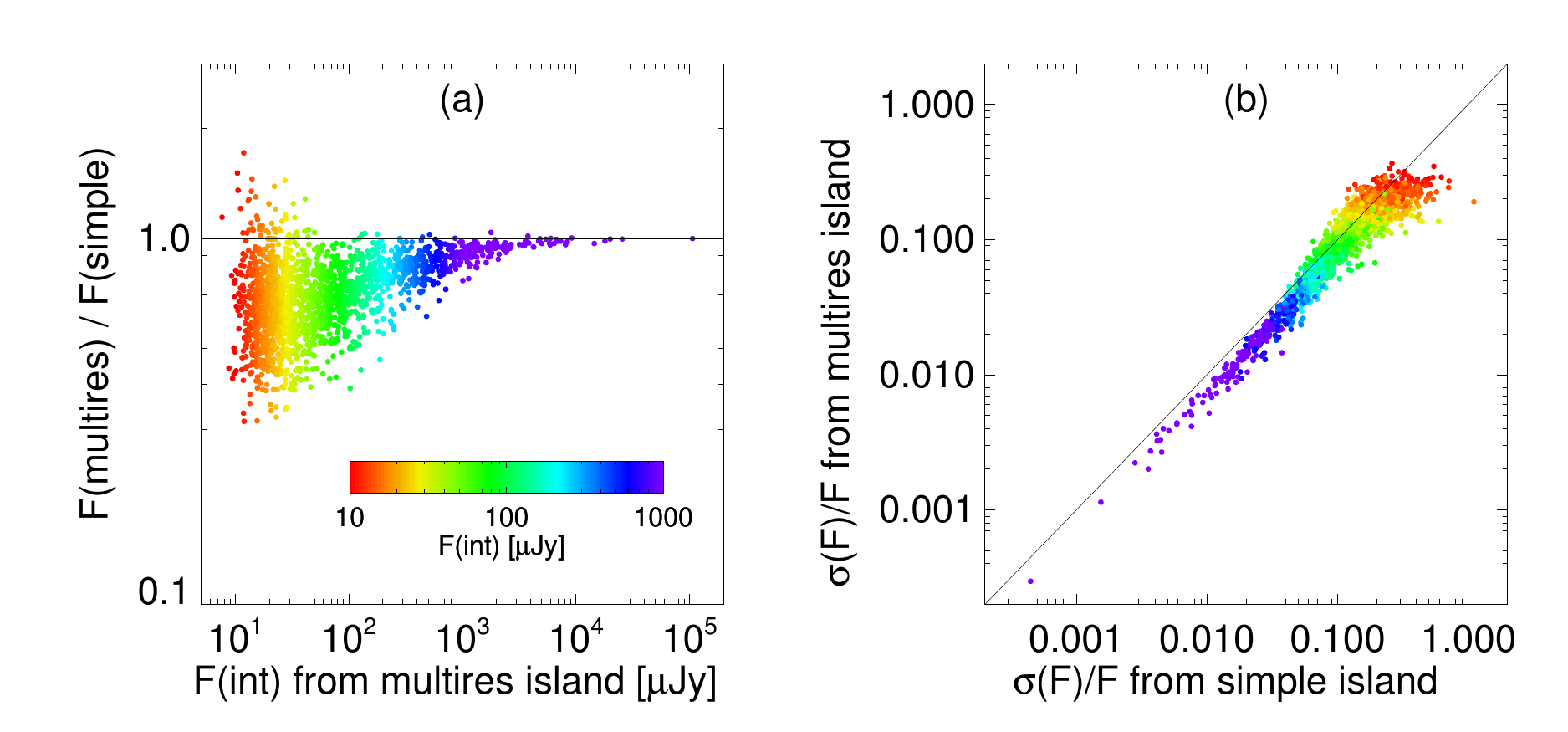}
	\caption{A comparison of the accuracy of flux densities determined using the multi-resolution island algorithm to flux densities
		computed using ``simple'' islands that are the same size at each resolution level.  The left panel (a) shows the ratio of flux \changed{densities} from the two 
		methods.  There is a systematic flux density underestimate by about $1\sigma$ for
		fainter sources due to the smaller island areas.  The right panel (b) compares the fractional noise in the flux densities.  The noise is about
		20\% smaller using the multi-resolution islands.  The color of the points indicates the source brightness.
	\label{fig:flux-accuracy}}
\end{figure}

Figure~\ref{fig:flux-accuracy} shows a comparison between the simple (single level) island flux
densities and our multi-resolution island flux densities (left panel), and a comparison of the
fractional flux density errors $\sigma(F)/F$ (right panel).  Sources are color-coded using the island flux.
There is a bias in the measured flux densities, with the multi-resolution values being
systematically lower for fainter sources.  This is not unexpected since the multi-resolution islands
are more tightly delineated, and some flux spills outside the summed regions.  On the other hand,
the right panel shows that the signal-to-noise in the measurements is about 20\% better in the
multi-resolution measurements.  The fractional flux density errors are systematically lower for both bright
and faint sources using the multi-resolution algorithm.

\begin{figure}
	\includegraphics[width=\linewidth]{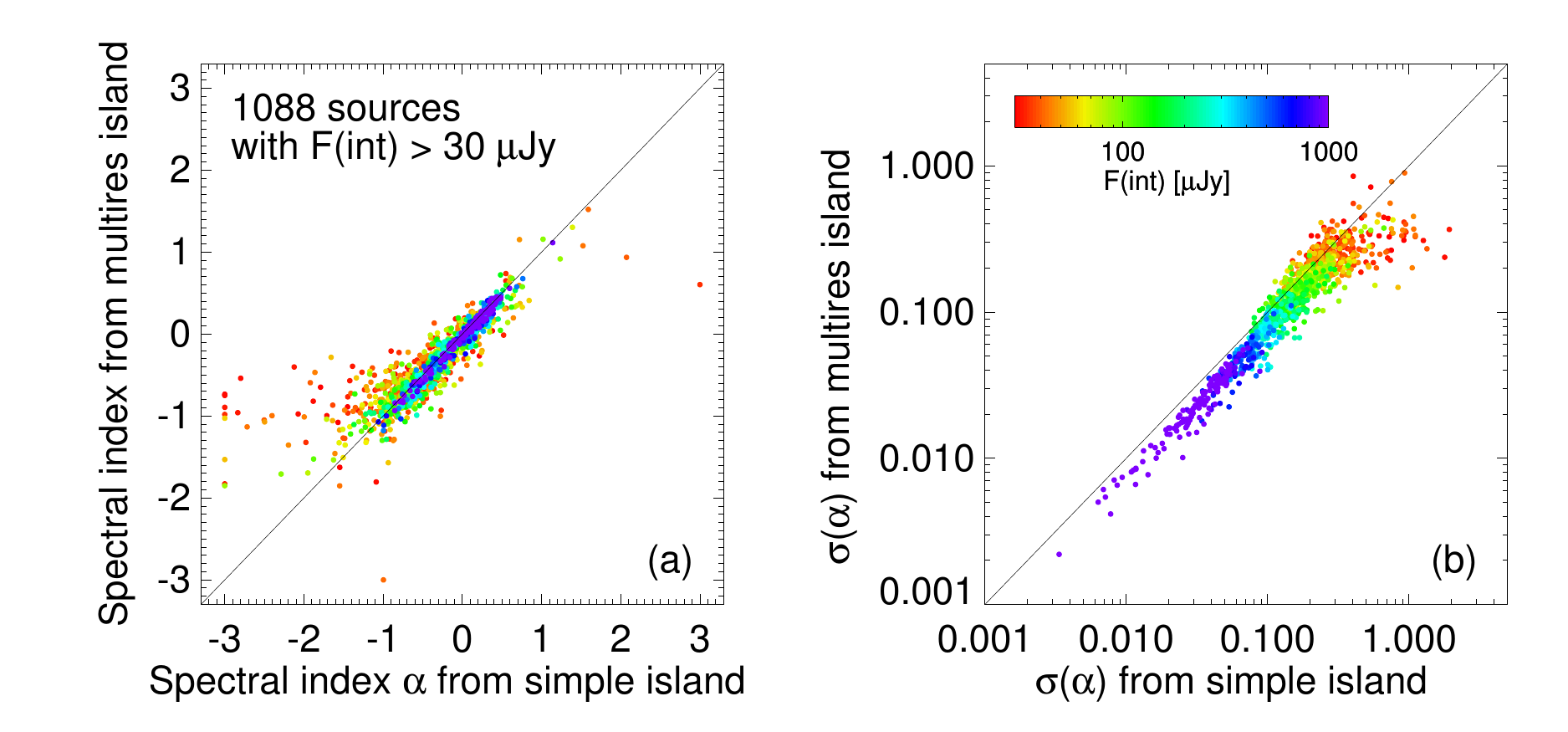}
	\caption{A comparison of the accuracy of spectral indices using the two island integration algorithms.
		The left panel (a) compares the spectral index $\alpha$ computed using ``simple'' islands (x-axis) to the value using the
		multi-resolution island.  The agreement is excellent, particular for brighter sources.  This plot includes only points with
		measurements at both 1.4 and 5~GHz and with
		flux densities greater than $30\,\mu\hbox{Jy}$ to exclude the faintest sources with poorly determined $\alpha$ values.  The right panel (b)
		compares the uncertainties $\sigma(\alpha)$ from the two methods. The multi-resolution spectral indices have 30\% smaller
		errors.  This improved accuracy led us to choose the multi-resolution algorithm for our catalog.
	\label{fig:spind-accuracy}}
\end{figure}

Figure~\ref{fig:spind-accuracy} shows the effect of the two algorithms on the spectral indices.  The
flux densities are computed using exactly the same islands for the (resolution-matched) 1.4~GHz and 5~GHz
images.  Each band is divided into two frequency channels, and a power law is fitted to the four
channels using a constrained algorithm that requires the spectral index $\alpha$ ($F_\nu \propto
\nu^{-\alpha}$) to lie in the range $-3 \le \alpha \le 3$.  The variances and covariances of the
fitted parameters are derived from the noise in the individual bands.
Figure~\ref{fig:spind-accuracy}(a) shows the spectral index from the simple island fluxes (x-axis)
versus the index from the multi-resolution fluxes (y-axis).  The sample of sources was restricted to
objects with $\sigma(\alpha) \le 0.1$ in the multi-resolution measurement.  As in
Figure~\ref{fig:flux-accuracy}, the points are color-coded by flux.  Clearly there is excellent
agreement, particularly for the brighter sources.  There is no significant bias between the two
spectral index estimates.

Figure~\ref{fig:spind-accuracy}(b) compares the noise in the spectral index for the two algorithms.
Here the advantage of the multi-resolution islands can be seen clearly, with the median noise being
about 30\% lower compared with the simple islands.

Our conclusion is that the multi-resolution islands lead to significantly higher signal-to-noise
measurements of the flux densities (although with some bias toward lower values) and also to more
accurate estimates of the spectral indices.  The algorithm is robust, giving reliable results even
in very crowded regions of the image.  We find that it has significant advantages compared to the
other algorithms that we tried.

\subsection{Future Improvements}

We foresee several possible areas for further improvement.  It might be effective to use a
continuous weighting scheme for pixels rather than the simple in-or-out approach used here.  Pixels
near the edges of islands could get a lower weight.  Pixels that have some but not much flux could
be weighted down.  A better weighting algorithm would probably reduce the bias and might ideally be
able match Gaussian fits in signal-to-noise while retaining the ability to robustly determine source
flux densities for complex objects that are poorly measured using Gaussian models.

Another idea would be to perform multi-resolution Gaussian fits at each level of the image stack.
That might have a couple of benefits: 1) the fitting process would converge more easily for the sharp
channel because sources are better separated, and 2) the resulting image could model the actual flux
distribution much more accurately using a sum of multiple Gaussians, while avoiding the difficulties
of simultaneously fitting overlapping Gaussians.

% \clearpage

% \bibliographystyle{aasjournal}

\bibliography{m33_radio}

\begin{thebibliography}{}
\expandafter\ifx\csname natexlab\endcsname\relax\def\natexlab#1{#1}\fi
\providecommand{\url}[1]{\href{#1}{#1}}

\bibitem[{{Asvarov}(2006)}]{asvarov06}
{Asvarov}, A.~I. 2006, \aap, 459, 519

\bibitem[{{Becker} {et~al.}(1995){Becker}, {White}, \& {Helfand}}]{becker95}
{Becker}, R.~H., {White}, R.~L., \& {Helfand}, D.~J. 1995, \apj, 450, 559

\bibitem[{{Bell}(2004)}]{bell04}
{Bell}, A.~R. 2004, \mnras, 353, 550

\bibitem[{{Bell} {et~al.}(2011){Bell}, {Schure}, \& {Reville}}]{bell11}
{Bell}, A.~R., {Schure}, K.~M., \& {Reville}, B. 2011, \mnras, 418, 1208

\bibitem[{{Berkhuijsen}(1986)}]{berkhuijsen96}
{Berkhuijsen}, E.~M. 1986, \aap, 166, 257

\bibitem[{{Berry}(2015)}]{berry15}
{Berry}, D.~S. 2015, Astronomy and Computing, 10, 22

\bibitem[{{Blair} \& {Long}(1997)}]{blair97}
{Blair}, W.~P., \& {Long}, K.~S. 1997, \apjs, 108, 261

\bibitem[{{Blair} {et~al.}(2012){Blair}, {Winkler}, \& {Long}}]{blair12}
{Blair}, W.~P., {Winkler}, P.~F., \& {Long}, K.~S. 2012, \apjs, 203, 8

\bibitem[{{Bozzetto} {et~al.}(2017){Bozzetto}, {Filipovi{\'c}}, {Vukoti{\'c}},
  {Pavlovi{\'c}}, {Uro{\v s}evi{\'c}}, {Kavanagh}, {Arbutina}, {Maggi},
  {Sasaki}, {Haberl}, {Crawford}, {Roper}, {Grieve}, \& {Points}}]{bozzetto17}
{Bozzetto}, L.~M., {Filipovi{\'c}}, M.~D., {Vukoti{\'c}}, B., {et~al.} 2017,
  \apjs, 230, 2

\bibitem[{{Chevalier} \& {Fransson}(2006)}]{chevalier06}
{Chevalier}, R.~A., \& {Fransson}, C. 2006, \apj, 651, 381

\bibitem[{{Chomiuk} \& {Wilcots}(2009)}]{chomiuk09}
{Chomiuk}, L., \& {Wilcots}, E.~M. 2009, \apj, 703, 370

\bibitem[{{Dennison} {et~al.}(1975){Dennison}, {Balonek}, {Terzian}, \&
  {Balick}}]{dennison75}
{Dennison}, B., {Balonek}, T.~J., {Terzian}, Y., \& {Balick}, B. 1975, \pasp,
  87, 83

\bibitem[{{Dickey} \& {Lockman}(1990)}]{dickey90}
{Dickey}, J.~M., \& {Lockman}, F.~J. 1990, \araa, 28, 215

\bibitem[{{D'Odorico} {et~al.}(1978){D'Odorico}, {Benvenuti}, \&
  {Sabbadin}}]{dodorico78}
{D'Odorico}, S., {Benvenuti}, P., \& {Sabbadin}, F. 1978, \aap, 63, 63

\bibitem[{{D'Odorico} {et~al.}(1980){D'Odorico}, {Dopita}, \&
  {Benvenuti}}]{dodorico80}
{D'Odorico}, S., {Dopita}, M.~A., \& {Benvenuti}, P. 1980, \aaps, 40, 67

\bibitem[{{D'Odorico} {et~al.}(1982){D'Odorico}, {Goss}, \&
  {Dopita}}]{dodorico82}
{D'Odorico}, S., {Goss}, W.~M., \& {Dopita}, M.~A. 1982, \mnras, 198, 1059

\bibitem[{{Dubus} {et~al.}(1997){Dubus}, {Charles}, {Long}, \&
  {Hakala}}]{dubus97}
{Dubus}, G., {Charles}, P.~A., {Long}, K.~S., \& {Hakala}, P.~J. 1997, \apjl,
  490, L47

\bibitem[{{Duschl} \& {Lesch}(1994)}]{duschl94}
{Duschl}, W.~J., \& {Lesch}, H. 1994, \aap, 286, 431

\bibitem[{{Elwood} {et~al.}(2019){Elwood}, {Murphy}, \&
  {D{\'{\i}}az-Rodr{\'{\i}}guez}}]{elwood19}
{Elwood}, B.~D., {Murphy}, J.~W., \& {D{\'{\i}}az-Rodr{\'{\i}}guez}, M. 2019,
  \mnras, 483, 4551

\bibitem[{{Fender} {et~al.}(2004){Fender}, {Belloni}, \& {Gallo}}]{fender04}
{Fender}, R.~P., {Belloni}, T.~M., \& {Gallo}, E. 2004, \mnras, 355, 1105

\bibitem[{{Filipovi{\'c}} {et~al.}(2005){Filipovi{\'c}}, {Payne}, {Reid},
  {Danforth}, {Staveley-Smith}, {Jones}, \& {White}}]{filipovic05}
{Filipovi{\'c}}, M.~D., {Payne}, J.~L., {Reid}, W., {et~al.} 2005, \mnras, 364,
  217

\bibitem[{{Foschini} {et~al.}(2004){Foschini}, {Rodriguez}, {Fuchs}, {Ho},
  {Dadina}, {Di Cocco}, {Courvoisier}, \& {Malaguti}}]{foschini04}
{Foschini}, L., {Rodriguez}, J., {Fuchs}, Y., {et~al.} 2004, \aap, 416, 529

\bibitem[{{Freedman} {et~al.}(2001){Freedman}, {Madore}, {Gibson}, {Ferrarese},
  {Kelson}, {Sakai}, {Mould}, {Kennicutt}, {Ford}, {Graham}, {Huchra},
  {Hughes}, {Illingworth}, {Macri}, \& {Stetson}}]{freedman01}
{Freedman}, W.~L., {Madore}, B.~F., {Gibson}, B.~K., {et~al.} 2001, \apj, 553,
  47

\bibitem[{{Gallo} {et~al.}(2005){Gallo}, {Fender}, {Kaiser}, {Russell},
  {Morganti}, {Oosterloo}, \& {Heinz}}]{gallo05}
{Gallo}, E., {Fender}, R., {Kaiser}, C., {et~al.} 2005, \nat, 436, 819

\bibitem[{{Galvin} \& {Filipovic}(2014)}]{galvin14}
{Galvin}, T.~J., \& {Filipovic}, M.~D. 2014, Serbian Astronomical Journal, 189,
  15

\bibitem[{{Galvin} {et~al.}(2014){Galvin}, {Filipovi{\'c}}, {Tothill},
  {Crawford}, {O'Brien}, {Seymour}, {Pannuti}, {Kosakowski}, \&
  {Sharma}}]{galvin14_ngc7793}
{Galvin}, T.~J., {Filipovi{\'c}}, M.~D., {Tothill}, N.~F.~H., {et~al.} 2014,
  \apss, 353, 603

\bibitem[{{Garofali} {et~al.}(2017){Garofali}, {Williams}, {Plucinsky},
  {Gaetz}, {Wold}, {Haberl}, {Long}, {Blair}, {Pannuti}, {Winkler}, \&
  {Gross}}]{garofali17}
{Garofali}, K., {Williams}, B.~F., {Plucinsky}, P.~P., {et~al.} 2017, \mnras,
  472, 308

\bibitem[{{Gebhardt} {et~al.}(2001){Gebhardt}, {Lauer}, {Kormendy}, {Pinkney},
  {Bower}, {Green}, {Gull}, {Hutchings}, {Kaiser}, {Nelson}, {Richstone}, \&
  {Weistrop}}]{gebhardt01}
{Gebhardt}, K., {Lauer}, T.~R., {Kormendy}, J., {et~al.} 2001, \aj, 122, 2469

\bibitem[{{Gordon} {et~al.}(1999){Gordon}, {Duric}, {Kirshner}, {Goss}, \&
  {Viallefond}}]{gordon99}
{Gordon}, S.~M., {Duric}, N., {Kirshner}, R.~P., {Goss}, W.~M., \&
  {Viallefond}, F. 1999, \apjs, 120, 247

\bibitem[{{Gordon} {et~al.}(1998){Gordon}, {Kirshner}, {Long}, {Blair},
  {Duric}, \& {Smith}}]{gordon98}
{Gordon}, S.~M., {Kirshner}, R.~P., {Long}, K.~S., {et~al.} 1998, \apjs, 117,
  89

\bibitem[{{Green}(2014)}]{green14}
{Green}, D.~A. 2014, Bulletin of the Astronomical Society of India, 42, 47

\bibitem[{{Grimm} {et~al.}(2007){Grimm}, {McDowell}, {Zezas}, {Kim}, \&
  {Fabbiano}}]{grimm07}
{Grimm}, H.-J., {McDowell}, J., {Zezas}, A., {Kim}, D.-W., \& {Fabbiano}, G.
  2007, \apjs, 173, 70

\bibitem[{{Hamilton} {et~al.}(1983){Hamilton}, {Sarazin}, \&
  {Chevalier}}]{hamilton83}
{Hamilton}, A.~J.~S., {Sarazin}, C.~L., \& {Chevalier}, R.~A. 1983, \apjs, 51,
  115

\bibitem[{{Hilditch} {et~al.}(2005){Hilditch}, {Howarth}, \&
  {Harries}}]{hilditch05}
{Hilditch}, R.~W., {Howarth}, I.~D., \& {Harries}, T.~J. 2005, \mnras, 357, 304

\bibitem[{{Johnston} {et~al.}(2008){Johnston}, {Taylor}, {Bailes}, {Bartel},
  {Baugh}, {Bietenholz}, {Blake}, {Braun}, {Brown}, {Chatterjee}, {Darling},
  {Deller}, {Dodson}, {Edwards}, {Ekers}, {Ellingsen}, {Feain}, {Gaensler},
  {Haverkorn}, {Hobbs}, {Hopkins}, {Jackson}, {James}, {Joncas}, {Kaspi},
  {Kilborn}, {Koribalski}, {Kothes}, {Landecker}, {Lenc}, {Lovell}, {Macquart},
  {Manchester}, {Matthews}, {McClure-Griffiths}, {Norris}, {Pen}, {Phillips},
  {Power}, {Protheroe}, {Sadler}, {Schmidt}, {Stairs}, {Staveley-Smith},
  {Stil}, {Tingay}, {Tzioumis}, {Walker}, {Wall}, \& {Wolleben}}]{johnston08}
{Johnston}, S., {Taylor}, R., {Bailes}, M., {et~al.} 2008, Experimental
  Astronomy, 22, 151

\bibitem[{{Lacey} \& {Duric}(2001)}]{lacey01}
{Lacey}, C.~K., \& {Duric}, N. 2001, \apj, 560, 719

\bibitem[{{Lee} \& {Lee}(2014{\natexlab{a}})}]{lee14}
{Lee}, J.~H., \& {Lee}, M.~G. 2014{\natexlab{a}}, \apj, 793, 134

\bibitem[{{Lee} \& {Lee}(2014{\natexlab{b}})}]{lee14b}
---. 2014{\natexlab{b}}, \apj, 786, 130

\bibitem[{{Lin} {et~al.}(2017){Lin}, {Hu}, {Kong}, {Gao}, {Zou}, {Wang},
  {Cheng}, {Fang}, {Lin}, \& {Wang}}]{lin17}
{Lin}, Z., {Hu}, N., {Kong}, X., {et~al.} 2017, \apj, 842, 97

\bibitem[{{Long} {et~al.}(1990){Long}, {Blair}, {Kirshner}, \&
  {Winkler}}]{long90}
{Long}, K.~S., {Blair}, W.~P., {Kirshner}, R.~P., \& {Winkler}, P.~F. 1990,
  \apjs, 72, 61

\bibitem[{{Long} {et~al.}(2018){Long}, {Blair}, {Milisavljevic}, {Raymond}, \&
  {Winkler}}]{long18}
{Long}, K.~S., {Blair}, W.~P., {Milisavljevic}, D., {Raymond}, J.~C., \&
  {Winkler}, P.~F. 2018, \apj, 193, 31

\bibitem[{{Long} {et~al.}(2002){Long}, {Charles}, \& {Dubus}}]{long02}
{Long}, K.~S., {Charles}, P.~A., \& {Dubus}, G. 2002, \apj, 569, 204

\bibitem[{{Long} {et~al.}(1981){Long}, {Dodorico}, {Charles}, \&
  {Dopita}}]{long81}
{Long}, K.~S., {Dodorico}, S., {Charles}, P.~A., \& {Dopita}, M.~A. 1981,
  \apjl, 246, L61

\bibitem[{{Long} {et~al.}(2014){Long}, {Kuntz}, {Blair}, {Godfrey},
  {Plucinsky}, {Soria}, {Stockdale}, \& {Winkler}}]{long14}
{Long}, K.~S., {Kuntz}, K.~D., {Blair}, W.~P., {et~al.} 2014, \apjs, 212, 21

\bibitem[{{Long} {et~al.}(2010){Long}, {Blair}, {Winkler}, {Becker}, {Gaetz},
  {Ghavamian}, {Helfand}, {Hughes}, {Kirshner}, {Kuntz}, {McNeil}, {Pannuti},
  {Plucinsky}, {Saul}, {T{\"u}llmann}, \& {Williams}}]{long10}
{Long}, K.~S., {Blair}, W.~P., {Winkler}, P.~F., {et~al.} 2010, \apjs, 187, 495

\bibitem[{{Maggi} {et~al.}(2016){Maggi}, {Haberl}, {Kavanagh}, {Sasaki},
  {Bozzetto}, {Filipovi{\'c}}, {Vasilopoulos}, {Pietsch}, {Points}, {Chu},
  {Dickel}, {Ehle}, {Williams}, \& {Greiner}}]{maggi16}
{Maggi}, P., {Haberl}, F., {Kavanagh}, P.~J., {et~al.} 2016, \aap, 585, A162

\bibitem[{{Massey} {et~al.}(2007){Massey}, {McNeill}, {Olsen}, {Hodge},
  {Blaha}, {Jacoby}, {Smith}, \& {Strong}}]{massey07}
{Massey}, P., {McNeill}, R.~T., {Olsen}, K.~A.~G., {et~al.} 2007, \aj, 134,
  2474

\bibitem[{{Massey} {et~al.}(2006){Massey}, {Olsen}, {Hodge}, {Strong},
  {Jacoby}, {Schlingman}, \& {Smith}}]{massey06}
{Massey}, P., {Olsen}, K.~A.~G., {Hodge}, P.~W., {et~al.} 2006, \aj, 131, 2478

\bibitem[{{Matonick} \& {Fesen}(1997)}]{matonick97}
{Matonick}, D.~M., \& {Fesen}, R.~A. 1997, \apjs, 112, 49

\bibitem[{{Middleton} {et~al.}(2011){Middleton}, {Sutton}, \&
  {Roberts}}]{middleton11}
{Middleton}, M.~J., {Sutton}, A.~D., \& {Roberts}, T.~P. 2011, \mnras, 417, 464

\bibitem[{{Norris} {et~al.}(2011){Norris}, {Hopkins}, {Afonso}, {Brown},
  {Condon}, {Dunne}, {Feain}, {Hollow}, {Jarvis}, {Johnston-Hollitt}, {Lenc},
  {Middelberg}, {Padovani}, {Prandoni}, {Rudnick}, {Seymour}, {Umana},
  {Andernach}, {Alexander}, {Appleton}, {Bacon}, {Banfield}, {Becker}, {Brown},
  {Ciliegi}, {Jackson}, {Eales}, {Edge}, {Gaensler}, {Giovannini}, {Hales},
  {Hancock}, {Huynh}, {Ibar}, {Ivison}, {Kennicutt}, {Kimball}, {Koekemoer},
  {Koribalski}, {L{\'o}pez-S{\'a}nchez}, {Mao}, {Murphy}, {Messias},
  {Pimbblet}, {Raccanelli}, {Randall}, {Reiprich}, {Roseboom},
  {R{\"o}ttgering}, {Saikia}, {Sharp}, {Slee}, {Smail}, {Thompson}, {Urquhart},
  {Wall}, \& {Zhao}}]{norris11}
{Norris}, R.~P., {Hopkins}, A.~M., {Afonso}, J., {et~al.} 2011, \pasa, 28, 215

\bibitem[{{Norris} {et~al.}(2013){Norris}, {Afonso}, {Bacon}, {Beck}, {Bell},
  {Beswick}, {Best}, {Bhatnagar}, {Bonafede}, {Brunetti}, {Budav{\'a}ri},
  {Cassano}, {Condon}, {Cress}, {Dabbech}, {Feain}, {Fender}, {Ferrari},
  {Gaensler}, {Giovannini}, {Haverkorn}, {Heald}, {Van der Heyden}, {Hopkins},
  {Jarvis}, {Johnston-Hollitt}, {Kothes}, {Van Langevelde}, {Lazio}, {Mao},
  {Mart{\'{\i}}nez-Sansigre}, {Mary}, {Mcalpine}, {Middelberg}, {Murphy},
  {Padovani}, {Paragi}, {Prandoni}, {Raccanelli}, {Rigby}, {Roseboom},
  {R{\"o}ttgering}, {Sabater}, {Salvato}, {Scaife}, {Schilizzi}, {Seymour},
  {Smith}, {Umana}, {Zhao}, \& {Zinn}}]{norris13}
{Norris}, R.~P., {Afonso}, J., {Bacon}, D., {et~al.} 2013, \pasa, 30, e020

\bibitem[{{Orosz} {et~al.}(2007){Orosz}, {McClintock}, {Narayan}, {Bailyn},
  {Hartman}, {Macri}, {Liu}, {Pietsch}, {Remillard}, {Shporer}, \&
  {Mazeh}}]{orosz07}
{Orosz}, J.~A., {McClintock}, J.~E., {Narayan}, R., {et~al.} 2007, \nat, 449,
  872

\bibitem[{{Pakull} \& {Mirioni}(2003)}]{pakull03}
{Pakull}, M.~W., \& {Mirioni}, L. 2003, in Revista Mexicana de Astronomia y
  Astrofisica Conference Series, Vol.~15, Revista Mexicana de Astronomia y
  Astrofisica Conference Series, ed. J.~{Arthur} \& W.~J. {Henney}, 197--199

\bibitem[{{Pannuti} {et~al.}(2002){Pannuti}, {Duric}, {Lacey}, {Ferguson},
  {Magnor}, \& {Mendelowitz}}]{pannuti02}
{Pannuti}, T.~G., {Duric}, N., {Lacey}, C.~K., {et~al.} 2002, \apj, 565, 966

\bibitem[{{Pietrzy{\'n}ski} {et~al.}(2013){Pietrzy{\'n}ski}, {Graczyk},
  {Gieren}, {Thompson}, {Pilecki}, {Udalski}, {Soszy{\'n}ski}, {Koz{\l}owski},
  {Konorski}, {Suchomska}, {Bono}, {Moroni}, {Villanova}, {Nardetto},
  {Bresolin}, {Kudritzki}, {Storm}, {Gallenne}, {Smolec}, {Minniti}, {Kubiak},
  {Szyma{\'n}ski}, {Poleski}, {Wyrzykowski}, {Ulaczyk}, {Pietrukowicz},
  {G{\'o}rski}, \& {Karczmarek}}]{pietrzynski13}
{Pietrzy{\'n}ski}, G., {Graczyk}, D., {Gieren}, W., {et~al.} 2013, \nat, 495,
  76

\bibitem[{{Plotkin} {et~al.}(2012){Plotkin}, {Markoff}, {Kelly}, {K{\"o}rding},
  \& {Anderson}}]{plotkin12}
{Plotkin}, R.~M., {Markoff}, S., {Kelly}, B.~C., {K{\"o}rding}, E., \&
  {Anderson}, S.~F. 2012, \mnras, 419, 267

\bibitem[{{Sarbadhicary} {et~al.}(2017){Sarbadhicary}, {Badenes}, {Chomiuk},
  {Caprioli}, \& {Huizenga}}]{sarbadhicary17}
{Sarbadhicary}, S.~K., {Badenes}, C., {Chomiuk}, L., {Caprioli}, D., \&
  {Huizenga}, D. 2017, \mnras, 464, 2326

\bibitem[{{Shimwell, T. W.} {et~al.}(2019){Shimwell, T. W.}, {Tasse, C.},
  {Hardcastle, M. J.}, {Mechev, A. P.}, {Williams, W. L.}, {Best, P. N.},
  {R\"ottgering, H. J. A.}, {Callingham, J. R.}, {Dijkema, T. J.}, {de
  Gasperin, F.}, {Hoang, D. N.}, {Hugo, B.}, {Mirmont, M.}, {Oonk, J. B. R.},
  {Prandoni, I.}, {Rafferty, D.}, {Sabater, J.}, {Smirnov, O.}, {van Weeren, R.
  J.}, {White, G. J.}, {Atemkeng, M.}, {Bester, L.}, {Bonnassieux, E.},
  {Br\"uggen, M.}, {Brunetti, G.}, {Chyzy, K. T.}, {Cochrane, R.}, {Conway, J.
  E.}, {Croston, J. H.}, {Danezi, A.}, {Duncan, K.}, {Haverkorn, M.}, {Heald,
  G. H.}, {Iacobelli, M.}, {Intema, H. T.}, {Jackson, N.}, {Jamrozy, M.},
  {Jarvis, M. J.}, {Lakhoo, R.}, {Mevius, M.}, {Miley, G. K.}, {Morabito, L.},
  {Morganti, R.}, {Nisbet, D.}, {Orr\'u, E.}, {Perkins, S.}, {Pizzo, R. F.},
  {Schrijvers, C.}, {Smith, D. J. B.}, {Vermeulen, R.}, {Wise, M. W.}, {Alegre,
  L.}, {Bacon, D. J.}, {van Bemmel, I. M.}, {Beswick, R. J.}, {Bonafede, A.},
  {Botteon, A.}, {Bourke, S.}, {Brienza, M.}, {Calistro Rivera, G.}, {Cassano,
  R.}, {Clarke, A. O.}, {Conselice, C. J.}, {Dettmar, R. J.}, {Drabent, A.},
  {Dumba, C.}, {Emig, K. L.}, {En\ss{}lin, T. A.}, {Ferrari, C.}, {Garrett, M.
  A.}, {G\'enova-Santos, R. T.}, {Goyal, A.}, {G\"urkan, G.}, {Hale, C.},
  {Harwood, J. J.}, {Heesen, V.}, {Hoeft, M.}, {Horellou, C.}, {Jackson, C.},
  {Kokotanekov, G.}, {Kondapally, R.}, {Kunert-Bajraszewska, M.}, {Mahatma,
  V.}, {Mahony, E. K.}, {Mandal, S.}, {McKean, J. P.}, {Merloni, A.}, {Mingo,
  B.}, {Miskolczi, A.}, {Mooney, S.}, {Nikiel-Wroczy\'{}nski, B.},
  {O\'{}Sullivan, S. P.}, {Quinn, J.}, {Reich, W.}, {Roskowi\'{}nski, C.},
  {Rowlinson, A.}, {Savini, F.}, {Saxena, A.}, {Schwarz, D. J.}, {Shulevski,
  A.}, {Sridhar, S. S.}, {Stacey, H. R.}, {Urquhart, S.}, {van der Wiel, M. H.
  D.}, {Varenius, E.}, {Webster, B.}, \& {Wilber, A.}}]{shimwell19}
{Shimwell, T. W.}, {Tasse, C.}, {Hardcastle, M. J.}, {et~al.} 2019, \aap, 622,
  A1.
\newblock \url{https://doi.org/10.1051/0004-6361/201833559}

\bibitem[{{Smol{\v c}i{\'c}} {et~al.}(2017){Smol{\v c}i{\'c}}, {Novak},
  {Bondi}, {Ciliegi}, {Mooley}, {Schinnerer}, {Zamorani}, {Navarrete},
  {Bourke}, {Karim}, {Vardoulaki}, {Leslie}, {Delhaize}, {Carilli}, {Myers},
  {Baran}, {Delvecchio}, {Miettinen}, {Banfield}, {Balokovi{\'c}}, {Bertoldi},
  {Capak}, {Frail}, {Hallinan}, {Hao}, {Herrera Ruiz}, {Horesh}, {Ilbert},
  {Intema}, {Jeli{\'c}}, {Kl{\"o}ckner}, {Krpan}, {Kulkarni}, {McCracken},
  {Laigle}, {Middleberg}, {Murphy}, {Sargent}, {Scoville}, \&
  {Sheth}}]{smolcic17}
{Smol{\v c}i{\'c}}, V., {Novak}, M., {Bondi}, M., {et~al.} 2017, \aap, 602, A1

\bibitem[{{Soderberg} {et~al.}(2005){Soderberg}, {Kulkarni}, {Berger},
  {Chevalier}, {Frail}, {Fox}, \& {Walker}}]{soderberg05}
{Soderberg}, A.~M., {Kulkarni}, S.~R., {Berger}, E., {et~al.} 2005, \apj, 621,
  908

\bibitem[{{Stark} {et~al.}(1992){Stark}, {Gammie}, {Wilson}, {Bally}, {Linke},
  {Heiles}, \& {Hurwitz}}]{stark92}
{Stark}, A.~A., {Gammie}, C.~F., {Wilson}, R.~W., {et~al.} 1992, \apjs, 79, 77

\bibitem[{{T{\"u}llmann} {et~al.}(2011){T{\"u}llmann}, {Gaetz}, {Plucinsky},
  {Kuntz}, {Williams}, {Pietsch}, {Haberl}, {Long}, {Blair}, {Sasaki},
  {Winkler}, {Challis}, {Pannuti}, {Edgar}, {Helfand}, {Hughes}, {Kirshner},
  {Mazeh}, \& {Shporer}}]{tuellmann11}
{T{\"u}llmann}, R., {Gaetz}, T.~J., {Plucinsky}, P.~P., {et~al.} 2011, \apjs,
  193, 31

\bibitem[{{Uro{\v{s}}evi{\'c}}(2014)}]{urovsevic14}
{Uro{\v{s}}evi{\'c}}, D. 2014, \apss, 354, 541

\bibitem[{{van Haarlem} {et~al.}(2013){van Haarlem}, {Wise}, {Gunst}, {Heald},
  {McKean}, {Hessels}, {de Bruyn}, {Nijboer}, {Swinbank}, {Fallows},
  {Brentjens}, {Nelles}, {Beck}, {Falcke}, {Fender}, {H{\"o}randel},
  {Koopmans}, {Mann}, {Miley}, {R{\"o}ttgering}, {Stappers}, {Wijers},
  {Zaroubi}, {van den Akker}, {Alexov}, {Anderson}, {Anderson}, {van Ardenne},
  {Arts}, {Asgekar}, {Avruch}, {Batejat}, {B{\"a}hren}, {Bell}, {Bell}, {van
  Bemmel}, {Bennema}, {Bentum}, {Bernardi}, {Best}, {B{\^i}rzan}, {Bonafede},
  {Boonstra}, {Braun}, {Bregman}, {Breitling}, {van de Brink}, {Broderick},
  {Broekema}, {Brouw}, {Br{\"u}ggen}, {Butcher}, {van Cappellen}, {Ciardi},
  {Coenen}, {Conway}, {Coolen}, {Corstanje}, {Damstra}, {Davies}, {Deller},
  {Dettmar}, {van Diepen}, {Dijkstra}, {Donker}, {Doorduin}, {Dromer}, {Drost},
  {van Duin}, {Eisl{\"o}ffel}, {van Enst}, {Ferrari}, {Frieswijk}, {Gankema},
  {Garrett}, {de Gasperin}, {Gerbers}, {de Geus}, {Grie{\ss}meier}, {Grit},
  {Gruppen}, {Hamaker}, {Hassall}, {Hoeft}, {Holties}, {Horneffer}, {van der
  Horst}, {van Houwelingen}, {Huijgen}, {Iacobelli}, {Intema}, {Jackson},
  {Jelic}, {de Jong}, {Juette}, {Kant}, {Karastergiou}, {Koers}, {Kollen},
  {Kondratiev}, {Kooistra}, {Koopman}, {Koster}, {Kuniyoshi}, {Kramer},
  {Kuper}, {Lambropoulos}, {Law}, {van Leeuwen}, {Lemaitre}, {Loose}, {Maat},
  {Macario}, {Markoff}, {Masters}, {McFadden}, {McKay-Bukowski}, {Meijering},
  {Meulman}, {Mevius}, {Middelberg}, {Millenaar}, {Miller-Jones}, {Mohan},
  {Mol}, {Morawietz}, {Morganti}, {Mulcahy}, {Mulder}, {Munk}, {Nieuwenhuis},
  {van Nieuwpoort}, {Noordam}, {Norden}, {Noutsos}, {Offringa}, {Olofsson},
  {Omar}, {Orr{\'u}}, {Overeem}, {Paas}, {Pandey-Pommier}, {Pandey}, {Pizzo},
  {Polatidis}, {Rafferty}, {Rawlings}, {Reich}, {de Reijer}, {Reitsma},
  {Renting}, {Riemers}, {Rol}, {Romein}, {Roosjen}, {Ruiter}, {Scaife}, {van
  der Schaaf}, {Scheers}, {Schellart}, {Schoenmakers}, {Schoonderbeek},
  {Serylak}, {Shulevski}, {Sluman}, {Smirnov}, {Sobey}, {Spreeuw}, {Steinmetz},
  {Sterks}, {Stiepel}, {Stuurwold}, {Tagger}, {Tang}, {Tasse}, {Thomas},
  {Thoudam}, {Toribio}, {van der Tol}, {Usov}, {van Veelen}, {van der Veen},
  {ter Veen}, {Verbiest}, {Vermeulen}, {Vermaas}, {Vocks}, {Vogt}, {de Vos},
  {van der Wal}, {van Weeren}, {Weggemans}, {Weltevrede}, {White}, {Wijnholds},
  {Wilhelmsson}, {Wucknitz}, {Yatawatta}, {Zarka}, {Zensus}, \& {van
  Zwieten}}]{vanhaarlem13}
{van Haarlem}, M.~P., {Wise}, M.~W., {Gunst}, A.~W., {et~al.} 2013, \aap, 556,
  A2

\bibitem[{{White} {et~al.}(1997){White}, {Becker}, {Helfand}, \&
  {Gregg}}]{white97}
{White}, R.~L., {Becker}, R.~H., {Helfand}, D.~J., \& {Gregg}, M.~D. 1997,
  \apj, 475, 479

\bibitem[{{White} \& {Long}(1991)}]{white91}
{White}, R.~L., \& {Long}, K.~S. 1991, \apj, 373, 543

\bibitem[{{Zaritsky} {et~al.}(1989){Zaritsky}, {Elston}, \&
  {Hill}}]{zaritsky89}
{Zaritsky}, D., {Elston}, R., \& {Hill}, J.~M. 1989, \aj, 97, 97

\end{thebibliography}

\begin{splitdeluxetable}{lllrrcccccBcrcrrrcrrccccc}
\tablecaption{M33 Radio Source Catalog \label{tab:radio_cat}}
\tablehead{\colhead{Name} & \colhead{RA} & \colhead{Dec} & \colhead{RAPeak} & \colhead{DecPeak} & \colhead{Wrn} & \colhead{H$\alpha(\mathrm{tot})$} & \colhead{H$\alpha(\mathrm{ave})$} & \colhead{F1.4GHz} & \colhead{$\alpha$} & \colhead{$\Fint$} & \colhead{$\nu_p$} & \colhead{$F_p$} & \colhead{Major} & \colhead{Minor} & \colhead{PA} & \colhead{nBands} & \colhead{Island} & \colhead{Sflag} & \colhead{SNRnm1} & \colhead{SNRnm2} & \colhead{SNRnm3} & \colhead{nXray} & \colhead{nHII} \\ 
\colhead{~}
& \colhead{(J2000)}
& \colhead{(J2000)}
& \colhead{(J2000)}
& \colhead{(J2000)}
& \colhead{~}
& \colhead{\pbox{5em}{\centering $(\mathrm{erg}\,\mathrm{cm}^{-2}$\break$\mathrm{s}^{-1})$}}
& \colhead{\pbox{5em}{\centering $(\mathrm{erg}\,\mathrm{cm}^{-2}$\break$\mathrm{s}^{-1}\mathrm{arcsec}^{-2})$}}
& \colhead{($\mu$Jy)}
& \colhead{~}
& \colhead{($\mu$Jy)}
& \colhead{(GHz)}
& \colhead{($\mu$Jy)}
& \colhead{(arcsec)}
& \colhead{(arcsec)}
& \colhead{(deg)}
& \colhead{~}
& \colhead{~}
& \colhead{~}
& \colhead{~}
& \colhead{~}
& \colhead{~}
& \colhead{~}
& \colhead{~}
}
\tabletypesize{\scriptsize}
\tablewidth{0pt}
\colnumbers
\startdata
W19-0669 & 01:32:27.188 & +30:25:34.97 & 23.11374 & 30.42654 & \nodata & 2.60e-14 & 9.54e-16 & \phn109.1 $\pm$ \phn9.8 & \phn0.889 $\pm$ 0.524 & \phn119.7 $\pm$ \phn8.6 & 1.554 & \phn84.8 $\pm$ \phn4.5 & 10.37 & 6.50 & 86.6 & 2 & 669 & 0 & \nodata & \nodata & \nodata & 0 & 1 \\
W19-0514 & 01:32:28.996 & +30:45:13.42 & 23.12105 & 30.75379 & \nodata & 1.89e-16 & 6.41e-18 & \phn140.9 $\pm$ 10.0 & -0.912 $\pm$ 0.183 & \phn122.9 $\pm$ \phn8.1 & 1.626 & \phn87.8 $\pm$ \phn3.5 & 9.12 & 5.80 & 84.1 & 4 & 514 & 0 & \nodata & \nodata & \nodata & 1 & 0 \\
W19-0705 & 01:32:29.035 & +30:27:43.42 & 23.12071 & 30.46156 & \nodata & 2.57e-15 & 8.70e-17 & \phn852.3 $\pm$ 31.2 & -0.220 $\pm$ 0.048 & \phn754.7 $\pm$ 19.0 & 2.433 & 224.9 $\pm$ \phn8.0 & 20.67 & 9.43 & 158.9 & 3 & 705 & 3 & L10-001 & \nodata & \nodata & 0 & 0 \\
W19-0974 & 01:32:29.420 & +30:36:15.02 & 23.12186 & 30.60490 & \nodata & 9.42e-15 & 2.96e-16 & \phn214.3 $\pm$ 17.2 & -0.253 $\pm$ 0.112 & \phn190.8 $\pm$ 11.7 & 2.215 & \phn59.5 $\pm$ \phn4.7 & 17.15 & 10.09 & 176.6 & 4 & 974 & 0 & \nodata & \nodata & \nodata & 2 & 1 \\
W19-0357 & 01:32:30.404 & +30:27:37.11 & 23.12781 & 30.45908 & \nodata & 6.71e-15 & 2.27e-16 & 3180.0 $\pm$ 48.5 & -0.323 $\pm$ 0.018 & 2621.4 $\pm$ 27.5 & 2.546 & 314.5 $\pm$ 10.7 & 18.76 & 16.90 & 119.9 & 4 & 357 & 3 & L10-001 & \nodata & \nodata & 0 & 0 \\
W19-0402 & 01:32:30.419 & +30:27:57.65 & 23.12650 & 30.46741 & \nodata & 4.90e-15 & 1.66e-16 & 2799.4 $\pm$ 53.1 & -0.254 $\pm$ 0.023 & 2408.5 $\pm$ 31.8 & 2.530 & 266.2 $\pm$ \phn8.6 & 26.32 & 18.91 & 85.7 & 4 & 402 & 11 & L10-001 & \nodata & \nodata & 0 & 0 \\
W19-0345 & 01:32:31.413 & +30:35:26.66 & 23.13029 & 30.59158 & \nodata & 3.85e-14 & 1.31e-15 & 1785.1 $\pm$ 47.8 & \phn0.282 $\pm$ 0.025 & 2319.1 $\pm$ 31.0 & 3.537 & 290.5 $\pm$ \phn6.4 & 32.85 & 19.46 & 64.0 & 4 & 345 & 9 & L10-002 & LL14-004 & \nodata & 1 & 3 \\
W19-0413 & 01:32:31.708 & +30:35:07.27 & 23.13192 & 30.58742 & \nodata & 1.86e-14 & 6.31e-16 & 1139.9 $\pm$ 32.7 & \phn0.455 $\pm$ 0.026 & 1798.1 $\pm$ 23.1 & 3.810 & 319.1 $\pm$ \phn7.1 & 31.02 & 19.89 & 170.8 & 4 & 413 & 0 & \nodata & \nodata & \nodata & 0 & 2 \\
W19-1104 & 01:32:32.541 & +30:35:19.21 & 23.13547 & 30.58854 & \nodata & 4.56e-14 & 1.54e-15 & \phn418.6 $\pm$ 16.3 & \phn0.330 $\pm$ 0.035 & \phn577.2 $\pm$ 10.6 & 3.710 & 192.7 $\pm$ \phn6.9 & 16.45 & 13.40 & 172.5 & 4 & 1104 & 0 & \nodata & \nodata & \nodata & 0 & 1 \\
W19-2430 & 01:32:33.371 & +30:35:09.19 & 23.13741 & 30.58549 & \nodata & 9.08e-15 & 3.33e-16 & \phn292.8 $\pm$ 20.0 & \phn0.419 $\pm$ 0.060 & \phn451.3 $\pm$ 13.3 & 3.928 & \phn98.0 $\pm$ \phn5.3 & 27.37 & 20.77 & 69.9 & 4 & 2430 & 3 & LL14-004 & \nodata & \nodata & 0 & 0 \\
W19-1906 & 01:32:34.180 & +30:27:32.60 & 23.14199 & 30.45855 & \nodata & 2.37e-14 & 8.68e-16 & \phn131.6 $\pm$ 12.7 & \phn0.175 $\pm$ 0.098 & \phn151.3 $\pm$ \phn8.5 & 3.108 & \phn80.1 $\pm$ \phn8.4 & 16.14 & 9.23 & 15.3 & 4 & 1906 & 0 & \nodata & \nodata & \nodata & 0 & 1 \\
W19-0461 & 01:32:34.388 & +30:27:11.89 & 23.14298 & 30.45189 & \nodata & 3.83e-14 & 1.20e-15 & \phn758.2 $\pm$ 36.7 & \phn0.078 $\pm$ 0.049 & \phn806.6 $\pm$ 23.2 & 3.078 & 122.8 $\pm$ \phn8.1 & 31.18 & 18.17 & 54.6 & 4 & 461 & 0 & \nodata & \nodata & \nodata & 0 & 3 \\
W19-0630 & 01:32:34.519 & +30:27:47.30 & 23.14423 & 30.46356 & \nodata & 3.24e-14 & 1.29e-15 & \phn713.1 $\pm$ 32.0 & \phn0.149 $\pm$ 0.047 & \phn798.4 $\pm$ 21.7 & 2.992 & 108.1 $\pm$ \phn7.2 & 21.92 & 14.93 & 149.5 & 4 & 630 & 0 & \nodata & \nodata & \nodata & 0 & 1 \\
W19-0588 & 01:32:34.890 & +30:30:29.66 & 23.14539 & 30.50856 & \nodata & 3.31e-14 & 1.04e-15 & \phn802.1 $\pm$ 31.3 & -0.053 $\pm$ 0.040 & \phn769.2 $\pm$ 17.9 & 3.086 & 123.8 $\pm$ \phn5.0 & 19.09 & 14.56 & 165.2 & 4 & 588 & 0 & \nodata & \nodata & \nodata & 0 & 1 \\
W19-2913 & 01:32:38.495 & +30:41:09.56 & 23.16036 & 30.68637 & \nodata & 1.24e-14 & 4.20e-16 & \phn109.6 $\pm$ 12.0 & -0.323 $\pm$ 0.128 & \phn\phn89.0 $\pm$ \phn6.3 & 2.670 & \phn20.0 $\pm$ \phn2.4 & 14.37 & 14.12 & 40.2 & 4 & 2913 & 0 & \nodata & \nodata & \nodata & 0 & 1 \\
W19-1415 & 01:32:39.095 & +30:40:41.36 & 23.16264 & 30.67805 & \nodata & 2.72e-14 & 1.09e-15 & \phn\phn58.0 $\pm$ 10.4 & -0.216 $\pm$ 0.199 & \phn\phn50.1 $\pm$ \phn5.9 & 2.762 & \phn31.0 $\pm$ \phn3.6 & 10.78 & 7.86 & 10.5 & 4 & 1415 & 0 & \nodata & \nodata & \nodata & 0 & 1 \\
W19-1037 & 01:32:39.492 & +30:31:56.18 & 23.16337 & 30.53332 & \nodata & 7.79e-16 & 2.64e-17 & \phn132.9 $\pm$ 16.1 & -0.409 $\pm$ 0.149 & \phn105.5 $\pm$ \phn9.2 & 2.462 & \phn47.4 $\pm$ \phn3.5 & 33.52 & 15.20 & 125.4 & 4 & 1037 & 11 & LL14-009 & \nodata & \nodata & 0 & 0 \\
W19-2377 & 01:32:39.755 & +30:22:27.86 & 23.16575 & 30.37444 & \nodata & 2.39e-14 & 8.08e-16 & \phn\phn26.1 $\pm$ \phn5.5 & -0.772 $\pm$ 1.985 & \phn\phn25.5 $\pm$ \phn5.2 & 1.442 & \phn28.4 $\pm$ \phn5.1 & 4.82 & 4.56 & 167.4 & 2 & 2377 & 0 & \nodata & \nodata & \nodata & 0 & 1 \\
W19-0406 & 01:32:39.849 & +30:24:30.52 & 23.16598 & 30.40860 & \nodata & 9.99e-15 & 3.38e-16 & \phn181.7 $\pm$ \phn8.9 & -0.739 $\pm$ 0.121 & \phn160.9 $\pm$ \phn7.2 & 1.650 & 109.3 $\pm$ \phn3.6 & 7.61 & 6.44 & 31.6 & 4 & 406 & 0 & \nodata & \nodata & \nodata & 0 & 1 \\
W19-2626 & 01:32:50.718 & +30:30:34.86 & 23.21116 & 30.50980 & W & 7.51e-16 & 2.54e-17 & \phn\phn15.1 $\pm$ \phn4.7 & -0.642 $\pm$ 0.467 & \phn\phn11.8 $\pm$ \phn3.0 & 2.062 & \phn12.2 $\pm$ \phn2.2 & 4.22 & 4.19 & 170.3 & 4 & 2626 & 0 & \nodata & \nodata & \nodata & 1 & 0
\\
\enddata
\end{splitdeluxetable}

\clearpage

\begin{splitdeluxetable}{lllrrcccccBrrrcccccc}
\tablecaption{Forced Photometry of M33 SNR Candidates \label{tab:forced}}
\tablehead{\colhead{Name} & \colhead{RA} & \colhead{Dec} & \colhead{$\rho$} & \colhead{D} & \colhead{Opt} & \colhead{Xray} & \colhead{Radio} & \colhead{H$\alpha(\mathrm{tot})$} & \colhead{H$\alpha(\mathrm{ave})$} & \colhead{F1.4GHz} & \colhead{$\alpha$} & \colhead{$\Fint$} & \colhead{$\nu_p$} & \colhead{nBands} & \colhead{Rflag} & \colhead{RadName1} & \colhead{RadName2} & \colhead{RadName3} \\ 
\colhead{~}
& \colhead{(J2000)}
& \colhead{(J2000)}
& \colhead{(kpc)}
& \colhead{(pc)}
& \colhead{~}
& \colhead{~}
& \colhead{~}
& \colhead{\pbox{5em}{\centering $(\mathrm{erg}\,\mathrm{cm}^{-2}$\break$\mathrm{s}^{-1})$}}
& \colhead{\pbox{5em}{\centering $(\mathrm{erg}\,\mathrm{cm}^{-2}$\break$\mathrm{s}^{-1}\mathrm{arcsec}^{-2})$}}
& \colhead{($\mu$Jy)}
& \colhead{~}
& \colhead{($\mu$Jy)}
& \colhead{(GHz)}
& \colhead{~}
& \colhead{~}
& \colhead{~}
& \colhead{~}
& \colhead{~}
}
\tabletypesize{\scriptsize}
\tablewidth{0pt}
\startdata
L10-001 & 01:32:30.37 & 30:27:46.9 & 6.46 & 126.7 & yes & yes & yes & 1.34e-13 & 1.65e-16 & 4920.2 $\pm$ 84.7 & -0.310 $\pm$ 0.019 & 3982.8 $\pm$ 44.8 & 2.771 & 4 & 9 & W19-0402 & W19-0357 & W19-0705 \\
L10-002 & 01:32:31.41 & 30:35:32.9 & 6.55 & 33.3 & yes & yes & yes & 2.89e-14 & 5.29e-16 & \phn282.0 $\pm$ 11.1 & -0.048 $\pm$ 0.043 & \phn272.7 $\pm$ \phn6.9 & 2.811 & 4 & 11 & W19-0345 & \nodata & \nodata \\
L10-003 & 01:32:42.54 & 30:20:58.9 & 6.11 & 104.4 & yes & no & yes & 1.09e-13 & 1.97e-16 & 2462.9 $\pm$ 49.9 & -1.028 $\pm$ 0.129 & 2260.2 $\pm$ 38.7 & 1.522 & 2 & 9 & W19-0422 & W19-0596 & W19-2050 \\
L10-004 & 01:32:44.83 & 30:22:14.6 & 5.83 & 42.8 & yes & no & yes & 1.04e-14 & 1.12e-16 & \phn272.5 $\pm$ 15.4 & -1.918 $\pm$ 0.467 & \phn250.7 $\pm$ 13.3 & 1.462 & 2 & 11 & W19-0691 & \nodata & \nodata \\
L10-005 & 01:32:46.73 & 30:34:37.8 & 5.2 & 49.1 & yes & yes & yes & 1.57e-14 & 1.30e-16 & \phn288.6 $\pm$ 16.7 & -0.486 $\pm$ 0.089 & \phn237.4 $\pm$ 10.8 & 2.094 & 4 & 11 & W19-0478 & \nodata & \nodata \\
L10-006 & 01:32:52.76 & 30:38:12.6 & 4.91 & 60.2 & yes & yes & yes & 2.39e-14 & 1.30e-16 & \phn333.1 $\pm$ 20.1 & -0.332 $\pm$ 0.083 & \phn283.1 $\pm$ 12.7 & 2.283 & 4 & 11 & W19-0277 & \nodata & \nodata \\
L10-007 & 01:32:53.36 & 30:48:23.1 & 6.32 & 77.0 & yes & yes & yes & 1.29e-14 & 4.25e-17 & \phn\phn79.3 $\pm$ 24.9 & -0.672 $\pm$ 0.452 & \phn\phn58.9 $\pm$ 14.3 & 2.177 & 4 & 0 & \nodata & \nodata & \nodata \\
L10-008 & 01:32:53.40 & 30:37:56.9 & 4.84 & 55.5 & yes & no & yes & 5.43e-14 & 3.41e-16 & \phn308.9 $\pm$ 17.4 & -0.160 $\pm$ 0.067 & \phn279.7 $\pm$ 10.6 & 2.601 & 4 & 3 & W19-0277 & \nodata & \nodata \\
L10-009 & 01:32:54.10 & 30:25:31.8 & 4.92 & 42.8 & yes & no & yes & 1.63e-14 & 1.79e-16 & \phn139.6 $\pm$ 11.1 & -0.406 $\pm$ 0.116 & \phn116.7 $\pm$ \phn7.1 & 2.174 & 4 & 11 & W19-1021 & \nodata & \nodata \\
L10-010 & 01:32:56.15 & 30:40:36.4 & 4.87 & 97.4 & yes & yes & yes & 8.99e-14 & 1.86e-16 & 1079.1 $\pm$ 34.9 & -0.165 $\pm$ 0.039 & \phn978.1 $\pm$ 22.1 & 2.544 & 4 & 9 & W19-1119 & W19-0827 & \nodata \\
L10-011 & 01:32:57.07 & 30:39:27.1 & 4.66 & 23.9 & yes & yes & yes & 2.47e-14 & 8.37e-16 & \phn376.6 $\pm$ \phn9.4 & -0.527 $\pm$ 0.043 & \phn317.3 $\pm$ \phn6.5 & 1.938 & 4 & 3 & W19-0206 & \nodata & \nodata \\
L10-012 & 01:33:00.15 & 30:30:46.2 & 4.11 & 56.2 & yes & yes & yes & 3.58e-13 & 2.28e-15 & \phn853.5 $\pm$ 22.5 & \phn0.048 $\pm$ 0.028 & \phn884.4 $\pm$ 14.4 & 2.926 & 4 & 9 & W19-0732 & W19-0508 & \nodata \\
L10-013 & 01:33:00.42 & 30:44:08.1 & 5.0 & 37.2 & yes & yes & yes & 9.03e-15 & 1.24e-16 & \phn308.3 $\pm$ \phn9.5 & -0.532 $\pm$ 0.049 & \phn255.8 $\pm$ \phn6.5 & 1.989 & 4 & 11 & W19-0208 & \nodata & \nodata \\
L10-014 & 01:33:00.67 & 30:30:59.3 & 4.07 & 49.6 & yes & no & yes & 2.49e-13 & 1.96e-15 & \phn579.0 $\pm$ 22.3 & \phn0.102 $\pm$ 0.039 & \phn628.6 $\pm$ 14.0 & 3.124 & 4 & 9 & W19-0449 & W19-0732 & \nodata \\
L10-015 & 01:33:01.51 & 30:30:49.6 & 4.01 & 32.0 & - & no & yes & 5.11e-14 & 9.77e-16 & \phn221.3 $\pm$ 13.3 & \phn0.126 $\pm$ 0.066 & \phn242.1 $\pm$ \phn9.2 & 2.852 & 4 & 9 & W19-0983 & W19-0449 & W19-0732 \\
L10-016 & 01:33:02.93 & 30:32:29.6 & 3.85 & 55.2 & yes & yes & yes & 3.72e-14 & 2.45e-16 & \phn124.7 $\pm$ 19.0 & -0.356 $\pm$ 0.196 & \phn102.7 $\pm$ 11.1 & 2.417 & 4 & 0 & \nodata & \nodata & \nodata \\
L10-017 & 01:33:03.57 & 30:31:20.9 & 3.84 & 36.6 & yes & yes & yes & 2.06e-14 & 2.93e-16 & \phn623.4 $\pm$ 14.5 & -0.395 $\pm$ 0.033 & \phn520.3 $\pm$ \phn9.3 & 2.211 & 4 & 11 & W19-0142 & \nodata & \nodata \\
L10-018 & 01:33:04.03 & 30:39:53.7 & 4.1 & 34.1 & yes & yes & yes & 4.17e-14 & 7.06e-16 & \phn505.9 $\pm$ 10.9 & -0.394 $\pm$ 0.033 & \phn431.9 $\pm$ \phn7.3 & 2.091 & 4 & 11 & W19-0188 & \nodata & \nodata \\
L10-019 & 01:33:07.55 & 30:42:52.5 & 4.19 & 74.8 & yes & yes & yes & 1.28e-13 & 4.45e-16 & \phn127.7 $\pm$ 20.4 & \phn0.473 $\pm$ 0.151 & \phn196.5 $\pm$ 16.2 & 3.479 & 4 & 11 & W19-0624 & \nodata & \nodata \\
L10-020 & 01:33:08.98 & 30:26:58.9 & 3.91 & 55.5 & yes & yes & yes & 2.32e-14 & 1.50e-16 & \phn\phn35.1 $\pm$ 14.0 & -0.142 $\pm$ 0.528 & \phn\phn32.7 $\pm$ \phn9.6 & 2.329 & 4 & 0 & \nodata & \nodata & \nodata
\\
\enddata
\end{splitdeluxetable}

\clearpage

\begin{splitdeluxetable}{lllrrrrrccrrBccccccccc}
\tablecaption{Forced Photometry of M33 X-ray Sources \label{tab:forced_xray}}
\tablehead{\colhead{Name} & \colhead{RA} & \colhead{Dec} & \colhead{D} & \colhead{$C_\mathrm{tot}$} & \colhead{$C_\mathrm{soft}$} & \colhead{$C_\mathrm{med}$} & \colhead{$C_\mathrm{hard}$} & \colhead{SNR} & \colhead{Radio} & \colhead{H$\alpha(\mathrm{tot})$} & \colhead{H$\alpha(\mathrm{ave})$} & \colhead{F1.4GHz} & \colhead{$\alpha$} & \colhead{$\Fint$} & \colhead{$\nu_p$} & \colhead{nBands} & \colhead{Rflag} & \colhead{RadName1} & \colhead{RadName2} & \colhead{RadName3} \\ 
\colhead{~}
& \colhead{(J2000)}
& \colhead{(J2000)}
& \colhead{(pc)}
& \colhead{(cts/s)}
& \colhead{(cts/s)}
& \colhead{(cts/s)}
& \colhead{(cts/s)}
& \colhead{~}
& \colhead{~}
& \colhead{\pbox{5em}{\centering $(\mathrm{erg}\,\mathrm{cm}^{-2}$\break$\mathrm{s}^{-1})$}}
& \colhead{\pbox{5em}{\centering $(\mathrm{erg}\,\mathrm{cm}^{-2}$\break$\mathrm{s}^{-1}\mathrm{arcsec}^{-2})$}}
& \colhead{($\mu$Jy)}
& \colhead{~}
& \colhead{($\mu$Jy)}
& \colhead{(GHz)}
& \colhead{~}
& \colhead{~}
& \colhead{~}
& \colhead{~}
& \colhead{~}
}
\tabletypesize{\scriptsize}
\tablewidth{0pt}
\startdata
T11-001 & 01:32:24.56 & 30:33:22.4 & 49.5 & 47.5 & 3.9 & 29.0 & 14.6 & no & no & 6.02e-15 & 1.22e-17 & \phn61.8 $\pm$ 31.8 & -3.000 $\pm$ 0.000 & \phn64.9 $\pm$ 33.3 & 1.378 & 4 & 0 & \nodata & \nodata & \nodata \\
T11-002 & 01:32:27.18 & 30:35:21.1 & 38.1 & 28.1 & 3.9 & 14.3 & 9.8 & no & no & 8.04e-15 & 2.72e-17 & -37.1 $\pm$ 21.6 & \phn0.777 $\pm$ 0.482 & -90.1 $\pm$ 17.1 & 4.382 & 4 & 0 & \nodata & \nodata & \nodata \\
T11-003 & 01:32:29.23 & 30:36:18.6 & 33.4 & 102.1 & 17.7 & 53.5 & 30.8 & no & yes & 3.91e-14 & 1.76e-16 & 197.4 $\pm$ 21.2 & -0.639 $\pm$ 0.207 & 165.3 $\pm$ 15.0 & 1.849 & 4 & 11 & W19-0974 & \nodata & \nodata \\
T11-004 & 01:32:29.31 & 30:45:14.0 & 36.6 & 32.8 & 7.9 & 22.1 & 2.8 & no & yes & 2.36e-15 & 8.88e-18 & 149.2 $\pm$ 25.3 & -0.186 $\pm$ 0.193 & 132.1 $\pm$ 15.0 & 2.687 & 4 & 11 & W19-0514 & \nodata & \nodata \\
T11-005 & 01:32:30.28 & 30:35:48.2 & 33.0 & 49.5 & 5.2 & 27.2 & 17.1 & no & yes & 9.88e-15 & 4.53e-17 & \phn\phn1.7 $\pm$ \phn0.4 & \phn3.000 $\pm$ 0.000 & 108.6 $\pm$ 22.3 & 5.575 & 4 & 11 & W19-0345 & \nodata & \nodata \\
T11-006 & 01:32:30.54 & 30:36:18.0 & 32.7 & 76.1 & 35.6 & 43.0 & -2.5 & no & no & 4.46e-14 & 2.05e-16 & \phn-0.7 $\pm$ \phn0.4 & \phn3.000 $\pm$ 0.000 & -41.9 $\pm$ 23.8 & 5.575 & 4 & 3 & W19-0974 & \nodata & \nodata \\
T11-007 & 01:32:32.84 & 30:40:27.9 & 26.3 & 23.8 & 0.9 & 12.4 & 10.5 & no & no & 4.98e-15 & 3.65e-17 & \phn21.3 $\pm$ 16.8 & -0.737 $\pm$ 1.327 & \phn16.7 $\pm$ 11.0 & 1.944 & 4 & 0 & \nodata & \nodata & \nodata \\
T11-008 & 01:32:33.76 & 30:47:26.8 & 40.7 & 25.6 & 25.4 & 5.5 & -5.3 & no & no & 5.73e-15 & 1.73e-17 & -58.8 $\pm$ 26.5 & -2.301 $\pm$ 3.352 & -53.9 $\pm$ 23.3 & 1.454 & 4 & 0 & \nodata & \nodata & \nodata \\
T11-009 & 01:32:36.35 & 30:40:45.4 & 22.8 & 13.1 & 4.5 & 7.5 & 1.1 & no & no & 3.97e-15 & 3.89e-17 & \phn-0.8 $\pm$ \phn6.4 & \phn2.241 $\pm$ 5.926 & -15.2 $\pm$ 10.0 & 5.151 & 4 & 0 & \nodata & \nodata & \nodata \\
T11-010 & 01:32:36.84 & 30:32:29.0 & 42.4 & 1424.2 & 129.9 & 628.0 & 666.2 & no & no & 1.01e-14 & 2.76e-17 & \phn32.2 $\pm$ 26.7 & -0.588 $\pm$ 1.204 & \phn25.2 $\pm$ 16.5 & 2.131 & 4 & 11 & W19-1897 & \nodata & \nodata \\
T11-011 & 01:32:40.62 & 30:37:21.3 & 21.1 & 17.5 & 7.5 & 10.6 & -0.6 & no & yes & 2.47e-15 & 2.72e-17 & \phn86.5 $\pm$ 17.1 & -0.960 $\pm$ 0.382 & \phn65.9 $\pm$ 10.9 & 1.858 & 4 & 0 & \nodata & \nodata & \nodata \\
T11-012 & 01:32:40.87 & 30:35:50.4 & 25.1 & 68.2 & 14.8 & 44.8 & 8.5 & no & no & 2.96e-15 & 2.33e-17 & \phn22.8 $\pm$ 15.8 & -3.000 $\pm$ 0.000 & \phn24.0 $\pm$ 16.6 & 1.378 & 4 & 0 & \nodata & \nodata & \nodata \\
T11-013 & 01:32:41.33 & 30:32:17.7 & 38.8 & 101.9 & 5.7 & 40.7 & 55.5 & no & no & 6.74e-15 & 2.23e-17 & -10.7 $\pm$ 20.9 & \phn0.916 $\pm$ 1.565 & -32.4 $\pm$ 15.9 & 4.719 & 4 & 0 & \nodata & \nodata & \nodata \\
T11-014 & 01:32:42.07 & 30:33:29.0 & 34.9 & 89.9 & 12.1 & 47.4 & 30.3 & no & yes & 8.37e-15 & 3.35e-17 & 228.9 $\pm$ 16.2 & \phn0.731 $\pm$ 0.060 & 505.0 $\pm$ 13.4 & 4.135 & 4 & 11 & W19-0115 & \nodata & \nodata \\
T11-015 & 01:32:42.46 & 30:48:15.1 & 39.6 & 117.4 & 22.5 & 64.2 & 30.7 & no & yes & 5.36e-15 & 1.71e-17 & \phn47.9 $\pm$ 21.7 & \phn0.576 $\pm$ 0.379 & \phn91.1 $\pm$ 14.7 & 4.272 & 4 & 11 & W19-1286 & \nodata & \nodata \\
T11-016 & 01:32:43.41 & 30:35:06.2 & 25.4 & 396.3 & 53.9 & 205.9 & 136.6 & no & yes & 9.08e-15 & 6.89e-17 & \phn15.4 $\pm$ 14.7 & \phn0.815 $\pm$ 0.800 & \phn38.5 $\pm$ 12.1 & 4.313 & 4 & 9 & W19-0521 & W19-0919 & \nodata \\
T11-017 & 01:32:44.17 & 30:35:59.7 & 22.1 & 20.6 & 0.4 & 10.8 & 9.3 & no & yes & 4.49e-14 & 4.30e-16 & \phn49.4 $\pm$ 12.6 & \phn0.442 $\pm$ 0.230 & \phn76.2 $\pm$ \phn9.2 & 3.739 & 4 & 11 & W19-1301 & \nodata & \nodata \\
T11-018 & 01:32:44.77 & 30:30:37.4 & 33.8 & 35.9 & 4.9 & 20.9 & 10.1 & no & yes & 2.65e-15 & 1.14e-17 & \phn87.6 $\pm$ 18.8 & -0.127 $\pm$ 0.217 & \phn79.1 $\pm$ \phn9.8 & 3.136 & 4 & 11 & W19-1200 & \nodata & \nodata \\
T11-019 & 01:32:45.06 & 30:39:11.3 & 16.7 & 26.2 & 7.3 & 11.1 & 7.8 & no & yes & 1.30e-13 & 2.20e-15 & 263.5 $\pm$ 11.0 & \phn0.263 $\pm$ 0.040 & 330.7 $\pm$ \phn7.7 & 3.313 & 4 & 11 & W19-0193 & \nodata & \nodata \\
T11-020 & 01:32:46.73 & 30:34:37.7 & 27.0 & 49.6 & 32.5 & 11.8 & 5.3 & L10-005 & yes & 1.86e-14 & 1.24e-16 & 319.6 $\pm$ 18.6 & -0.478 $\pm$ 0.089 & 263.2 $\pm$ 12.0 & 2.102 & 4 & 11 & W19-0478 & \nodata & \nodata
\\
\enddata
\end{splitdeluxetable}

\clearpage

\begin{deluxetable}{lcl}
\tablecaption{Radio Source Catalog Column Descriptions\label{tab:radio_cat_columns}}
\tablehead{\colhead{Column Name} & \colhead{Units} & \colhead{Description} }
% \tabletypesize{\scriptsize}
\tablewidth{0pt}
\startdata
Name         &        & Name of source (= W19-island number) \\
RA           & hh:mm:ss.sss & J2000 RA (flux-weighted centroid) \\
Dec          & ${+}$dd:mm:ss.ss & J2000 Declination (flux-weighted centroid \\
RAPeak       & deg    & J2000 RA (peak) \\
DecPeak      & deg    & J2000 Declination (peak) \\
Wrn          &        & W indicates source is below detection threshold\tablenotemark{a} \\
$H\alpha(\mathrm{tot})$ & $\mathrm{erg\, cm^{-2} s^{-1}}$ & Integrated H-alpha flux \\
$H\alpha(\mathrm{ave})$ & $\mathrm{erg\, cm^{-2} s^{-1} arcsec^{-2}}$ & H-alpha surface brightness \\
% $H\alpha(\mathrm{tot})$ & \pbox{5em}{\centering $\mathrm{erg}\,\mathrm{cm}^{-2}$\break$\mathrm{s}^{-1}$} & Integrated H-alpha flux \\
% $H\alpha(\mathrm{ave})$ & \pbox{5em}{\centering $\mathrm{erg}\,\mathrm{cm}^{-2}$\break$\mathrm{s}^{-1}\mathrm{arcsec}^{-2}$} & H-alpha surface brightness \\
F1.4GHz      & \uJy   & Integrated flux at frequency 1.4 GHz, $\Fint$ \\
F1.4Err      & \uJy   & Error on F1.4GHz \\
Spind        &        & Spectral index $\alpha$, $F_\nu = \Fint (\nu/\nu_0)^\alpha$ (clipped to range -3 to 3) \\
SpErr        &        & Error on Spind (zero for points at Spind clip limits) \\
Fint         & \uJy   & Flux at pivot frequency pFreq, $\Fint$ \\
FiErr        & \uJy   & Error on Fint \\
pFreq        & GHz    & Pivot frequency $\nu_0$ where signal-to-noise is maximized \\
Fpeak        & \uJy   & Peak pixel in island from detection image, $F_p$ \\
FpErr        & \uJy   & Noise in Fpeak \\
Major        & arcsec & Major axis FWHM (includes Gaussian beam with FWHM = 5.9 arcsec) \\
Minor        & arcsec & Minor axis FWHM (includes Gaussian beam with FWHM = 5.9 arcsec) \\
PA           & deg    & Position angle of major axis \\
nBands       &        & Number of frequency bands with data (1 to 4 of 1.3775, 1.8135, 4.679, 5.575 GHz) \\
Island       &        & Island number \\
Sflag        &        & SNR detection flag\tablenotemark{b} \\
SNRname[1-3] &        & Names of associated SNR(s) in \cite{long18}\tablenotemark{c} \\
nXray        &        & Number of associated X-ray sources in \cite{tuellmann11} \\
nHII         &        & Number of associated H II regions in \cite{lin17} \\
\enddata
\tablecomments{Column descriptions for Table~\ref{tab:radio_cat}.}
\tablenotetext{a}{2868 sources are detected at $4\sigma$ or greater.
7 sources marked as 'W' are below the detection threshold but have SNR, HII or X-ray counterparts.}
\tablenotetext{b}{The detection flag (Sflag) is a bit flag where the bits indicate
whether the association between the radio source and the SNR is unambiguous:
1 = this source may have a match in the SNR catalog;
2 = unambiguous match: this source matches only one object from the SNR catalog;
8 = mutually good match: this source is best for the other object, and the other object is best for this source.
The most reliable matches will have flag bit 8 set.  All of those will have bit 1 set as well, and most of them
will also have bit 2 set.
See Table~\ref{tab:flags} for more information.}
\tablenotetext{c}{Multiple SNR matches are listed in order of decreasing island overlap.}
\end{deluxetable}

\clearpage

\begin{deluxetable}{lcl}
\tablecaption{SNR and X-ray Forced Photometry Catalog Column Descriptions\label{tab:forced_columns}}
\tablehead{\colhead{Column Name} & \colhead{Units} & \colhead{Description} }
% \tabletypesize{\scriptsize}
\tablewidth{0pt}
\startdata
Name         &        & Name of SNR \citep{long18} or X-ray source \citep{tuellmann11}.\\
RA           & hh:mm:ss.sss & J2000 RA from external catalog \\
Dec          & ${+}$dd:mm:ss.ss & J2000 Declination from external catalog \\
$\rho$       & kpc    & Galactocentric distance \\
$D$     	 & pc    & Diameter \\
Opt          &        & yes means SNR has \sii:\ha\ $> 0.4$ \\
Xray         &        & yes means SNR has XMM or Chandra X-ray detection \\
Radio        &        & yes means radio flux is detected $>3\sigma$ \\
$H\alpha(\mathrm{tot})$ & $\mathrm{erg\, cm^{-2} s^{-1}}$ & Integrated H-alpha flux \\
$H\alpha(\mathrm{ave})$ & $\mathrm{erg\, cm^{-2} s^{-1} arcsec^{-2}}$ & H-alpha surface brightness \\
F1.4GHz      & \uJy   & Integrated flux at frequency 1.4 GHz, $\Fint$ \\
F1.4Err      & \uJy   & Error on F1.4GHz \\
Spind        &        & Spectral index $\alpha$, $F_\nu = \Fint (\nu/\nu_0)^\alpha$ (clipped to range -3 to 3) \\
SpErr        &        & Error on Spind (zero for points at Spind clip limits) \\
Fint         & \uJy   & Flux at pivot frequency pFreq, $\Fint$ \\
FiErr        & \uJy   & Error on Fint \\
pFreq        & GHz    & Pivot frequency $\nu_0$ where signal-to-noise is maximized \\
nBands       &        & Number of frequency bands with data (1 to 4 of 1.3775, 1.8135, 4.679, 5.575 GHz) \\
Rflag        &        & Radio detection flag\tablenotemark{a} \\
Radname[1-3] &        & Names of associated radio source(s)\tablenotemark{b} \\
$C_{tot}$    & cts/s  & Chandra X-ray total count rate \\
$C_{soft}$   & cts/s  & Chandra X-ray count rate in soft band (0.35--1.0 keV)\\
$C_{med}$    & cts/s  & Chandra X-ray count rate in medium band (1.0--2.0 keV)\\
$C_{hard}$   & cts/s  & Chandra X-ray count rate in hard band (2.0--8.0 keV)\\
SNR          &        & yes means X-ray source has match in SNR catalog \\
\enddata
\tablecomments{Column descriptions for Table~\ref{tab:forced} and~\ref{tab:forced_xray}.}
\tablenotetext{a}{The detection flag (Rflag) is a bit flag where the bits indicate
whether the association between the radio source and the SNR or X-ray source is unambiguous:
1 = this source may have a match in the radio catalog;
2 = unambiguous match: this source matches only one object from the radio catalog;
8 = mutually good match: this source is best for the other object, and the other object is best for this source.
The most reliable matches will have flag bit 8 set.  All of those will have bit 1 set as well, and most of them
will also have bit 2 set.
See Table~\ref{tab:flags} for more information.}
\tablenotetext{b}{Multiple radio matches are listed in order of decreasing island overlap.}
\end{deluxetable}

\clearpage



\end{document}